\documentclass[lettersize,journal]{IEEEtran}
\usepackage{amsmath,amsfonts}
\usepackage{algorithmic}
\usepackage{algorithm}
\usepackage{array}
\usepackage[caption=false,font=normalsize,labelfont=sf,textfont=sf]{subfig}
\usepackage{textcomp}
\usepackage{stfloats}
\usepackage{url}
\usepackage{verbatim}
\usepackage{graphicx}
\usepackage{multirow}
\usepackage{cite}
\usepackage{xcolor}
\usepackage{arydshln}
\usepackage {booktabs}
\definecolor{EMgray}{gray}{0.45}

\usepackage{hyperref}

\def\BibTeX{{\rm B\kern-.05em{\sc i\kern-.025em b}\kern-.08em
    T\kern-.1667em\lower.7ex\hbox{E}\kern-.125emX}}

\def\orcid#1{\kern .08em\href{https://orcid.org/#1}{\includegraphics[bb=0 0 128 128,keepaspectratio,width=0.8em]{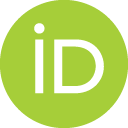}}}

\begin{document}

%\title{Untitled}
\title{Masked Modeling Duo: Towards a Universal Audio Pre-training Framework}

\author{Daisuke~Niizumi\orcid{0000-0002-5063-0508}, Daiki~Takeuchi, Yasunori~Ohishi\orcid{0000-0002-7856-248X},~\IEEEmembership{Member,~IEEE,} Noboru~Harada\orcid{0000-0002-1759-4533},~\IEEEmembership{Senior~Member,~IEEE,} and~Kunio~Kashino,~\IEEEmembership{Senior~Member,~IEEE}
        % <-this % stops a space
\thanks{Manuscript received December 21, 2021; revised September 18, 2022.}%
\thanks{The authors are with Communication Science  Laboratories, Nippon Telegraph and Telephone Corporation,  Atsugi 243-0198, Japan (e-mail: daisuke.niizumi@ntt.com; d.takeuchi@ntt.com; yasunori.ohishi@ntt.com; harada.noboru@ntt.com; kunio.kashino@ntt.com)}
}

\markboth{Journal of \LaTeX\ Class Files,~Vol.~14, No.~8, August~2021}%
{Shell \MakeLowercase{\textit{et al.}}: A Sample Article Using IEEEtran.cls for IEEE Journals}

\IEEEpubid{0000--0000/00\$00.00~\copyright~2021 IEEE}

\maketitle

\begin{abstract}
%マスク予測による自己教師あり学習手法により、汎用音響信号表現は大きな進歩を遂げた。
%Self-supervised learning methods based on masked prediction have significantly advanced general-purpose audio representation.
%マスク予測による自己教師あり学習により、特にMasked Autoencoders (MAE)以降、汎用音響信号表現が飛躍的に進歩した。MAEはマスクされた入力信号の復元タスクを通じて効果的に学習する一方、復元に使うことで表現を間接的に学習する。
%MAE effectively learns by reconstructing the masked input signal, thus indirectly learning the representation through the reconstruction.
%本研究はマスク予測による事前学習を改善する手法Masked Modeling Duo (M2D)を提案する。M2Dは教師信号となるマスクされた入力信号の表現を予測することにより学習する。従来と異なり、マスク部分だけ表現にエンコードして教師信号を得るため、構成する２つのネットワークの両方に入力信号のモデル化を促すことが特徴である。
%M2Dは汎用表現の性能を向上させる一方、実際の応用では私有データ専用表現が必要だと考える。産業分野や医療分野など実際の応用タスクでは、データは非公開所有であることが多い。例えば工場で稼働する工作機械の音は、汎用表現に利用される大規模データセットのデータ分布とは通常異なる。このため、汎用表現を適用できる範囲には限界があり、特に実応用を考慮するときに顕著であると考えられる。
%For example, a machine tool operating sound in a factory typically differs from the data distribution of a large dataset used for general-purpose representation.
%そこで、本研究はM2Dを拡張し応用タスク専用表現の事前学習を可能にするM2D for X (M2D-X)を提案する。M2D-Xは入力データに背景ノイズを導入し、ノイズ除去と蒸留タスクをM2Dのマスク予測タスクと組み合わせる。これらによりドメイン知識からの学習や小さなデータの事前学習に対応し、そのため幅広いタスク専用表現の事前学習を可能にする。
%我々は3つの観点から評価し、汎用音響信号表現、競争の激しい音声分野専用の表現、小さなデータ専用表現、全てstate-of-the-art性能が得られることを確認した。これらにより、提案法は汎用音の事前学習フレームワークとしての有効性を示す。更に、公開するコードと事前学習済みモデルにより将来の応用研究に貢献する。
Self-supervised learning (SSL) using masked prediction has made great strides in general-purpose audio representation. %, especially since the advent of Masked Autoencoders.
This study proposes Masked Modeling Duo (M2D), an improved masked prediction SSL, which learns by predicting representations of masked input signals that serve as training signals. Unlike conventional methods, M2D obtains a training signal by encoding only the masked part, encouraging the two networks in M2D to model the input.
While M2D improves general-purpose audio representations, a specialized representation is essential for real-world applications, such as in industrial and medical domains. The often confidential and proprietary data in such domains is typically limited in size and has a different distribution from that in pre-training datasets. 
Therefore, we propose M2D for X (M2D-X), which extends M2D to enable the pre-training of specialized representations for an application X.
M2D-X learns from M2D and an additional task and inputs background noise. We make the additional task configurable to serve diverse applications, while the background noise helps learn on small data and forms a denoising task that makes representation robust. With these design choices, M2D-X should learn a representation specialized to serve various application needs.
Our experiments confirmed that the representations for general-purpose audio, specialized for the highly competitive AudioSet and speech domain, and a small-data medical task achieve top-level performance, demonstrating the potential of using our models as a universal audio pre-training framework.
Our code is available online for future studies.
\end{abstract}
%\footnote{\url{https://github.com/nttcslab/m2d}}


\section{Introduction}
%マスク予測による自己教師あり学習手法は飛躍的な進歩を遂げ、自然言語や画像分野で高性能なモデルの事前学習を実現した[BERT,RoBERTa,画像SSL, MAE]。これらは入力信号の非マスク部分をエンコードした表現からマスクした部分を予測するタスクを通じて事前学習する。マスク部分をより精度よく予測するために、非マスク部分から効率よく入力信号の情報を表現にエンコードすることが必要となるため、従って入力信号のモデル化が効果的に促される。
\IEEEPARstart{S}ELF-SUPERVISED learning (SSL) methods based on masked prediction have made great strides, enabling pre-training of high-performance models in the natural language processing (NLP) and image domains\cite{bert,liu2019roberta,he2022masked,tao2022MIM:SIM,assran2022MIM:MSN,chen2022MIM:CAE,elnouby2021MIM:SplitMask}.
They are pre-trained through the task of predicting masked portions from representations that encode unmasked (visible) portions of the input signal.
To better predict the masked portions, the information about the input signal from the unmasked portions should be efficiently encoded into a representation to effectively encourage the modeling of the input signal.

%音声分野では、幅広い音声タスクで性能を発揮する音声自己教師あり学習モデルが多数提案されている[wav2vec2.0, HuBERT, WavLM, XLSR]。音響分野では、音声に加えて環境音や音楽を含む多様な音のタスクで有効なモデルが提案されている[SSAST, MAE-AST, MSM-MAE, M2D他]。
In the speech domain, a number of masked prediction-based speech SSL models have been proposed that perform well on a wide range of speech tasks\cite{baevski2020wav2vec2,Hsu2021HuBERT,conneau2021xlsr,Chen2022WavLM}. In the audio domain, many SSL models have been proposed that perform effectively in diverse audio tasks, including environmental sounds and music in addition to speech \cite{gong2022ssast,huang2022audiomae,Baade2022MAE-AST,chong2022maskspec,niizumi2022msm-mae}.

%特に、画像分野のMasked Autoencoders (MAE)は大きな流れを巻き起こし、音響信号分野ではMAEをベースにした手法が多数続いた[MAE-AST他]。MAEは入力信号の大部分(75%)をマスクし、残り少量の可視部分(25%)をエンコードした表現を使って、マスク部分を復元する。入力信号の25%から効率よく情報をvision transformer (ViT)を使って抽出することで、効果的な表現を学習する。しかしながらMAEは復元誤差を用いて学習するため、表現の最適化を間接的に行うと考えられる。
In particular, masked autoencoders (MAE) in the image domain sparked a major trend, followed by a number of MAE-based methods in the audio domain \cite{huang2022audiomae,Baade2022MAE-AST,chong2022maskspec,niizumi2022msm-mae}.
MAE masks a large portion (75\%) of the input signal and reconstructs the masked portion using an encoded representation of the remaining small amount (25\%) of the visible portion.
It learns an effective representation by efficiently extracting information from the 25\% of the input signal using a vision transformer (ViT)\cite{ViT}.
However, MAE learns from reconstruction errors and is thus considered to optimize the representation indirectly.

%我々は損失の計算に表現を巻き込むことで、より表現を効果的に最適化できると考える。入力信号を表現にエンコードし、そのうちマスク部分を教師信号として利用する手法[data2vec,画像SSL該当手法]は、このことを実現している。しかしながら、これらの手法は入力信号全体をエンコードしているため、得られる教師信号の表現には入力のモデル化が促されないと考えられる。なぜなら、予測のためのソース（可視）とターゲット（マスク）の情報が利用可能である限り、トレーニング信号は信号をモデル化するのではなく、予測の答えを汚染する可能性があるからだ。
We believe that representations can be optimized more effectively by involving them in the loss calculation. Methods\cite{tao2022MIM:SIM,assran2022MIM:MSN,baevski2022data2vec} that encode the input into a representation and use the masked portion of it as a training signal accomplish this.
However, since these methods encode the entire input signal, the resulting training signal representation is not always encouraged to model the input. This is because if both the source (visible) and target (masked) information for prediction is available, it may be easier for the training signal to contaminate the target information than to model the signal.

%我々の仮説として、入力信号全体ではなくマスク部分のみを利用してエンコードすることで、その表現に入力のモデル化が促されやすいと考える。例えば2つの要素(S1とS2)で構成される心音を入力信号とするとき、S1全体がマスクされる場合を考える。このとき、マスクされていないS2部分の表現がS2の情報しか含まなければS1部分の予測が困難になる一方、心音全体の一部としてモデル化された情報であれば、S1部分を予測しやすいと考える。一方で、教師信号となるマスク部分のS1は、心音全体としてのエンコードされた表現であれば予測に一致しやすいと考える。
We hypothesize that encoding only the masked portion of the input signal, rather than the entire signal, would encourage the representation to model the input. For example, consider a heart sound consisting of two elements (S1 and S2) as an input signal, where the entire S1 is masked. In this case, if the representation of the unmasked S2 contains only S2 information, it will be difficult to predict S1, but if the information is modeled as part of the whole heart sound, the prediction will be easy. On the other hand, the masked S1 part representation, which is the training signal, is more likely to match the prediction if it is encoded as part of the whole heart sound.

\IEEEpubidadjcol

\begin{figure}[tbp]
  \vspace{-5pt}
  \centering
  \includegraphics[width=0.99\columnwidth]{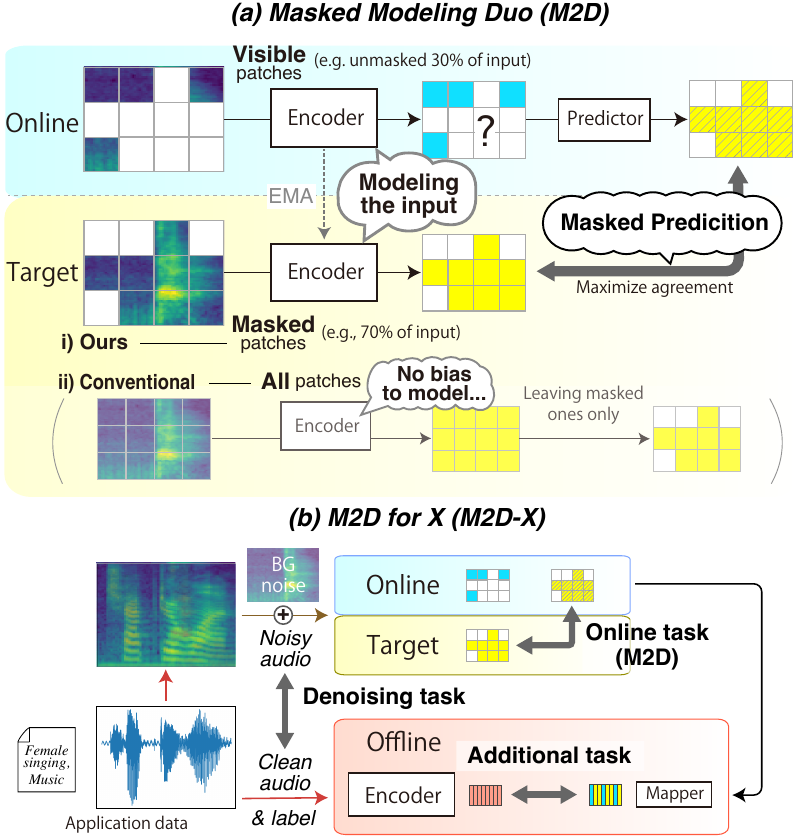}
  \vspace{-10pt}
  %(a) M2D overview: ランダムにマスクした入力信号のうち、Offline networkは可視部分を、Target networkはマスク部分をエンコードする。M2DはOnline表現を使ったTarget表現の予測タスクを通じて、両方の表現に入力信号のモデル化を促す。(b)M2D-X overview: 背景ノイズとofflineネットワークを追加することで、M2D-Xはドメイン知識の学習と少データでの学習を可能にし、幅広いタスク専用表現の事前学習を可能にする。
  \caption{(a) M2D overview. The online network encodes the visible patches of the randomly masked input signal, while the target network encodes the masked patches. M2D encourages both representations to model the input signal through a task to predict the target representation using the online representation. (b) M2D-X overview. Introducing background noise and offline network allows M2D-X to learn from M2D, denoising, and an additional task, enabling pre-training of specialized representations for diverse applications.}
  \label{fig:overview}
  \vspace{-5pt}
\end{figure}

%我々は仮説を実現する新たな手法Masked Modeling Duo (M2D)を提案する。M2Dは入力信号を分割したマスク部分、非マスク部分それぞれを別々にエンコードすることで、両方の表現に入力信号のモデル化を促す。またこうすることで、教師信号となるマスク部分の表現には、予測に利用する非マスク部分と直接的な情報の重複がなく、マスク予測タスクがより困難になる。ViTをエンコーダーに利用するなど、M2Dはその他MAEの構成を踏襲する。我々の提案はシンプルだが、M2DはMAEを改善し、より効果的に表現を学習すると考える。
We propose a new method, Masked Modeling Duo (M2D), that implements our hypothesis by encoding the masked and unmasked portions of the input signal separately, thereby encouraging both representations to model the input signal [Fig. \ref{fig:overview}(a)]. M2D also facilitates the task of masked prediction since the representation of the masked part, which serves as the training signal, has no direct information overlap with the unmasked part used for prediction. M2D follows other MAE configurations, such as using ViT as the encoder. Although our proposed method is simple, it should provide an improvement over MAE and learn representations more effectively.

%本研究は現実的な応用を考慮し、下流タスク専用の表現の事前学習までスコープとする。従来の手法は、大規模データセットを事前学習した汎用表現を多様なタスクで構成されるベンチマークで評価する[AST,BYOL-A,Universal,BEATs]。一般的には汎用表現をfine-tuneして下流タスクを解く一方で、タスクのデータ分布が事前学習と異なると十分な性能が得られない。実際な応用では、例えば産業分野や医療分野などにおいて私有データが使われることが多く、大規模データセットのデータ分布に近いことは期待できない。一方で、応用タスクと同じドメインのデータやタスクのデータで学習することが有効であることが知られており[NTT]、我々はこれが現実的な解決策と考える。
Considering realistic applications, the scope of this study extends to the pre-training of representations specialized for application tasks.
Conventional methods evaluate a general-purpose audio representation pre-trained on a large dataset with benchmarks of various tasks\cite{gong2021ast,niizumi2023byol-a,chen2022beats}. While a general-purpose representation is usually fine-tuned to solve the application tasks, the performance can be insufficient if the data distribution of the task differs from that of the pre-training. For example, private and confidential data are often used in the industrial and medical fields, and we cannot expect the data distribution of such data to be close to that of a public pre-training dataset. On the other hand, previous studies\cite{Junyi2023Parameter,Ashihara2023JPSSL} showed that training on the task data or on data from the same domain is effective, and we consider such training a practical solution.

%そこで我々は、応用タスクの専用表現を学習するため、M2Dを拡張したM2D for X (M2D-X)を提案する(Fig. XX)。十分な性能を得るには、ラベルなどの情報やドメインテクニックから学習する必要があるかもしれない。一方で、特に産業や医療分野のようにデータサイズが限られるとき、マスク予測による学習が困難になる場合が考えられる[x]。M2D-Xはドメインに特化した追加学習タスクをM2Dにアドオンし、背景ノイズを導入することでデノイズタスクも追加するマルチタスク学習フレームワークである。追加タスクとして、アプリケーションラベルの教師あり学習、ドメインモデルの蒸留、またはM2Dの過学習を抑える正則化など、様々なアプリケーションのニーズに応えられる。背景ノイズはデータ拡張効果で小さなデータの学習を可能にし、デノイズの学習に繋がることで表現を頑健にして性能向上を助ける。結果として、M2D-Xは多様な音のタスクに利用できる、ユニバーサルな音の事前学習フレームワークとして機能すると考えられる。
Therefore, we propose M2D for X (M2D-X), an extension of M2D, to learn specialized representations of application tasks [Fig. \ref{fig:overview}(b)].
The application task may require learning from information such as labels and domain techniques to gain sufficient performance. In addition, the application data may be small, as in the industrial and medical domains, making learning by masked prediction challenging\cite{Xie2023OnDataScalingMasked,zhang2023mae-survey}.
M2D-X is a multitask learning framework that adds an additional domain-specific learning task and a denoising task using background noise to M2D.
To accommodate various application demands, we leave freedom with the choice of additional learning tasks, such as supervised learning of application labels, distillation of domain models, or regularizing M2D from overfitting.
Background noise helps in learning from small data with data augmentation effects and forms a denoising task that makes the representation robust.
As a result, M2D-X should serve as a universal audio pre-training framework that addresses the diverse needs of audio applications.

%大規模データセットによる一般の音、競争の激しい音声、小規模データの医療アプリケーション、これら3つのシナリオによる実験で、M2Dはトップレベルの性能を示す表現を学習することを示し、M2D-Xは表現を分野に特化して性能を伸ばすことを示した。特に小さなデータによる実験結果は、産業用途などで大規模データの取得が困難であったり、かつ一般のデータ分布と大きく異なる場合でも、M2D-Xが効果的な事前学習を可能にすることを示した。
Experiments with three settings--general audio with large-scale datasets, competitive speech representation, and medical applications with small data--confirmed that M2D and M2D-X achieve top-level performance, demonstrating their potential to serve various applications as a pre-training framework. We find that encoding only the masked parts of the input to obtain a training signal improves masked prediction-based SSL, that combining supervised learning with M2D SSL enhances specialized performance on top of the general-purpose performance, and that M2D-X configured for regularization using background noise enables successful further pre-training on small data.
% 貢献は以下の通りである:
Our contributions include:

%- マスク予測を改善した新たな自己教師あり学習M2Dの提案。M2Dは最先端性能の汎用音響信号表現を事前学習できることを示す。
%- M2Dを拡張したM2D-Xを提案、ドメインやタスク専用の表現の事前学習を可能にすることを示す。音響分野では、応用データを使った事前学習についての研究は、我々の知る限り、これまで見られない。
%- M2D-Xによる少データの表現学習は、背景ノイズの利用と蒸留タスクにより可能になっていることを示す。
%- 少データの表現学習を少リソース・計算機環境設定で行う現実的な例を提供する。
\begin{itemize}
\item We propose a new masked prediction-based SSL, M2D, and show that it learns general-purpose audio representations with top-level performance.
\item We propose M2D-X that extends M2D and succeeds in learning representations specialized to applications with top-level performance. To our knowledge, this is the first study to conduct pre-training on application data in addition to general-purpose data in the audio domain.
%\item Demonstrate that the use of background noise and the distillation task in M2D-X enable pre-training on small data.
\item We provide a realistic example of small data under a restricted data/computational resource setting with our code and the pre-trained weights for future studies\footnote{\url{https://github.com/nttcslab/m2d}}.
\end{itemize}

%Fig. YYは本論文で想定する3つの事前学習シナリオを示す。(i)は大規模データセットから汎用音響表現を学習する場合、(ii)は応用ドメインやタスクの十分なデータで専用表現を学習する場合、(iii)は応用タスクの小さいデータで汎用表現をfurther pre-trainingして専用表現を学習する場合である。
%Fig. \ref{fig:strategy} shows the three pre-training scenarios we assume in this paper. We validate our methods in these scenarios in the experiments.

%%%%%%%%%%%%%%%%%%%%%%%%%%%%%%%%%%%%%%%%%%%%%%%%%%%%%%%%%%%%%%%%%%%%%%%
%%%%%%%%%%%%%%%%%%%%%%%%%%%%%%  SECTION  %%%%%%%%%%%%%%%%%%%%%%%%%%%%%%
\section{Related Work}

%%%%%%%%%%%%%%%%%%%%% SUB SECTION %%%%%%%%%%%%%%%%%%%%%%
\subsection{Relationship with our previous work}
We introduced M2D\cite{niizumi2022M2D} and M2D for Speech (M2D-S)\cite{niizumi2023m2d4speech} in our previous work for learning general-purpose audio and speech representations.
This paper redefines the M2D family and explores its potential as an application-agnostic audio pre-training framework.

%我々はM2Dを改良し、fine-tuningでvariable-length audioを取り扱えるようにpositional encodingのinterpolationを導入(Section XX)する。またGeneral-purpose audio representationの実験を改善し、表IIに記載するfine-tuningのパラメータとしてSpecAugment、Adjust positional encoding、Freeze embedding layerを学習安定性を目的として導入する(Section XXX)。
We refine M2D and introduce positional encoding interpolation (Section \ref{sec:method:xfer-model-4-audio}) to handle variable-length audio in fine-tuning. We also improve and stabilize the experiments by using the statistics of the pre-training dataset (e.g., AudioSet\cite{gemmeke2017audioset}) for standardizing spectrograms instead of that of downstream tasks, and by introducing fine-tuning parameters SpecAugment\cite{specaugment}, Adjust positional encoding, and Freeze embedding layer as listed in Table \ref{tab:exp-general:ft-parameters} (Section \ref{sec:eval-general}).
%我々はM2D-Sをapplication-agnostic frameworkに拡張し、M2D for X (M2D-X)と再定義する(Section XX)。M2D-XはM2D-SのOffline networkを、様々なアプリケーションニーズに応えるため任意のタスクによる学習を可能にするように拡張する。これにより、M2D-Xは幅広いタスクに専用音響表現の事前学習を行うためのフレームワークとして提案する。
We extend M2D-S to an application-agnostic framework and redefine it as M2D for X (M2D-X) (Section \ref{sec:method:m2d-x}). M2D-X redesigns the M2D-S offline network, making it possible to configure an additional learning task to meet various application needs. %We thus propose M2D-X as a framework for pre-training audio representations specialized for desired application tasks.

%%%%%%%%%%%%%%%%%%%%% SUB SECTION %%%%%%%%%%%%%%%%%%%%%%
\vspace{-0.3cm}
\subsection{Background: Image SSL}
%本研究は効果的なマスク予測手法としてMasked Autoencoders (MAE)に、また、ターゲットネットワークを使い潜在表現を直接学習するフレームワークとしてBootstrap Your Own Latent (BYOL) にそれぞれインスパイアされた。MAEは入力データの再構成タスクで学習するのに対して、本研究はマスクされた潜在表現の予測タスクで学習する。BYOLはデータ拡張に不変な表現の学習を行うフレームワークであり、それに対して我々はデータ拡張を利用しない。
Our M2D was inspired by MAE\cite{he2022masked} for an effective masked prediction-based SSL and Bootstrap Your Own Latent\cite{grill2020byol} (BYOL) as a framework for directly learning latent representations using a target network.
MAE learns to reconstruct the input data, whereas our M2D learns to predict masked latents, and BYOL does not learn the masked prediction task.
%BYOL learns representations invariant to data augmentation, unlike M2D.

%ターゲットネットワークを利用する方法
SIM\cite{tao2022MIM:SIM}, MSN\cite{assran2022MIM:MSN}, and data2vec\cite{baevski2022data2vec} learn to predict masked patch representations using a target network, but, unlike ours, the target network inputs all patches.
CAE\cite{chen2022MIM:CAE} and SplitMask\cite{elnouby2021MIM:SplitMask} encode target representations using only masked patches like ours does, but without the use of a target network.
All methods above learn image representations, and data2vec also learns speech and audio representations.

%%%%%%%%%%%%%%%%%%%%% SUB SECTION %%%%%%%%%%%%%%%%%%%%%%
\vspace{-0.3cm}
\subsection{Speech and Audio SSL using Transformer}
%masked predictionでtransformerを学習させる自己教師あり学習モデルは、様々な分野で有望な性能を示している。進歩を遂げている音声表現モデルのうち、spectrogramを入力とするものにはMockingjayやTERAが挙げられ、TERAは時間ステップに加えて周波数ビンもマスクするmasking strategyで性能を伸ばした。特に一方入力音声波形からの音響特徴抽出も学習させるwav2vec2、data2vec， HuBERT、WavLMなどのモデルは、ベクトル量子化やクラスタリングを用いて離散化した疑似ラベルを利用することで、SOTA性能を更新してきた。特にWavLMはmasked speech denoisingも行うことでASR以外のタスクでも性能を伸ばした。
SSL methods that train transformers with masked prediction have shown promising performance in various domains.
Speech representation models, such as Mockingjay\cite{Liu2020Mockingjay} and TERA\cite{Liu2021TERA}, take spectrograms as input. Similar to general-purpose models, TERA employs a masking strategy for splitting both frequency bins and time steps.
SOTA models, such as wav2vec2.0\cite{baevski2020wav2vec2}, BigSSL\cite{zhang2022bigssl}, data2vec\cite{baevski2022data2vec}, HuBERT\cite{Hsu2021HuBERT}, and WavLM\cite{Chen2022WavLM}, take speech waveforms as input to learn representations. Notably, models such as wav2vec2.0, HuBERT, and WavLM effectively learn by using discretized pseudo-labels with vector quantization or clustering of pre-trained model features. %In addition, WavLM has improved performance on non-ASR tasks through additional masked speech denoising.
% HuBERTやWavLMは事前学習モデルの表現をクラスタリングすることで教師信号を得る。
%我々のM2D-Xに近い手法として、MT4SSLはネットワーク構成が、DistillHuBERTはHuBERTの蒸留という点で、RobustDistillerは蒸留とノイズ除去を行う点が似ている。これらは音声表現を学習する。
Methods similar to our M2D-X include MT4SSL\cite{ma2022mt4ssl} and M-CTRL\cite{Choi2023m-ctrl} for network configuration, DistillHuBERT\cite{DistilHuBERT} for distillation, WavLM and Wang et al.\cite{Wang2022NoiseContrastive}
%and Zhu et al.\cite{Zhu2023JointSESSL}
 for denoising, and RobustDistiller\cite{guimares2023robustdistiller} for distillation and denoising; these methods learn speech representations.

%汎用音響表現としてSSAST，MAE-AST，MSM-MAE，AudioMAE，Masked Modeling Duo, BEATsなどが挙げられ、SOTA性能を示した。これらは典型的に音響特徴として入力するspectrogramを時間周波数軸両方でパッチに分割し、Vision Transformer (ViT)を学習させる。また、これらはドメインに特化したテクニックを利用しないことも特徴である。例えばmasking strategyについて音声表現では連続する時間ステップをマスクするのに対して、入力信号不可知としてランダムにマスクする。
General-purpose representation models, such as SSAST\cite{gong2022ssast}, ATST\cite{Li2022ATST}, MAE-AST\cite{Baade2022MAE-AST}, MaskSpec\cite{chong2022maskspec}, MSM-MAE\cite{niizumi2022msm-mae}, Audio-MAE\cite{huang2022audiomae}, and BEATs\cite{chen2022beats}, have shown SOTA performance.
They typically take spectrograms as acoustic feature input, split input on both the time and frequency axes, and train a Vision Transformer (ViT)\cite{ViT}, but differ from ours in that they do not learn from predicting the representations encoded from masked patches only.
%音を学習するその他のSSL手法にはAAとBBが挙げられ、特にCC、DD、EEはBYOLを学習フレームワークとして学習する。
Other audio SSL methods using non-ViT encoders, e.g., BYOL-A\cite{niizumi2023byol-a}, Wang et al.\cite{wang2022universal}, and DeLoRes\cite{Ghosh2022DeLoRes} do not learn from masked prediction.

%事前学習済みモデルを使って疑似ラベルを作るまたはモデルを蒸留する研究にはHuBERT、WavLM、MT4SSL、DistillHuBERTが挙げられる。特にMT4SSLはネットワーク構成が近く、DistillHuBERTはHuBERTの蒸留という点が似ている。
%Previous works that created pseudo-labels or distill models include BEATs, HuBERT, WavLM, DistilHuBERT \cite{DistilHuBERT}, MT4SSL \cite{ma2022mt4ssl}, and RobustDistiller \cite{guimares2023robustdistiller}. In particular, concurrent works MT4SSL and RobustDistiller learn multitasks, similar to ours, and RobustDistiller auxiliary learns to denoise speech in addition to distillation.

%%%%%%%%%%%%%%%%%%%%% SUB SECTION %%%%%%%%%%%%%%%%%%%%%%
\vspace{-0.2cm}
\subsection{Learning Specialized Representations}

%本論文に近い先行研究として、Melms et al.とBYOL-Sは汎用音響表現BYOL-Aを、それぞれ医療、音声に特化している。
In previous work similar to that described in this paper, Melms et al. \cite{le2023lungsound} and BYOL-S \cite{scheidwasserclow2021serab} have specialized a general-purpose model BYOL-A \cite{niizumi2023byol-a} in medical and speech applications.
SSAST has adapted the patch size and pre-training dataset and compared results with speech models on SUPERB.
In the NLP domain, LIBERT \cite{lauscher-etal-2020-specializing} specializes BERT \cite{bert} using an additional task for pre-training a lexically informed BERT, a multitask learning setting similar to ours.

%目標タスクやそのドメインのデータを使ったFurther pre-trainingはNLP分野で研究され、Sun et al.は目標タスクのデータが小さい場合タスクのドメインのデータを使ったfurther pre-trainingが有効であると報告する一方、Zhu et al.は目標タスクのデータが大きい場合further pre-trainingは必ずしも有効ではないと報告している。画像分野ではLee et al.がMAEを使ってfurther pre-trainingしており、overfittingを避けるため蒸留を組み合わせる構成が我々のM2D-Xに近い。一方で我々は背景ノイズとノイズ除去タスクを導入し、音のfurther pre-trainingを行う。
Further pre-training using data from the target task or its domain has been studied in the NLP domain; note that it is different from continuous learning, where a model learns a new task without forgetting the previously learned tasks. Sun et al.\cite{Sun2019FurPT} reported that further pre-training using data from the task domain is effective when the target task data is small. In contrast, Zhu et al.\cite{zhu-etal-2021-pre} reported that further pre-training using data from the task domain is not always effective when the target task data is sufficient.
In the image domain, Lee et al.\cite{lee2023selfdistillation} use MAE for further pre-training, and their configuration of combining distillation to avoid overfitting is similar to our M2D-X. Meanwhile, we introduce the use of background noise and a denoising task for further pre-training of sound.

%教師あり学習手法としては、AST、EAT、PaSST、HTS-ATがSOTA性能を示している。
%For supervised learning, AST\cite{gong2021ast}, EAT\cite{gazneli2022EAT}, PaSST\cite{Koutini2022passt}, and HTS-AT\cite{Chen2022HTS-AT} have shown SOTA performance.

%%%%%%%%%%%%%%%%%%%%%%%%%%%%%%%%%%%%%%%%%%%%%%%%%%%%%%%%%%%%%%%%%%%%%%%
%%%%%%%%%%%%%%%%%%%%%%%%%%%%%%  SECTION  %%%%%%%%%%%%%%%%%%%%%%%%%%%%%%
\section{Proposed Methods}

%我々は、ViTを利用するマスク予測を改善したSSL手法Masked Modeling Duo (M2D)と、望む下流タスクXに特化した表現学習を行うための手法M2D for Xを提案する。
We propose Masked Modeling Duo (M2D), an improved masked prediction SSL, and M2D for X (M2D-X), a method for learning representations specialized for an application $X$.

%%%%%%%%%%%%%%%%%%%%% SUB SECTION %%%%%%%%%%%%%%%%%%%%%%
\subsection{Masked Modeling Duo}
\begin{figure}[tbp]
  \vspace{-5pt}
  \centering
  \includegraphics[width=0.85\columnwidth]{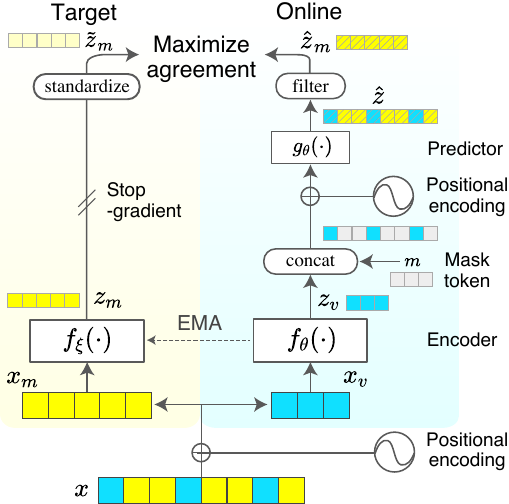}
  \vspace{-3pt}
  \caption{M2D pre-training flow. The online network encodes the visible patch $x_v$ and predicts the representation $\tilde{z}_m$ of the masked patch $x_m$ encoded in the target network. M2D learns the representation by maximizing the agreement between the prediction $\hat{z}_m$ and $\tilde{z}_m$.}
  %onlineネットワークは可視パッチ x_vをエンコードし、targetネットワークでエンコードされたマスクパッチx_mの表現z~_mを予測する。
  \label{fig:system}
  \vspace{-5pt}
\end{figure}

%M2Dは非マスク(可視)パッチの表現を使ってマスクパッチの表現を予測することで学習する自己教師あり学習手法である。図Y(a)に示すM2Dの構成のうち、オンラインネットワークは可視パッチを表現にエンコードし、マスクパッチの表現を予測する。一方ターゲットネットワークは可視パッチの情報を利用せずに、マスクパッチのみを教師信号としての表現にエンコードする。従来の手法がターゲットネットワークに全てのパッチを与えるのに対して、我々は各ネットワークに与えられる情報は排他的である。
M2D is an SSL method that learns by predicting the representation of masked patches from the representation of unmasked (visible) patches. As shown in Fig.~\ref{fig:system}, M2D consists of online and target networks. The online network encodes the visible patches into a representation and predicts the masked patch representation. Meanwhile, the target network does not use the information from the visible patches but encodes only the masked patches into the representation as training signals. While conventional methods give all patches to the target network, the information we provide to each network is exclusive.

\vspace{0.2cm}
\subsubsection{Pre-training} \label{sec:m2d-pre-training}
%図Xに示されるM2Dはまず、入力をマスク・可視パッチに分割、それぞれを２つのネットワークで処理、損失を計算し、最後にネットワーク全体を更新する。
%As shown in Figure \ref{fig:system}, 
M2D first splits the input into masked and visible patches, processes them in two networks, calculates the loss, and finally updates the entire network.

\vspace{0.1cm}
\noindent\textbf{Processing input.}\hspace{0.2cm}
%まず、フレームワークは入力データX(オーディオスペクトログラムや画像など)を格子状のパッチに分割、学習可能な線形層(patch embedding層)で写像し、位置エンコーディングを付与し、そのうちマスク率に応じた個数をランダムに選んでマスクパッチXMとし、残りを可視パッチXVとして割り当てる。
The framework partitions the input data $x$ (audio spectrogram, image, etc.) into a grid of patches, maps them with a learnable linear layer (patch embedding layer), adds positional encoding, and randomly selects a number of patches according to a masking ratio as masked patches $x_m$ (e.g., 70\% of the input) and the rest as visible patches $x_v$ (e.g., the remaining 30\%).
%Random sampling is structure-agnostic, as in MAE\cite{he2022masked}.
While we use the same positional encoding as MAE\cite{he2022masked}, we tested various masking ratios as discussed in Section \ref{sec:exp-general:mask-ratio}.

%\vspace{0.1cm}
\noindent\textbf{Online and target networks.}\hspace{0.2cm}
%オンラインネットワークは、パラメーター集合thetaで定義され、エンコーダFSを用いて可視パッチXVを潜在表現ZVにエンコードする。
The online network, defined by a set of weights $\theta$, uses the online encoder $f_\theta$ to encode the visible patches $x_v$ into the representation $z_v = f_\theta(x_v)$.
%ZVにマスクパッチMを追加、位置エンコーディング$p$を足し合わせ、予測器GSを使って入力全体の潜在表現ZZを予測する。
It concatenates shared, learnable masked tokens $m$ to $z_v$, adds the positional encoding $p$, and predicts  $\hat{z}$, representations of all input patches, using the predictor $g_\theta$.
\begin{equation}
  \hat{z} = g_\theta(\text{concat}(z_v, m) + p)
\end{equation}

%その後、予測結果をフィルタリングしてマスクパッチに対応する表現のみをZZMとして出力する。
It then filters the prediction result $\hat{z}$ to output $\hat{z}_m = \{\, \hat{z}[i] \mid i \in I_M \,\}$ containing only masked patch representations, where $I_M$ is the set of indices of the masked patches.
%It then outputs the representation $\hat{z}_m = \{\, \hat{z}[i] \mid i \in V \,\}$, corresponding to the masked patches among the prediction results $\hat{z}$, where $V$ is the set of indices of the masked patches.
%\hat{z}_m = \{\hat{z}[i]\}, i \in V,

%ターゲットネットワークはパラメータXIで定義され、パラメータ以外はオンラインエンコーダと同一のエンコーダFXのみを持つ。ネットワークはFXを用いてマスクパッチXMをエンコードした表現ZMを出力する。
The target network is defined by parameter $\xi$ and consists only of momentum encoder $f_\xi$, which is identical to the online encoder except for the parameter.
The network encodes masked patches $x_m$ using $f_\xi$ to output the representation $z_m = f_\xi(x_m)$.
%我々はMAEと同様にZMを標準化する。ターゲットの標準化することで、学習が安定することを実験的に確認した。
We then standardize $z_m$ to $\tilde{z}_m = ({z_m - \text{mean}{(z_m)}})/{\sqrt{\text{var}{(z_m)}}}$. We do this to stabilize the training, which we empirically confirmed in preliminary experiments, rather than for performance gain as in MAE.
%We then calculate $\tilde{z}_m$, a standardized $z_m$, for stabilizing training which we empirically confirmed in the preliminary experiments, rather than for performance gain as in MAE.
%\begin{equation}
%  \tilde{z}_m = ({z_m - \text{mean}{(z_m)}})/{\sqrt{\text{var}{(z_m)}}},
%\end{equation}

\vspace{0.1cm}
\noindent\textbf{Loss function.}\hspace{0.2cm}
%損失はオンラインの予測出力ZZMに対してターゲットの出力ZMを教師信号として計算される。
The loss is calculated using the standardized target output $\tilde{z}_m$ as a training signal against the online prediction output $\hat{z}_m$.
%BYOLに則り、L2ノーマライズしたZZMとHZMのmean square error (MSE)により損失Lを計算する。
Inspired by BYOL\cite{grill2020byol}, we calculate the loss $L_\text{m2d}$ by the mean square error (MSE) of $l_2$-normalized $\hat{z}_m$ and $\tilde{z}_m$.
\begin{equation}
L_\text{m2d} = ||l_2(\hat{z}_m) - l_2(\tilde{z}_m)||^2_2 = 2 - 2 \cdot \frac{\langle \hat{z}_m, \tilde{z}_m \rangle }{||\hat{z}_m||_2 \cdot ||\tilde{z}_m||_2},
\label{eq:eq-byol-mse}
\end{equation}
where $\langle\cdot, \cdot\rangle$ denotes the inner product.

\vspace{0.1cm}
\noindent\textbf{Updating network parameters.}\hspace{0.2cm}
Our framework updates parameters $\theta$ and $\xi$ after each training step. It updates $\theta$ only by minimizing the loss $L_\text{m2d}$ as depicted by the stop-gradient in Fig. \ref{fig:system}, whereas updating $\xi$ is based on a slowly moving exponential average of $\theta$ with a decay rate $\tau$:
\begin{equation}
    \xi \leftarrow \tau \xi + (1 - \tau) \theta
\end{equation}

%オンラインネットワークのゆっくりとした移動平均をbootstrapすることで、不十分な解に崩壊することを避け、有益な表現の学習を導けることが経験的に示されている。
It has been empirically shown that stop-gradient operation helps avoid collapsing to an uninformative solution and that the moving-average behavior may lead to learning effective representations\cite{grill2020byol,Xinlei2021SimSham}.
%提案手法は、モメンタムターゲットでエンコードする表現とオンラインでエンコードする表現の表現の一致が最大化するよう表現を学習する。
%Our method learns representations by maximizing the agreement between the representations encoded at the momentum target and those encoded at the online networks.
After the training, we transfer only the $f_\theta$ as a pre-trained model.

\vspace{0.2cm}
\subsubsection{Transferring Models to Audio Tasks} \label{sec:method:xfer-model-4-audio}
%モデルの入力時間は固定であるが、可変長の音声を特徴量にエンコードする2つの方法を提供することで、事前に訓練されたモデルを任意の音声長の入力で下流のタスクに使用できるようにする。またこのとき、モデルの出力する特徴量を時間フレームごと一つの特徴量ベクトルのシーケンスに変換する。
While the model's input duration is fixed, we provide two ways to encode variable-length audio into features so that we can transfer the pre-trained models to tasks with any audio length. In addition, we convert the model's per-patch output features into a sequence of time-framed feature vectors.

\vspace{0.1cm}
\noindent\textbf{Encoding variable-length audio.}\hspace{0.2cm}
%入力取扱い方法. 一つめの方法は、モデルの入力音長ごとに入力音を分割し、それぞれを個別にエンコードたあと一つにつなげる(Split&Collectと呼ぶ)。２つ目は一般的に行われる方法で、モデルの位置エンコーディングをタスクの音データの長さに合わせて補間することで、モデルが長い音の入力を一度にエンコードできるようにする(Positional Completionと呼ぶ)。前者の方法はモデルをフリーズしても学習した知識をそのまま活かすことができる一方、後者は補間した位置エンコーディングに合わせてfine-tuningすることで学習した知識を転移利用できる。
The first way splits the input audio data by the model's input duration, encodes each separately, and then concatenates them. % (Split\&Collect)
The second is a common practice, where we interpolate the model's positional encoding to match the audio length of the task so that the model can encode long audio at once.
The former method can exploit the learned knowledge even when the model is frozen, while the latter can transfer the learned knowledge by fine-tuning it to the interpolated positions.
% (CompletePos)
%殆どの場合は前者の方法を適用し、事前学習に近い音の長さのタスクをFine-tuningする場合のみに後者を適用した。
We applied the former in most cases and the latter only when fine-tuning with an audio length close to that of pre-training.

\vspace{0.1cm}
\noindent\textbf{Converting feature into a sequence.}\hspace{0.2cm}
%我々はMSM-MAEの特徴量計算を利用してViTの出力を時間フレームごとに取り扱えるようにする。
We adopt the MSM-MAE\cite{niizumi2022msm-mae} feature calculation to enable the processing of features on a time-frame basis. The pseudo-code for the calculation follows:
\begin{equation}
\begin{split}
z' = & z.\text{reshape}(B,  N_F, N_T, D)\\
    & .\text{transpose}(1, 2)\\
    &.\text{reshape}(B, N_T, N_F D), \label{eq:msm-mae}
\end{split}    
\end{equation}
where $z \in R^{B \times N_F N_T \times D}$ is the encoder output, $B$ is batch size, $N_F$ is the number of patches along frequency, $N_T$ is the number of patches along time, $D$ is a feature dimension, and $z' \in R^{B \times N_T \times N_F D}$ is the calculation result.
%This calculation summarizes encoded features of a time frame for all frequencies into a single vector. For example, the feature dimension of $z'$ will be $3840$ when $D=768$ and $N_F=5$, which is used in our experiments.

%この計算により、その後時間フレームごとの平均など統計量を計算するのが容易になる。
This calculation makes it easier to subsequently compute statistics along time frames, e.g., temporal average pooling.
We summarize an audio clip-level feature $z'' = 1/N_T\sum_{t=1}^{N_T} z'[t]$, averaging $z'$ over time in the experiments. For example, $z''$~becomes a 3840-d feature, where $D=768$ and $N_F=5$.

\begin{figure*}[tbp]
  \vspace{-5pt}
  \centering
  \includegraphics[width=0.95\textwidth]{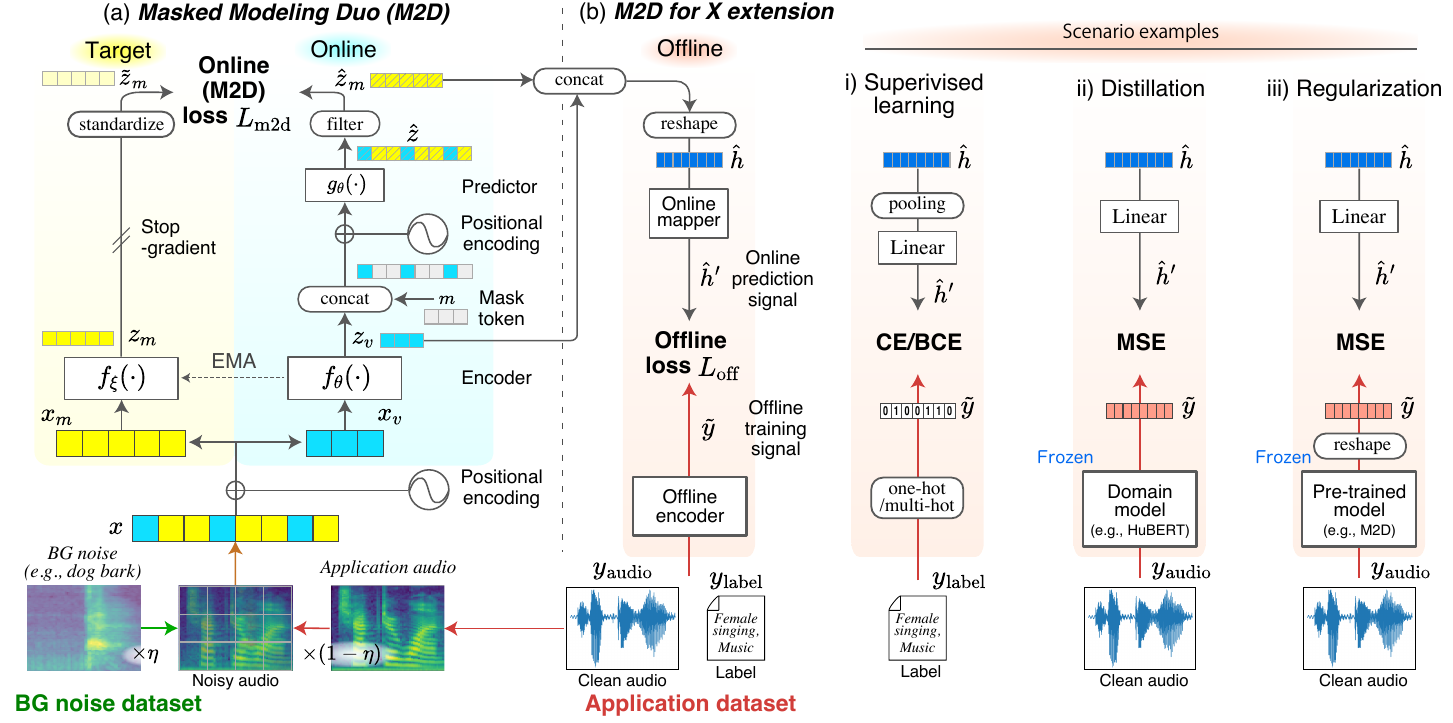}
  \vspace{-5pt}
  \caption{M2D-X pre-training flow and application scenarios. M2D-X adds an offline network to M2D and inputs background noise, forming a multitask of M2D, a denoising task, and an additional task. Scenario examples show the offline network configurations to compose an additional task for each purpose: supervised learning of task labels, distilling a domain pre-trained model, or regularizing to prevent overfitting M2D to a small dataset.}
  \label{fig:system-x}
  \vspace{-5pt}
\end{figure*}

%%%%%%%%%%%%%%%%%%%%% SUB SECTION %%%%%%%%%%%%%%%%%%%%%%
\subsection{Masked Modeling Duo for X}\label{sec:method:m2d-x}

%M2D-Xは、M2Dの効果的な表現学習に加えて追加タスクと背景ノイズを用いて、小データの場合も含め、アプリケーションのデータ分布に特化した表現を学習できるように設計された事前学習フレームワークである。
M2D-X is a pre-training framework designed to learn a representation specialized to the data distribution of an application $X$, even with small data.
%図XX(b)に示すように、offline networkはアプリケーションのデータ(オーディオとラベル)を入力する一方、M2DはApplication audioにbackground noiseを混ぜ合わせたnoisy audioを入力する。
As shown in Fig.~\ref{fig:system-x}, it extends M2D with an offline network and adds background noise input.
%Offline networkで学習する追加タスクはアプリケーションの様々な学習ニーズに対応して設定できるようにした。
The additional task in the offline network is configurable to accommodate the various application needs.
To form a denoising task, the offline network inputs clean application data (audio and labels), while M2D inputs noisy audio mixed with application audio and background noise.

%背景ノイズを取り入れることで、アプリケーションデータ分布にフォーカスした学習を促進できるようにした。背景ノイズによりデータ拡張の効果とデノイズタスクを生み、アプリケーションのデータ分布にフォーカスしたロバストな表現の学習に寄与し、特に小さなデータセットに有効である。
We incorporate background noise to enable successful pre-training on the application data.
Background noise creates the effect of data augmentation, which is especially helpful for small datasets, and the denoising task, which contributes to learning a robust representation focused on the application data distribution.

%これにより、Offline networkでのアプリケーションに特化したタスクとM2Dの表現学習のマルチタスク学習に加え、背景ノイズの導入によるデノイジングタスクやデータ拡張の効果を組み合わせ、効果的な事前学習を行う。
%These allow for pre-training by combining the multi-task learning of an application-specific task in the offline network and M2D representation learning, as well as the denoising task and data augmentation effects of introducing background noise.

\vspace{0.2cm}
\subsubsection{Offline Network}
%Offline networkは、M2Dでエンコードしたaudio featureとアプリケーションデータから得られる教師信号から損失を計算する。M2Dブランチからz_vとzh_mをつなぎ合わせ、式4のreshape操作を行い、音の特徴量hhを得る。アプリケーションデータy_audioやy_labelをOffline encoderで変換した教師信号yhを得る。オフライン損失Loffは、hhをOnline mapperで変換したhhdを得て、yhとの間で計算する。損失関数は、シナリオに応じて例えば式2のMSE損失や交差エントロピー損失などを用いる。
As illustrated in Fig.~\ref{fig:system-x}(b), the offline network computes the loss from the audio features encoded in M2D and the training signal obtained from the application data.
The $z_v$ and $\hat{z}_m$ from the M2D branch are concatenated and reshaped using \eqref{eq:msm-mae} to obtain the audio feature $\hat{h}$ per time frame.
The offline encoder converts the application data $y_\text{audio}$ and/or $y_\text{label}$ into the training signal $\tilde{y}$.
The offline loss $L_\text{off}$ is calculated using $\tilde{y}$ and $\hat{h'}$ converted from $\hat{h}$ by the online mapper.
The actual loss function depends on the application scenario, e.g., MSE loss in \eqref{eq:eq-byol-mse} or cross-entropy loss.

%このネットワークブランチはノイジーオーディオを入力するM2Dの特徴量を入力し、クリーンなオーディオとラベルとの間で損失を計算することで、自然とデノイズタスクを形成する。結果として、ネットワーク全体はM2D、オフラインロスのタスク、そしてデノイズタスクから学習する。
This branch network naturally forms a denoising task by calculating the loss between the noisy audio M2D features and the clean application data (audio/labels). As a result, the entire network learns from a multi-task of M2D, an additional task using the offline loss, and a denoising task.

\vspace{0.2cm}
\subsubsection{Noisy Audio}
%noisy audioは、目標データセットA、背景データセットBそれぞれのlog-melスペクトログラムに対して、データセットノイズ比率Aを使って次の式でミックスして得られる。
The noisy audio $x_\text{noisy}$ is obtained by mixing the target log-mel spectrogram data $x_\text{targ}$ and background log-mel spectrogram data $x_\text{bg}$ with the dataset noise ratio $\eta$. This process is expressed by the following equation as in \cite{niizumi2021byol-a}.
\begin{equation}
x_\text{noisy} = \log{\left((1 - \eta) \exp(x_\text{targ}) + \eta \exp(x_\text{bg})\right)}
\label{eq:noisy-audio}
\end{equation}

%マスク予測による学習はデータセットのサイズが大きいことが必要だが[A]、AudioSetなど大規模データセットをミックスすることでデータ拡張の効果が生まれ、小さい目標データセットでも有効な表現の学習を可能にする。
Although pre-training with masked prediction requires a large dataset\cite{Xie2023OnDataScalingMasked,zhang2023mae-survey}, mixing large datasets such as AudioSet leads to the effect of data augmentation, allowing the pre-training of effective representations even with a small target dataset.
%x_Tとx_Bはそれぞれ元のサンプルからランダムに切り出すので、その組み合わせであるx_Nは非常に多くのバリエーションを生み出す。
%The number of possible $x_\text{noisy}$ variations is enormous because $x_\text{targ}$ and $x_\text{bg}$ are randomly cropped from each original sample.

\vspace{0.2cm}
\subsubsection{M2D-X Pre-training}
%M2Dは通常の学習プロセスを実行し損失Lm2dを得る一方、Offline networkでは並行してLoffが得られる。M2D-X全体の損失は、Lm2dとLoffの重み付け和として計算する。
In the M2D-X framework, M2D performs its training process described in Section \ref{sec:m2d-pre-training} and calculates loss $L_\text{m2d}$, while the offline network calculates its loss, $L_\text{off}$.
The overall M2D-X loss $L_\text{m2dX}$ is then calculated by combining $L_\text{m2d}$ and $L_\text{off}$:
%\vspace{-0.1cm}
\begin{equation}
L_\text{m2dX} = \lambda_\text{m2d} L_\text{m2d} + \lambda_\text{off} L_\text{off},
\label{eq:eq-m2ds-loss}
\end{equation}
where the loss weights $\lambda_\text{m2d}$ and $\lambda_\text{off}$ control the contribution.
%Lm2dXを全体に逆伝播して全てのネットワークの重みを更新する。
M2D-X updates the weights of the entire network by backpropagating $L_\text{m2dX}$.

\vspace{0.2cm}
\subsubsection{Specialized Learning Scenarios}
%図4(b)のシナリオ例のように、M2D-Xフレームワークはオンラインマッパ、オフラインエンコーダ、オフライン損失関数を用途に合わせて選ぶことで、アプリケーションに特化した事前学習が可能である。
As shown in the scenario examples in Fig.~\ref{fig:system-x}(b), the M2D-X framework enables pre-training specialized to the application by adopting an online mapper, an offline encoder, and an offline loss function for each application.

\vspace{0.1cm}
\noindent\textbf{i) Supervised Learning.}\hspace{0.2cm}
%i) 教師あり学習 このシナリオは応用の教師信号に適した表現を事前学習する。Online mapperではtemporal pooling(例 average pooling)とLinear層を組み合わせ、hhをhhdに変換する。one-hotまたはmulti-hot encodingしたラベルとの間でcross-entropy(CE)またはbinary cross-entropy(BCE)損失を計算することで、offline networkは通常の教師あり学習としてタスクを形成する。この損失は各クラスラベル毎に音の表現を分布させる。X節はこのシナリオを評価する。
This scenario pre-trains a suitable representation for the application's training signal.
The online mapper combines temporal pooling (e.g., average pooling) and a linear layer to convert $\hat{h}$ into $\hat{h'}$.
The offline network forms a task of conventional supervised learning by calculating the cross-entropy (CE) or binary cross-entropy (BCE) loss between one-hot or multi-hot encoded labels.
These loss functions should help build a distribution of the audio representation suitable for representing the application's class definition.
Section \ref{sec:eval-audioset} evaluates this scenario.

\vspace{0.1cm}
\noindent\textbf{ii) Distillation.}\hspace{0.2cm}
%ii) 蒸留 このシナリオはドメインに特化したモデルが既に存在する場合に、学習済みの知識を蒸留する。Offline encoderにdomain modelを利用することで可能となる。オンラインマッパは線形層を、オフラインエンコーダにはドメインの事前学習モデルを、式2同様のMSE損失を使うことで、トラメインモデルの特徴量を蒸留できる。X節はこのシナリオを評価する。
This scenario distills the learned knowledge when a domain-specific model already exists.
The domain model features can be distilled by using a linear layer for the online mapper, a frozen domain pre-trained model for the offline encoder, and MSE loss, for example.
Section \ref{sec:eval-speech} evaluates this scenario.

\vspace{0.1cm}
\noindent\textbf{iii) Regularization.}\hspace{0.2cm}
%iii) 正則化 このシナリオは小さなデータをM2Dで事前学習するときの過学習を防ぐ。事前学習済モデルをアプリケーションデータで追加事前学習するとき、Offline encoderに重みを固定した同じモデルを利用して、前のシナリオ同様蒸留を形成する。固定されたモデルは事前学習済みの分布に従って特徴量を出力するため、これを蒸留することで小さなデータセットに過学習することを防ぐことができる。X節はこのシナリオを評価する。
Problems typically found in confidential and proprietary data in domains such as industrial and medical domains are a small dataset and a data distribution different from that of a general large pre-training dataset. In the regularization scenario, we address these problems by using an enhanced further pre-training strategy based on distillation.
Further pre-training is a strategy where a pre-trained model is pre-trained on a target dataset \cite{lee2023selfdistillation} to make the model effective for the target while utilizing the learned knowledge in the previous pre-training.

We use the original pre-trained model as a frozen offline encoder and its weights to initialize M2D online encoder weights, forming a distillation, the same as in the previous scenario.
Then, the offline encoder with frozen weights maintains its features to follow the original pre-training data distribution even with the target data input and helps prevent the M2D online encoder from overfitting on a small dataset while further pre-training.
Section \ref{sec:eval-icbhi} evaluates this scenario.

%はじめの２つのシナリオは従来の教師あり学習と蒸留をM2Dの効果的な学習と組み合わせることで強化し、正則化のシナリオは追加事前学習を強化する。これらの例に挙げられるように、M2D-Xは多様なアプリケーションの汎用事前学習フレームワークとして役立つ。
\vspace{0.1cm}
The first two scenarios enhance conventional supervised learning and distillation combined with effective masked prediction-based learning of M2D, while the regularization scenario enhances further pre-training. As described in these examples, our aim is for M2D-X to serve as a universal audio pre-training framework for diverse applications.

\begin{table*}[tbp]
\caption{Evaluation \ref{sec:eval-general} General-purpose audio representation: Downstream task details.}
\label{tab:list-ds}
\centering
\vspace{-5pt}
\resizebox{1.0\textwidth}{!}{%
\begin{tabular}{rccccccccccc}\toprule
& \multicolumn{4}{c}{Environmental sound tasks} & \multicolumn{4}{c}{Speech tasks} & \multicolumn{3}{c}{Music tasks} \\
\cmidrule(lr){2-5} \cmidrule(lr){6-9} \cmidrule(lr){10-12}  
 & AS2M\cite{gemmeke2017audioset} & AS20K\cite{gemmeke2017audioset} & ESC-50\cite{piczak2015esc50} & US8K\cite{salamon2014urbansound} & SPCV2\cite{speechcommandsv2} & VC1\cite{voxceleb} & VF\cite{voxceleb} & CRM-D\cite{cao2014cremad} & GTZAN\cite{gt2013gtzan} & NSynth\cite{nsynth2017} & Surge\cite{turian2021torchsynth} \\
\midrule
Evaluation protocol$^\dagger$ & FT & FT & FT / LE & LE & FT / LE & FT / LE & LE & LE & LE & LE & LE\\
 \addlinespace[0.1cm]
No. of training samples & 2,005,132 & 21,940 & \multirow{3}{*}{\shortstack{5 folds\\ 2000}} & \multirow{3}{*}{\shortstack{10 folds\\ 8732}} & 84,843 & 138,361 & 121,281 & 5155 & 443 & 289,205 & 148,896\\
% \addlinespace[0.01cm]
No. of validation samples & - & - & & &  9981 & 6904 & 26,684 & 732 & 197 & 12,678 & 17,160 \\
No. of test samples & 20,178 & 20,178 & & & 11,005 & 8251 & 28,463 & 1551 & 290 & 4096 & 17,336 \\
 \addlinespace[0.02cm]
No. of classes & 527 & 527 & 50 & 10 & 35 & 1251 & 6 & 6 & 10 & 11 & 88 \\
% Tags & Tags & ESCs & ASCs & words & IDs & languages & emotions & genres & families & notes \\
Average duration & 10.0 s & 10.0 s & 5.0 s & 4.0 s & 1.0 s & 8.2 s & 5.8 s & 2.5 s & 30.0 s & 4.0 s &  4.0 s \\
\bottomrule
\addlinespace[0.05cm]
\multicolumn{12}{l}{$^{\dagger}$ FT and LE stand for fine-tuning and linear evaluation, respectively.}\\
%\addlinespace[0.05cm]
%\multicolumn{10}{l}{$^\ast$ Surge\cite{turian2021torchsynth} is a new dataset that evaluates itself using COALA\cite{favory2020coala} and OpenL3\cite{cramer2019openl3}.}\\
\end{tabular}
}
\vspace{-10pt}
\end{table*}

\begin{table}[tb!]
\vspace{-5pt}
\caption{Evaluation \ref{sec:eval-general} General-purpose audio representation: Fine-tuning settings}
\label{tab:exp-general:ft-parameters}
\vspace{-5pt}
\centering
\resizebox{1.0\columnwidth}{!}{%
\begin{tabular}{llllll}
\toprule
Parameter & AS2M & AS20K & ESC-50 & SPCV2 & VC1 \\
\midrule
Learning rate & 2.0 & 0.5 & 0.5 & 0.5 & 0.0005 \\
Batch size & 64 & 64 & 128 & 128 & 64 \\
Optimizer & LARS & SGD & SGD & SGD & AdamW \\
Mixup ratio & 0.5 & 0.3 & 0.0 & 0.3 & 0.0 \\
Random resize crop & - & \checkmark & \checkmark & \checkmark & - \\
SpecAugment$^\sharp$\cite{specaugment} & 30/192 & 30/192 & 15/48 & 30/48 & 30/48 \\
Training epochs (total) & 70 & 200 & 200 & 200 & 50 \\
Training epochs (warm-up) & 15 & 5 & 5 & 5 & 5 \\
Structured Patchout\cite{Koutini2022passt} ratio & 0.5 & 0.5 & 0.5 & 0.5 & 0.0 \\
Adjust positional encoding & \checkmark & \checkmark & - & - & \checkmark \\
Freeze embedding layer\cite{kumar2022finetune} & - & - & \checkmark & - & - \\
\bottomrule
\addlinespace[0.05cm]
\multicolumn{6}{l}{$^{\sharp}$ The frequency/time masking parameters.}\\
\end{tabular}
}
\vspace{-10pt}
\end{table}

%%%%%%%%%%%%%%%%%%%%%%%%%%%%%%%%%%%%%%%%%%%%%%%%%%%%%%%%%%%%%%%%%%%%%%%
%%%%%%%%%%%%%%%%%%%%%%%%%%%%%%  SECTION  %%%%%%%%%%%%%%%%%%%%%%%%%%%%%%
\section{Experiments}
%実験では、M2Dによる汎用的な音声表現の学習と、M2DおよびM2D-Xによる特化した音声表現の学習を評価した。実験は、一般音声、音声、医療アプリケーションの3つの音声設定で行った。大規模なデータセットを持つ一般音声のシナリオでは、M2Dは汎用音声表現(セクションAA)とAudioSetへの特化型表現(セクションAAA)を学習する。競合領域での性能を評価するため、M2DとM2D-Xは特化した音声表現を学習する(セクションBBB)。小規模データ領域への適用性を評価するため、M2DとM2D-Xは医療アプリケーションに特化した表現を学習する(セクションCCC)。これらの実験を通して、普遍的な音声事前学習フレームワークとしての我々の手法の有効性を検証する。
In the experiments, we evaluated learning of a general-purpose audio representation with M2D, followed by learning of specialized representations with M2D and M2D-X.
Experiments included three audio settings: general audio, speech, and a medical application sound.
In a scenario of general audio with a large-scale dataset, M2D learns a general-purpose audio representation (Section \ref{sec:eval-general}), and M2D-X learns a specialized representation of AudioSet (Section \ref{sec:eval-audioset}).
To assess the performance in a highly competitive domain, M2D and M2D-X learn specialized speech representations (Section \ref{sec:eval-speech}).
To assess applicability to a small-data regime, M2D and M2D-X learn specialized representations for medical applications (Section \ref{sec:eval-icbhi}) under a further pre-training strategy.
Through these experiments, we validated the effectiveness of our methods as a universal audio pre-training framework.

%%%%%%%%%%%%%%%%%%%%% SUB SECTION %%%%%%%%%%%%%%%%%%%%%%
\subsection{Basic Experimental Setup}\label{sec:exp:basic}
%我々は以下の基本的な実装や設定を利用して本研究の実験を行った。
We conducted experiments using the following basic implementations and settings.
%先行研究との比較を容易にするため、MAEの実装をベースに最小限の変更を加える形でM2Dを実装した。我々はMAE学習フレームワークにターゲットネットワークを追加実装し、MAEのデコーダをそのまま我々のPredictorとして流用した。我々はすべての実験でエンコーダにMAEで実装されるバニラViT-Baseを利用し、入力音長をATSTと同じ6秒に、パッチサイズを多くの研究と同じ16x16に固定した。マスク率には予備実験で良い性能を示した0.6, 0.7を利用した。
To facilitate comparison with previous studies, we implemented M2D based on the MAE implementation with minimal modifications; specifically, we implemented an additional target network on top of the MAE code, adopted the MAE decoder as our predictor $g_\theta$ without changes, and used vanilla ViT-Base\cite{ViT} with a 768-d output feature as our encoders ($f_\theta$ and $f_\xi$) for all experiments.
Our basic settings are: input audio length of 6 s, the same as in ATST\cite{Li2022ATST}; patch size of $16\times 16$, the same as in most studies\cite{Baade2022MAE-AST, chong2022maskspec, gong2022ssast, huang2022audiomae}; and the masking ratio of 0.6 and 0.7, which performed well in preliminary experiments.

%エンコーダにはViTをそのまま利用する一方で、出力特徴量の計算はN章で説明した音に最適化した方法を実装した。位置エンコーディングの補間はMAEの実装を利用した。ファインチューニングによる一部のタスクを用いた評価において、学習の安定化を図るためpatch embedding layerの重みを固定した。
While we used ViT as is for the encoder, we implemented the feature calculation optimized for audio, as described in Section \ref{sec:method:xfer-model-4-audio}.
%, and we used the MAE implementation for interpolating the positional encoding.
In the fine-tuning evaluation, the weights of the patch embedding layer were fixed in order to stabilize the fine-tuning\cite{kumar2022finetune} for some tasks.

%オーディオ処理
We preprocessed audio samples to a log-scaled mel spectrogram with a sampling frequency of 16,000 Hz, window size of 25 ms, hop size of 10 ms, and mel-spaced frequency bins $F=80$ in the range of 50 to 8000 Hz.
We standardized spectrograms with the statistics of the pre-training dataset (e.g., average $-7.1$ and standard deviation $4.2$ for AudioSet).
%デフォルトの入力音長6sの入力スペクトログラムのサイズは80x608 (Freq. bins x Time frames)となる。
The input spectrogram with the default input duration of 6 s has a size of $80 \times 608$ (Freq. bins $\times$ Time frames), making $N_F=5$ and $N_T=38$ with patch size of $16\times 16$.

%\vspace{0.1cm}
%\noindent\textbf{Pre-training details.}\hspace{0.2cm}
%事前学習には、我々はエポック数300、warm upエポック数20、バッチサイズ2048、base learning rate 3e-4とした他はlearning rate schedulingや最適化を含めすべてMAEと同じ設定とした。ターゲットネットワーク更新のEMAパラメータTAUは学習開始時0.99995から終了時0.99999まで線形補間した。
For pre-training, we set the number of epochs as 300, the number of warm-up epochs as 20, batch size as 2048, and the base learning rate as 3e-4. All other settings were the same as in the MAE, including the learning rate scheduling and optimizer. The EMA decay rate $\tau$ for the target network update was linearly interpolated from 0.99995 at the start of training to 0.99999 at the end.
Appendix \ref{appendix:resources} describes the required computational resources.

For linear evaluation and fine-tuning, we used typical random seeds (e.g., 42), ran each evaluation six times, and averaged the results with 95\% confidence intervals.

%%%%%%%%%%%%%%%%%%%%% SUB SECTION %%%%%%%%%%%%%%%%%%%%%%
\subsection{Evaluation: General-purpose Audio Representation}\label{sec:eval-general}

%この実験では、M2DをAudioSet上で事前学習させ、汎用音響信号表現を学習させ、様々なタスクで[BYOL-A]に準じて評価プラットフォームEVARの実装を利用して、線形評価とファインチューニングによる評価を実施した。
In this experiment, we validated that M2D learns a general-purpose audio representation that performs well.
We pre-trained M2D on a large-scale general audio dataset, AudioSet, and evaluated it in linear evaluation and fine-tuning.

%モデルの事前学習はSection Xに記載した基本設定で行った。
We pre-trained models with the basic setup described in Section \ref{sec:exp:basic} on AudioSet\cite{gemmeke2017audioset} with 2,005,132 samples (5569 h) of 10-s audio from the balanced and unbalanced train segments. We randomly cropped 6-s audio from a 10-s sample.
%評価に利用した下流タスクを表XXに一覧する。これらは様々な音の分類タスクで構成され、評価結果は全て正解率またはmean average precision (mAP)である。タスクのデータセットに可変長の音が含まれる場合、各サンプルからデータの平均長をランダムクロップまたはゼロパディングして長さが揃った音のバッチとしてモデルに入力した。各データサンプルはデータセットのlog-melスペクトログラムの統計量で標準化することで、事前学習のデータ分布に合わせた。
Table \ref{tab:list-ds} lists the downstream tasks used in the evaluation. These are various sound classification tasks, and results are accuracies or mean average precision (mAP). When a task's dataset contained variable-length audio, we randomly cropped or zero-padded the average audio length for each sample to create a batch of length-aligned audio fed into the model.% We standardized data samples after processing them into log-mel spectrograms with the statistics of AudioSet.

\begin{itemize}
    \item AudioSet\cite{gemmeke2017audioset}: a multi-label audio event classification task with two settings. AS2M uses all 2M training samples, whereas AS20K uses 21K samples from the balanced train segments for training.
    \item ESC-50\cite{piczak2015esc50}: an environmental sound classification task with 50 classes. We follow standard leave-one-out cross-validation (LOOCV) with the official five folds.
    \item UrbanSound8K\cite{salamon2014urbansound} (US8K): an urban sound classification task. We follow the official ten-fold LOOCV.
    \item Speech Commands V2\cite{speechcommandsv2} (SPCV2): a speech command word classification task.
    \item VoxCeleb1\cite{voxceleb} (VC1): a speaker identification task with interviews of 1251 celebrities.
    \item VoxForge\cite{voxforge} (VF): a language identification task.% with six languages.
    \item CREMA-D\cite{cao2014cremad} (CRM-D): a speech emotion recognition task for 91 speakers. We follow \cite{niizumi2023byol-a} for the data splits.
    \item GTZAN\cite{gt2013gtzan}: a music genre recognition task. We follow fault-filtered partitioning\cite{kereliuk2015music}\cite{sturm2013gtzansplit}.
    \item NSynth\cite{nsynth2017}: an instrument family classification task.% of musical notes.
    \item Pitch Audio Dataset (Surge synthesizer)\cite{turian2021torchsynth} (Surge): a pitch audio classification task composed of 88 MIDI note classes. We follow \cite{niizumi2023byol-a} for the data splits.
%    \item FSD50K\cite{fonseca2020fsd50k}: a multi-label classification SER task with 200 classes drawn from AudioSet ontology that covers a wide variety of sound events.
\end{itemize}

We followed BYOL-A\cite{niizumi2023byol-a} to use a unified evaluation platform, EVAR\footnote{\url{https://github.com/nttcslab/eval-audio-repr}}, for linear evaluation and fine-tuning.
%比較するMAEには、4層、6ヘッド、384次元の埋め込みによる小さなデコーダーにすることで音響信号に最適化したMSM-MAEを利用した。予備実験により標準的なMAEの性能を改善できることを確認している。
We used the MSM-MAE\cite{niizumi2022msm-mae}, an MAE variant for audio, as a baseline MAE for comparison.
We transferred the pre-trained MSM-MAE in the same way as with M2D described in Section \ref{sec:method:xfer-model-4-audio} for fair comparison while keeping pre-training parameters the same as in the original MAE, such as the masking ratio of 0.75.
Preliminary experiments verified that the MSM-MAE outperforms a vanilla MAE.
%All these settings were common in the M2D and MSM-MAE pre-training, except for the EMA decay rate.

%なお、表XXX,TTに示す実験結果は[M2D]で報告済みの結果と比べて少し改善している。これはN章で説明した入力の取り扱いの違いとコードの改善によるものである。
%Note that the experimental results shown in Tables \ref{tab:exp-general:sota-le} and \ref{tab:exp-general:sota-ft} show a slight improvement over the results already reported in previous M2D\cite{niizumi2022M2D}. This is due to the different input handling described in Section \ref{sec:method:xfer-model-4-audio} and code improvements.

\vspace{0.1cm}
\noindent\textbf{Linear evaluation details.}\hspace{0.2cm}
%線形評価の手順はBYOL-Aに従った。
We followed BYOL-A\cite{niizumi2023byol-a} for the linear evaluation procedure.
The evaluation pipeline first converts downstream task samples into features using a \textit{frozen} pre-trained model, trains a linear layer with task labels, and then conducts tests to get the results. 
We used the validation set for early stopping with a patience of 20 epochs and trained the linear layer for up to 200 epochs using the Adam optimizer and a learning rate of 0.00003.

\vspace{0.1cm}
\noindent\textbf{Fine-tuning details.}\hspace{0.2cm}
%我々は先行研究で共通して使われるタスクを用いて性能を評価する。構成するタスクはLinear evaluationと同じESC-50, SPCV2, VC1に加えてAS2MとAS20Kである。
We used the tasks commonly used in previous studies, namely ESC-50, SPCV2, and VC1, the same as in the linear evaluation, plus AS20K and AS2M. %AudioSet20K (AS20K) and a full set of AudioSet (AS2M).
%Fine-tuningパイプラインはATSTとAudio-MAEに則る。線形分類器を事前学習済みモデルの上に追加し、全体を学習する。我々はデータ拡張にはSpecAugmentとMixupとRRC、更にPaSSTで提案されたStructured Patchoutを利用し、オプティマイザにはSGD、学習率はCosine annealingでスケジューリングした。AS-2Mでは、Audio-MAE同様1エポックごとweighted samplerを用いてreplacementなしに200Kデータをサンプリングして学習した。タスクごとに探索したハイパーパラメータを表Xに一覧する。
The fine-tuning pipeline follows ATST\cite{Li2022ATST} and Audio-MAE\cite{huang2022audiomae}. A linear classifier was added on top of the pre-trained model to train the entire network.
We used SpecAugment\cite{specaugment}, Mixup\cite{zhang2018mixup,niizumi2023byol-a}, and Random Resize Crop (RRC)\cite{niizumi2023byol-a} for data augmentation, along with Structured Patchout proposed in PaSST\cite{Koutini2022passt} that masks patches during training. We used SGD, LARS, and AdamW for the optimizer and Cosine annealing\cite{loshchilov2016sgdr} for learning rate scheduling.
For AS-2M, we used a weighted sampler as in Audio-MAE, sampling 200K data without replacement for each epoch. We used the hyperparameters for each task as listed in Table \ref{tab:exp-general:ft-parameters}.
When using the Structured Patchout, we used 768-d features calculated by averaging all ViT outputs because \eqref{eq:msm-mae} is not applicable to the masked patches.

\begin{table*}[tb!]
\vspace{-10pt}
\caption{Evaluation \ref{sec:eval-general} General-purpose audio representation: Linear evaluation results (\%) with 95\% CI.}
\vspace{-5pt}
\label{tab:exp-general:sota-le}
\centering
\resizebox{0.9\textwidth}{!}{%
\begin{tabular}{llllllllllll} \toprule
 &&  \multicolumn{2}{c}{Env. sound tasks} & \multicolumn{4}{c}{Speech tasks} & \multicolumn{3}{c}{Music tasks} \\
\cmidrule(lr){3-4} \cmidrule(lr){5-8} \cmidrule(lr){9-11} 
Model (/masking ratio) & \# Params &    ESC-50 &    US8K &    SPCV2 & VC1 &  VF &  CRM-D &  GTZAN &  NSynth &  Surge & Avg. \\
\midrule
%Wav2Vec2\cite{baevski2020wav2vec2}$^{\dagger}$  &  \underline{57.6} {\fontsize{6pt}{6pt}\selectfont $\pm$ 0.8} &  \underline{66.9} {\fontsize{6pt}{6pt}\selectfont $\pm$ 0.4} & \textbf{\underline{96.6}} {\fontsize{6pt}{6pt}\selectfont $\pm$ 0.0}&  \underline{40.9} {\fontsize{6pt}{6pt}\selectfont $\pm$ 0.6} & \textbf{\underline{99.2}} {\fontsize{6pt}{6pt}\selectfont $\pm$ 0.1}& \underline{65.5} {\fontsize{6pt}{6pt}\selectfont $\pm$ 1.7}&  \underline{57.8} {\fontsize{6pt}{6pt}\selectfont $\pm$ 1.3} &  \underline{56.6} {\fontsize{6pt}{6pt}\selectfont $\pm$ 0.6} &  \underline{15.2} {\fontsize{6pt}{6pt}\selectfont $\pm$ 0.9} \\
\multicolumn{10}{l}{\textit{(Previous SSL methods)}}  \\
DeLoRes-M\cite{Ghosh2022DeLoRes} & 5.3M  & - & 82.7 & 89.7 & 45.3 & 88.0 & - & - & 75.0 & - \\
SF NFNet-F0\cite{wang2022universal} & 63M & 91.1 & - & 93.0 & 64.9 & 90.4 & - & - & \textbf{78.2} &- \\
%OpenL3\cite{cramer2019openl3} $^{\dagger}$  & 79.8 & 79.3 & \underline{87.9} & \underline{40.7} & \underline{90.1} & \underline{60.4} & \underline{73.3} & \underline{75.6} & \underline{36.4} \\
BYOL-A\cite{niizumi2023byol-a}   & 5.3M &  83.2 {\fontsize{6pt}{6pt}\selectfont $\pm$ 0.6} &  79.7 {\fontsize{6pt}{6pt}\selectfont $\pm$ 0.5} &  93.1 {\fontsize{6pt}{6pt}\selectfont $\pm$ 0.4} & 57.6 {\fontsize{6pt}{6pt}\selectfont $\pm$ 0.2}&  93.3 {\fontsize{6pt}{6pt}\selectfont $\pm$ 0.3} &  63.8 {\fontsize{6pt}{6pt}\selectfont $\pm$ 1.0} &  70.1 {\fontsize{6pt}{6pt}\selectfont $\pm$ 3.6} &  73.1 {\fontsize{6pt}{6pt}\selectfont $\pm$ 0.8} &  37.6 {\fontsize{6pt}{6pt}\selectfont $\pm$ 0.3} & 72.4\\
ATST Base\cite{Li2022ATST}$^{\dagger}$ & 86M &\textbf{94.1 {\fontsize{6pt}{6pt}\selectfont $\pm$ 0.6}}& 85.8 & 95.1 & 72.0 & 97.6 {\fontsize{6pt}{6pt}\selectfont $\pm$ 0.0} & 68.8 {\fontsize{6pt}{6pt}\selectfont $\pm$ 1.3} & 78.9 {\fontsize{6pt}{6pt}\selectfont $\pm$ 3.5} & 76.2 & 32.8 {\fontsize{6pt}{6pt}\selectfont $\pm$ 0.0} & 77.9\\
BEATs$_\text{iter3}$\cite{chen2022beats}$^{\dagger}$ & 90M & 86.9 {\fontsize{6pt}{6pt}\selectfont $\pm$ 1.4} & 84.8 {\fontsize{6pt}{6pt}\selectfont $\pm$ 0.1} & 89.4 {\fontsize{6pt}{6pt}\selectfont $\pm$ 0.1} & 41.4 {\fontsize{6pt}{6pt}\selectfont $\pm$ 0.7} & 94.1 {\fontsize{6pt}{6pt}\selectfont $\pm$ 0.3} & 64.7 {\fontsize{6pt}{6pt}\selectfont $\pm$ 0.8} & 72.6 {\fontsize{6pt}{6pt}\selectfont $\pm$ 4.3} & 75.9 {\fontsize{6pt}{6pt}\selectfont $\pm$ 0.2} & 39.3 {\fontsize{6pt}{6pt}\selectfont $\pm$ 0.4} & 72.1\\
\midrule

\multicolumn{10}{l}{\textit{(i) Baseline: MAE variant}}  \\
MSM-MAE/0.75\cite{niizumi2022msm-mae} & 86M & 89.2 {\fontsize{6pt}{6pt}\selectfont $\pm$ 0.9} & 87.4 {\fontsize{6pt}{6pt}\selectfont $\pm$ 0.2} & 96.0 {\fontsize{6pt}{6pt}\selectfont $\pm$ 0.1} & 73.6 {\fontsize{6pt}{6pt}\selectfont $\pm$ 0.2} & 97.8 {\fontsize{6pt}{6pt}\selectfont $\pm$ 0.2} & 71.2 {\fontsize{6pt}{6pt}\selectfont $\pm$ 0.4} & 79.2 {\fontsize{6pt}{6pt}\selectfont $\pm$ 0.9} & 74.6 {\fontsize{6pt}{6pt}\selectfont $\pm$ 0.9} &\textbf{43.3 {\fontsize{6pt}{6pt}\selectfont $\pm$ 0.3}}& 79.1\\
%\addlinespace[-0.02cm] \hdashline \addlinespace[0.05cm]
\multicolumn{10}{l}{\textit{(ii) Conventional: M2D variants that feed all patches to the target network}}  \\
M2D/0.6 {\scriptsize (all patches$\rightarrow$target)} & 86M & 90.5 {\fontsize{6pt}{6pt}\selectfont $\pm$ 0.4} & 87.0 {\fontsize{6pt}{6pt}\selectfont $\pm$ 0.3} & 96.0 {\fontsize{6pt}{6pt}\selectfont $\pm$ 0.1} &\textbf{76.1 {\fontsize{6pt}{6pt}\selectfont $\pm$ 0.2}}& 97.8 {\fontsize{6pt}{6pt}\selectfont $\pm$ 0.1} & 72.7 {\fontsize{6pt}{6pt}\selectfont $\pm$ 0.8} & 81.3 {\fontsize{6pt}{6pt}\selectfont $\pm$ 2.7} & 75.2 {\fontsize{6pt}{6pt}\selectfont $\pm$ 0.1} & 41.2 {\fontsize{6pt}{6pt}\selectfont $\pm$ 0.2} & 79.7\\
M2D/0.7 {\scriptsize (all patches$\rightarrow$target)} & 86M & 91.0 {\fontsize{6pt}{6pt}\selectfont $\pm$ 0.2} & 86.7 {\fontsize{6pt}{6pt}\selectfont $\pm$ 0.3} & 96.2 {\fontsize{6pt}{6pt}\selectfont $\pm$ 0.1} & 72.5 {\fontsize{6pt}{6pt}\selectfont $\pm$ 0.2} & 98.2 {\fontsize{6pt}{6pt}\selectfont $\pm$ 0.0} & 72.5 {\fontsize{6pt}{6pt}\selectfont $\pm$ 1.1} &\textbf{85.2 {\fontsize{6pt}{6pt}\selectfont $\pm$ 2.4}}& {76.8 {\fontsize{6pt}{6pt}\selectfont $\pm$ 0.2}}& 41.9 {\fontsize{6pt}{6pt}\selectfont $\pm$ 0.2} & 80.1\\
%\addlinespace[-0.02cm] \hdashline \addlinespace[0.05cm]
\multicolumn{10}{l}{\textit{(iii) Ours}}  \\
M2D/0.6 & 86M & 91.6 {\fontsize{6pt}{6pt}\selectfont $\pm$ 0.5} & 87.2 {\fontsize{6pt}{6pt}\selectfont $\pm$ 0.3} &\textbf{96.2 {\fontsize{6pt}{6pt}\selectfont $\pm$ 0.1}}& 75.0 {\fontsize{6pt}{6pt}\selectfont $\pm$ 0.3} & 98.2 {\fontsize{6pt}{6pt}\selectfont $\pm$ 0.1} & 71.4 {\fontsize{6pt}{6pt}\selectfont $\pm$ 0.9} & 83.4 {\fontsize{6pt}{6pt}\selectfont $\pm$ 3.6} & 76.1 {\fontsize{6pt}{6pt}\selectfont $\pm$ 0.1} & 41.7 {\fontsize{6pt}{6pt}\selectfont $\pm$ 0.2} & 80.1\\
M2D/0.7 & 86M & 91.3 {\fontsize{6pt}{6pt}\selectfont $\pm$ 0.6} &\textbf{87.6 {\fontsize{6pt}{6pt}\selectfont $\pm$ 0.2}}& 96.0 {\fontsize{6pt}{6pt}\selectfont $\pm$ 0.1} & 73.4 {\fontsize{6pt}{6pt}\selectfont $\pm$ 0.2} &\textbf{98.3 {\fontsize{6pt}{6pt}\selectfont $\pm$ 0.0}}&\textbf{73.0 {\fontsize{6pt}{6pt}\selectfont $\pm$ 0.7}}& 84.1 {\fontsize{6pt}{6pt}\selectfont $\pm$ 2.7} & 75.7 {\fontsize{6pt}{6pt}\selectfont $\pm$ 0.1} & 42.1 {\fontsize{6pt}{6pt}\selectfont $\pm$ 0.2} & \textbf{80.2}\\

\midrule
\multicolumn{10}{l}{\textit{(Reference supervised learning methods and SSL with )}}  \\
%\textcolor{gray}{ConformerXL-P Non-RA\cite{zhang2022bigssl}} & \textcolor{gray}{-} & \textcolor{gray}{-} & \textcolor{gray}{\textbf{97.5}} & \textcolor{gray}{50.3} & \textcolor{gray}{\textbf{99.7}} & \textcolor{gray}{\textbf{88.2}} & \textcolor{gray}{-} & \textcolor{gray}{-} & \textcolor{gray}{-} \\
\textcolor{gray}{AST\cite{gong2021ast}$^{\dagger}$} & \textcolor{gray}{86M}       &\textcolor{gray}{{93.5 {\fontsize{6pt}{6pt}\selectfont $\pm$ 0.4}}} & \textcolor{gray}{{85.5 {\fontsize{6pt}{6pt}\selectfont $\pm$ 0.2}}} &  \textcolor{gray}{71.8 {\fontsize{6pt}{6pt}\selectfont $\pm$ 0.4}} &  \textcolor{gray}{16.5 {\fontsize{6pt}{6pt}\selectfont $\pm$ 0.4}} &  \textcolor{gray}{81.2 {\fontsize{6pt}{6pt}\selectfont $\pm$ 0.2}} &  \textcolor{gray}{57.9 {\fontsize{6pt}{6pt}\selectfont $\pm$ 0.6}} & \textcolor{gray}{{84.3 {\fontsize{6pt}{6pt}\selectfont $\pm$ 1.8}}} &  \textcolor{gray}{73.2 {\fontsize{6pt}{6pt}\selectfont $\pm$ 0.2}} &  \textcolor{gray}{25.8 {\fontsize{6pt}{6pt}\selectfont $\pm$ 0.2}} & \textcolor{gray}{65.5}\\
\textcolor{gray}{AST-Fusion\#5\#12\cite{niizumi2022composing}} & \textcolor{gray}{86M} &  \textcolor{gray}{{94.2}} &  \textcolor{gray}{85.5} &  \textcolor{gray}{80.4} &  \textcolor{gray}{24.9} &  \textcolor{gray}{87.6} &  \textcolor{gray}{60.7} &  \textcolor{gray}{{82.9}} &  \textcolor{gray}{77.6} &  \textcolor{gray}{34.6} & \textcolor{gray}{69.8} \\
\textcolor{gray}{HTS-AT\cite{Chen2022HTS-AT}$^{\dagger}$} & \textcolor{gray}{31M} &\textcolor{gray}{\textbf{95.7 {\fontsize{6pt}{6pt}\selectfont $\pm$ 0.7}}} & \textcolor{gray}{83.8 {\fontsize{6pt}{6pt}\selectfont $\pm$ 0.1}} & \textcolor{gray}{82.1 {\fontsize{6pt}{6pt}\selectfont $\pm$ 0.3}} & \textcolor{gray}{18.1 {\fontsize{6pt}{6pt}\selectfont $\pm$ 0.4}} & \textcolor{gray}{82.3 {\fontsize{6pt}{6pt}\selectfont $\pm$ 0.3}} & \textcolor{gray}{56.2 {\fontsize{6pt}{6pt}\selectfont $\pm$ 0.6}} & \textcolor{gray}{85.1 {\fontsize{6pt}{6pt}\selectfont $\pm$ 0.5}} & \textcolor{gray}{73.3 {\fontsize{6pt}{6pt}\selectfont $\pm$ 0.8}} & \textcolor{gray}{26.3 {\fontsize{6pt}{6pt}\selectfont $\pm$ 0.5}} & \textcolor{gray}{67.0}\\
\textcolor{gray}{BEATs$_\text{iter3+}$\cite{chen2022beats}$^{\dagger}$} & \textcolor{gray}{90M} & \textcolor{gray}{95.5 {\fontsize{6pt}{6pt}\selectfont $\pm$ 0.3}} & \textcolor{gray}{{87.6 {\fontsize{6pt}{6pt}\selectfont $\pm$ 0.3}}} & \textcolor{gray}{86.7 {\fontsize{6pt}{6pt}\selectfont $\pm$ 0.1}} & \textcolor{gray}{37.0 {\fontsize{6pt}{6pt}\selectfont $\pm$ 0.2}} & \textcolor{gray}{92.5 {\fontsize{6pt}{6pt}\selectfont $\pm$ 0.1}} & \textcolor{gray}{67.6 {\fontsize{6pt}{6pt}\selectfont $\pm$ 1.5}} & \textcolor{gray}{84.6 {\fontsize{6pt}{6pt}\selectfont $\pm$ 0.5}} & \textcolor{gray}{73.1 {\fontsize{6pt}{6pt}\selectfont $\pm$ 0.4}} & \textcolor{gray}{35.7 {\fontsize{6pt}{6pt}\selectfont $\pm$ 0.3}} & \textcolor{gray}{73.4}\\
%\textcolor{gray}{BYOL-S\cite{scheidwasserclow2021serab}} & \textcolor{gray}{-} & \textcolor{gray}{-} & \textcolor{gray}{-} & \textcolor{gray}{-} & \textcolor{gray}{-} & \textcolor{gray}{76.9} & \textcolor{gray}{-} & \textcolor{gray}{-} & \textcolor{gray}{-}\\
\bottomrule
\addlinespace[0.05cm]
\multicolumn{10}{l}{$^{\dagger}$
The results with CI value were obtained in \cite{niizumi2023byol-a}, \cite{niizumi2022M2D}, and this study using publicly available pre-trained models.}\\
\end{tabular}
}
\vspace{-10pt}
\end{table*}

\begin{table}[tb!]
\vspace{-5pt}
\caption{Evaluation \ref{sec:eval-general} General-purpose audio representation: Fine-tuning results with 95\% CI.}
\label{tab:exp-general:sota-ft}
\vspace{-5pt}
\centering
\resizebox{1.0\columnwidth}{!}{%
\begin{tabular}{llllll}
\toprule
 & AS2M & AS20K &     ESC-50 &  SPCV2 &       VC1\\
\vspace{-1pt} Model (/masking ratio){\scriptsize , \# Params}  & mAP & mAP &  acc(\%) &  acc(\%) &    acc(\%)\\
\midrule
\multicolumn{6}{l}{\textit{(Previous SSL methods)}}  \\
%Conformer\cite{Srivastava2022conformer} & 27.6 & 88.0 & - & - \\
DeLoRes-M\cite{Ghosh2022DeLoRes}{\scriptsize , 5.3M} & - & - & - & 96.0 & 62.0 \\
MAE-AST\cite{Baade2022MAE-AST}{\scriptsize , 86M} & - & 30.6 & 90.0 & 98.0 & 63.3 \\
MaskSpec {\scriptsize /-small}\cite{chong2022maskspec}{\scriptsize , 86M} & 47.1 & 32.3 & 90.7 & 97.7 & - \\
SSAST {\scriptsize 250/400/Frame}\cite{gong2022ssast}{\scriptsize , 89M} & - & 31.0 & 88.8 & 98.2 & 80.8 \\
data2vec\cite{baevski2022data2vec}{\scriptsize , 94M} & - & 34.5 & - & - & - \\
%Audio-MAE (global)\cite{huang2022maskedlisten} & 36.6 & 93.6 & 98.3 & 94.1 \\
Audio-MAE {\scriptsize (local)}\cite{huang2022audiomae}{\scriptsize , 86M} & 47.3 & 37.1 & 94.1 & 98.3 & 94.8 \\
ATST Base\cite{Li2022ATST}{\scriptsize , 86M} & - & {37.4} & - & 98.0 & 94.3 \\
BEATs$_\text{iter3}$\cite{chen2022beats}{\scriptsize , 90M} & \textbf{48.0} & {38.3} & {95.6} & 98.3 & - \\
\midrule
\multicolumn{6}{l}{\textit{(i) Baseline: MAE variant}}  \\
%MSM-MAE/0.75\cite{niizumi2022msm-mae}  &  47.0 {\fontsize{6pt}{6pt}\selectfont $\pm$ 0.1} &  37.8 {\fontsize{6pt}{6pt}\selectfont $\pm$ 0.1} &  95.0 {\fontsize{6pt}{6pt}\selectfont $\pm$ 0.1} &  98.4 {\fontsize{6pt}{6pt}\selectfont $\pm$ 0.0} &  96.5 {\fontsize{6pt}{6pt}\selectfont $\pm$ 0.2} \\
MSM-MAE/0.75\cite{niizumi2022msm-mae}{\scriptsize , 86M} & 47.4 {\fontsize{6pt}{6pt}\selectfont $\pm$ 0.1} & 37.9 {\fontsize{6pt}{6pt}\selectfont $\pm$ 0.0} & 95.4 {\fontsize{6pt}{6pt}\selectfont $\pm$ 0.1} & 98.4 {\fontsize{6pt}{6pt}\selectfont $\pm$ 0.0} & 96.6 {\fontsize{6pt}{6pt}\selectfont $\pm$ 0.1} \\
\multicolumn{6}{l}{\textit{(ii) Conventional: M2D variants that feed all patches to the target network}}  \\
%M2D/0.6 (all patches$\rightarrow$target) &  46.9 {\fontsize{6pt}{6pt}\selectfont $\pm$ 0.1} &  37.5 {\fontsize{6pt}{6pt}\selectfont $\pm$ 0.1} &  94.9 {\fontsize{6pt}{6pt}\selectfont $\pm$ 0.3} &  98.4 {\fontsize{6pt}{6pt}\selectfont $\pm$ 0.0} &\textbf{96.6 {\fontsize{6pt}{6pt}\selectfont $\pm$ 0.2}}\\
%M2D/0.7 (all patches$\rightarrow$target) &  46.9 {\fontsize{6pt}{6pt}\selectfont $\pm$ 0.1} &  38.3 {\fontsize{6pt}{6pt}\selectfont $\pm$ 0.1} &  95.5 {\fontsize{6pt}{6pt}\selectfont $\pm$ 0.4} &  98.4 {\fontsize{6pt}{6pt}\selectfont $\pm$ 0.1} &  96.1 {\fontsize{6pt}{6pt}\selectfont $\pm$ 0.1} \\
M2D/0.6 {\scriptsize (all patches$\rightarrow$target), 86M} & 47.4 {\fontsize{6pt}{6pt}\selectfont $\pm$ 0.2} & 37.6 {\fontsize{6pt}{6pt}\selectfont $\pm$ 0.1} & 95.4 {\fontsize{6pt}{6pt}\selectfont $\pm$ 0.3} & 98.4 {\fontsize{6pt}{6pt}\selectfont $\pm$ 0.0} &\textbf{96.6 {\fontsize{6pt}{6pt}\selectfont $\pm$ 0.3}}\\
M2D/0.7 {\scriptsize (all patches$\rightarrow$target), 86M} & 47.6 {\fontsize{6pt}{6pt}\selectfont $\pm$ 0.0} & 38.3 {\fontsize{6pt}{6pt}\selectfont $\pm$ 0.0} & 95.8 {\fontsize{6pt}{6pt}\selectfont $\pm$ 0.2} & 98.4 {\fontsize{6pt}{6pt}\selectfont $\pm$ 0.0} & 96.0 {\fontsize{6pt}{6pt}\selectfont $\pm$ 0.1} \\
\multicolumn{6}{l}{\textit{(iii) Ours}}  \\
%M2D/0.6 &  46.9 {\fontsize{6pt}{6pt}\selectfont $\pm$ 0.1} &  38.1 {\fontsize{6pt}{6pt}\selectfont $\pm$ 0.1} &  95.2 {\fontsize{6pt}{6pt}\selectfont $\pm$ 0.1} &\textbf{98.5 {\fontsize{6pt}{6pt}\selectfont $\pm$ 0.1}}&  96.5 {\fontsize{6pt}{6pt}\selectfont $\pm$ 0.3} \\
%M2D/0.7 & {47.1 {\fontsize{6pt}{6pt}\selectfont $\pm$ 0.1}}&\textbf{38.5 {\fontsize{6pt}{6pt}\selectfont $\pm$ 0.1}}& {95.5 {\fontsize{6pt}{6pt}\selectfont $\pm$ 0.1}}&  98.4 {\fontsize{6pt}{6pt}\selectfont $\pm$ 0.1} &  96.3 {\fontsize{6pt}{6pt}\selectfont $\pm$ 0.2} \\
M2D/0.6{\scriptsize , 86M} & 47.7 {\fontsize{6pt}{6pt}\selectfont $\pm$ 0.2} & 38.4 {\fontsize{6pt}{6pt}\selectfont $\pm$ 0.1} & 95.6 {\fontsize{6pt}{6pt}\selectfont $\pm$ 0.1} &\textbf{98.5 {\fontsize{6pt}{6pt}\selectfont $\pm$ 0.1}}& 96.5 {\fontsize{6pt}{6pt}\selectfont $\pm$ 0.1} \\
M2D/0.7{\scriptsize , 86M} & {47.9 {\fontsize{6pt}{6pt}\selectfont $\pm$ 0.0}}&\textbf{38.6 {\fontsize{6pt}{6pt}\selectfont $\pm$ 0.1}}&\textbf{96.0 {\fontsize{6pt}{6pt}\selectfont $\pm$ 0.2}}& 98.4 {\fontsize{6pt}{6pt}\selectfont $\pm$ 0.1} & 96.3 {\fontsize{6pt}{6pt}\selectfont $\pm$ 0.2} \\

\midrule
\multicolumn{6}{l}{\textit{(Reference supervised learning method results)}}  \\
\textcolor{gray}{AST {\scriptsize (Single),-P/S}\cite{gong2021ast}{\scriptsize , 86M}}   & \textcolor{gray}{45.9} & \textcolor{gray}{34.7} & \textcolor{gray}{95.6} & \textcolor{gray}{98.11} & \textcolor{gray}{-} \\
\textcolor{gray}{EAT-S/M\cite{gazneli2022EAT}{\scriptsize , 25.5M}}  & \textcolor{gray}{42.6} & \textcolor{gray}{-} & \textcolor{gray}{96.3} & \textcolor{gray}{98.15} & \textcolor{gray}{-} \\
\textcolor{gray}{PaSST\cite{Koutini2022passt}{\scriptsize , 86M}} & \textcolor{gray}{47.1} & \textcolor{gray}{-} & \textcolor{gray}{96.8} & \textcolor{gray}{-} & \textcolor{gray}{-}\\
\textcolor{gray}{HTS-AT\cite{Chen2022HTS-AT}{\scriptsize , 31M}} & \textcolor{gray}{47.1} & \textcolor{gray}{-} & \textcolor{gray}{{97.0}} & \textcolor{gray}{98.0} & \textcolor{gray}{-}\\
\textcolor{gray}{BEATs$_\text{iter3+}$\cite{chen2022beats}{\scriptsize , 90M}} & \textcolor{gray}{\textbf{48.6}} & \textcolor{gray}{\textbf{41.8}} & \textcolor{gray}{\textbf{98.1}} & \textcolor{gray}{98.1} & \textcolor{gray}{-} \\
\bottomrule\\
\end{tabular}
}
\vspace{-20pt}
\end{table}

\vspace{0.1cm}
\subsubsection{Comparison with Baselines} \label{sec:exp-general:baseline}
%提案手法のベースとなったMAE、ターゲットネットワークに全てのパッチを与える従来の方法、これらに対してM2Dの結果を比較する。表III, IVに記載の(iii), (i), (ii)がそれぞれの結果に対応する。これらは全て同じコードベースで事前学習、評価され、違いは学習アルゴリズムのみである。
We compare the results for (i) MAE on which our methods are based, (ii) the conventional method that gives all patches to the target network, and (iii) our M2D in Tables \ref{tab:exp-general:sota-le} and \ref{tab:exp-general:sota-ft}.
We used the same codebase to obtain these results; only the learning algorithm differed.

%\vspace{0.1cm}
%\noindent{Linear evaluation:}\hspace{0.1cm}
%IMAEと比べてIIIM2Dは平均+0.7または+1.0性能を向上し、全体的に性能を上回っている。事前学習において入力信号の再構成の代わりに表現を予測することの有効性を示す。一方でピッチの分類タスクSurgeではMAEが+1.2性能を上回っており、周波数に関する情報をより抽出できていると考えられる。IICONVに対するIIIM2Dの性能向上は+0.2または+0.5と小さいが、全体的に性能を改善している。ターゲットネットワークに全ての入力信号を与えて教師信号の表現を得るのではなく、マスクパッチのみを与える提案により学習効果を改善したことを示す。
Table \ref{tab:exp-general:sota-le} shows linear evaluation results.
Compared to (i) MAE, (iii) M2D outperforms MAE with about a $+1.0$ average performance improvement on most tasks, demonstrating the effectiveness of predicting the representation instead of reconstructing the input signal during pre-training. On the other hand, MAE outperforms ours on the pitch classification task Surge, suggesting that MAE extracts more frequency-related information.
The average performance improvement of (iii) M2D over (ii) Conventional is $+0.4$ at masking ratios of $0.6$ and $+0.1$ at $0.7$, which is small, but indicates an overall performance improvement, supporting the effectiveness of the proposed method to feed only masked patches instead of providing all input signals to the target network.% to obtain a representation of the training signal.

%\vspace{0.1cm}
%\noindent{Fine-tuning:}\hspace{0.1cm}
%IIIはIやIIと比べて同等か少し上回る性能を示しており、提案手法が過去の手法を改善していることを示した。
Table \ref{tab:exp-general:sota-ft} shows fine-tuning results.
(iii) M2D performed better than (i) MAE and (ii) Conventional on most tasks, while underperforming others on VC1, which requires detailed information to differentiate the 1251 speakers. We think the reason is similar to that for the performance drop on Surge in Table \ref{tab:exp-general:sota-le}; M2D may have learned to model more abstract information than the details.
%全体的に、これらの結果は、提案法の有効性を裏付ける。
Overall, these results support the effectiveness of M2D.

\vspace{0.1cm}
\subsubsection{Comparison with SOTA Models} \label{sec:exp-general:sota}
M2D achieves the best results on four out of nine tasks in linear evaluation (Table \ref{tab:exp-general:sota-le}) and three out of five in fine-tuning (Table \ref{tab:exp-general:sota-ft}), confirming its effectiveness in learning general-purpose audio representations.
While achieving good performance on average, M2D underperforms with ATST on ESC-50 in linear evaluation and BEATs on AS2M in fine-tuning. ATST learns representations invariant to data augmentation, and BEATs learn from vector quantized representations. They may benefit from learning a representation robust to minor differences in a sound to perform well on these tasks.
%M2DはSOTAモデルと比較して同等、あるいは上回る結果を示し、汎用音響信号表現の事前学習において提案するM2Dの有効性を裏付ける。特に表IIのlinear evaluationの結果においてM2Dは、US8K、CRM-D、GTZAN、Surgeにおいて約3パーセントポイント以上他の結果を上回ることを示す。表IVのfine-tuningでは、VC1においてM2Dは1.5ポイント以上他のSOTA結果を上回ることを示す。
%In particular, the linear evaluation results in Table \ref{tab:exp-general:sota-le} show that M2D outperforms the other methods in US8K, CRM-D, GTZAN, and Surge by about 3 percentage points or more.
%The fine-tuning in Table \ref{tab:exp-general:sota-ft} shows that M2D outperforms other SOTA methods by more than 1.5 percentage points in VC1.

\begin{figure}[tbp]
  %\vspace{-5pt}
  \centering
  \includegraphics[width=0.9\columnwidth]{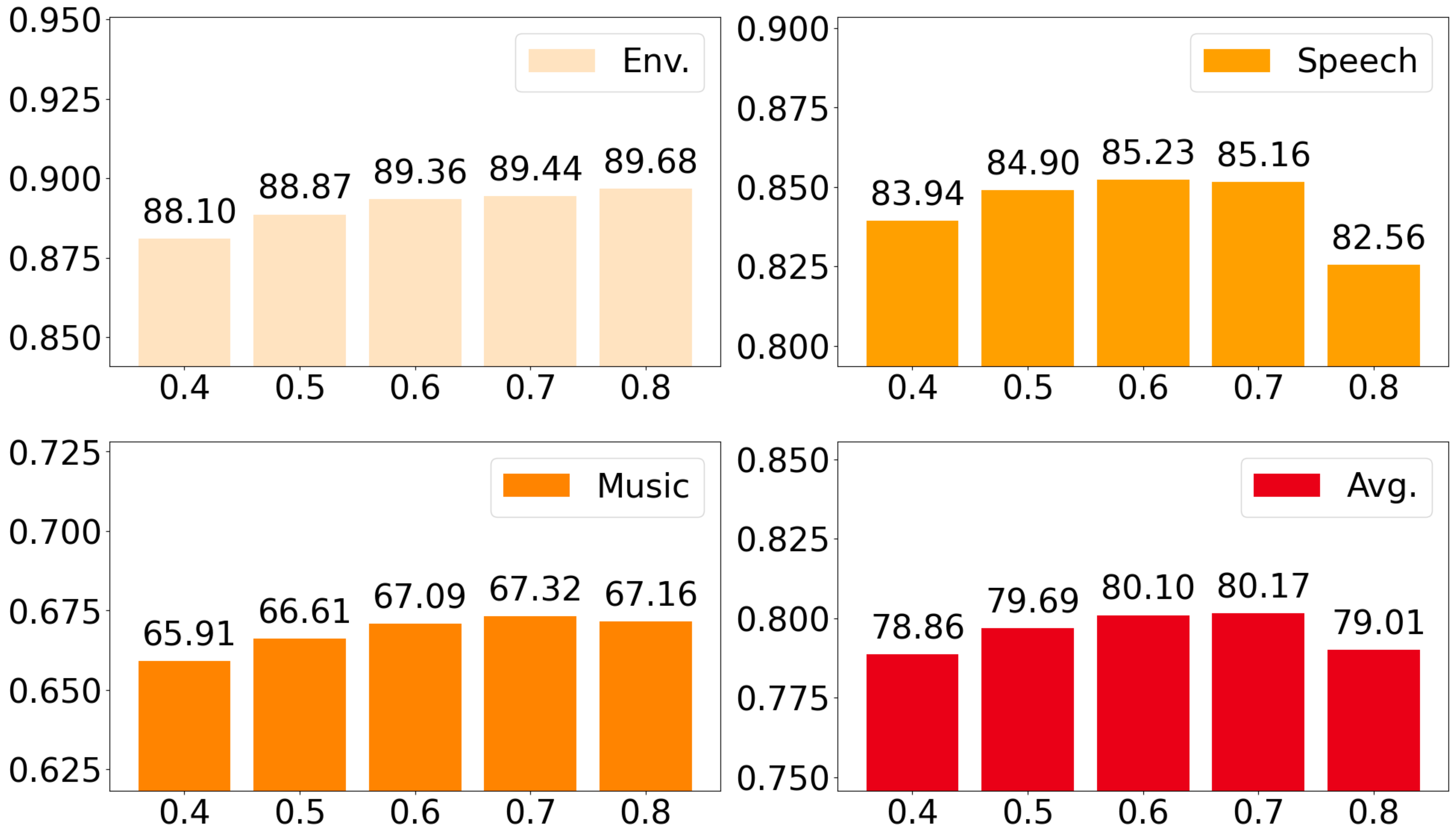} 
  \vspace{-5pt}
  \caption{Evaluation \ref{sec:eval-general} General-purpose audio representation: Masking ratio ablation average linear evaluation results (\%).}
  \label{fig:mask-ratio-results}
    \vspace{-10pt}
\end{figure}

%%%%%%%%%%%%%%%%%
\vspace{0.1cm}
\subsubsection{Masking Ratio Ablations} \label{sec:exp-general:mask-ratio}
%%%%%%%%%%%%%%%%%
%様々なマスク率でのLinear evaluation性能を評価する。
We evaluated the impact of various masking ratios using linear evaluation.
%図4に3つのタスクグループの結果と全体の平均を示す。平均結果はマスク率0.7が最も良いことを示している。
Fig. \ref{fig:mask-ratio-results} shows the results for the three task groups and the overall average. The best average result (Avg.) was obtained at 0.7.
%しかしながら、3つのグループはそれぞれ違った傾向を示しており、最適なマスク率がタスクによって異なることを示している。環境音タスクでは0.8で最も良い結果となり、音声タスクでは0.6がピークである。これらに対して音楽タスクでは0.8に最大値を持つ。このように、一つの共通な最適値を設定するのが困難である。
However, the three groups show different trends, indicating that the optimal masking ratio depends on the task. The environmental sound tasks show the best result at 0.8 (or more), while the speech and music tasks have a peak at 0.6 and 0.7, respectively. %Thus, it is difficult to set a single common optimal value.

%我々は最適なマスク率の違いは、MAEで議論されたように目的とする情報の密度の違いによるものと考えられる。音声タスクの音は連続する音素から成り、多くのマスクは予測が困難になる一方、環境音や音楽タスクは連続する長い音(エアコンの音や長い音符の楽器音など)も含まれるためマスク率が高くても予測がしやすいと考えられる。
We suspect that these optimal masking ratios are due to differences in the density of the target information, as discussed in the MAE paper.
The sounds in the speech tasks consist of successive phonemes, making many masks difficult to predict, while the environmental and music tasks include long continuous sounds (e.g., the sound of an air conditioner or long notes of musical instruments), making them easier to predict even at higher masking ratios.
%実験結果は学習される表現の持つ様々な音に対する有用性がマスク率により変化する、更には制御できることを示していると考えられる。
The results may indicate that the usefulness of the learned representations for various sounds can be varied and even controlled by masking ratios.

\begin{table*}[tb!]
\vspace{-5pt}
\caption{Evaluation \ref{sec:eval-audioset} Specialized representation for AudioSet: Linear evaluation results (\%) with 95\% CI.}
\vspace{-10pt}
\label{tab:exp-audioset:sota-le}
\centering
\resizebox{0.9\textwidth}{!}{%
\begin{tabular}{llllllllllll} \toprule
 &&  \multicolumn{2}{c}{Env. sound tasks} & \multicolumn{4}{c}{Speech tasks} & \multicolumn{3}{c}{Music tasks} \\
\cmidrule(lr){3-4} \cmidrule(lr){5-8} \cmidrule(lr){9-11} 
Model (/masking ratio) & \# Params &    ESC-50 &    US8K &    SPCV2 & VC1 &  VF &  CRM-D &  GTZAN &  NSynth &  Surge & Avg. \\
\midrule
AST\cite{gong2021ast}$^{\dagger}$    & 86M   & {93.5 {\fontsize{6pt}{6pt}\selectfont $\pm$ 0.4}}& {85.5 {\fontsize{6pt}{6pt}\selectfont $\pm$ 0.2}}&  71.8 {\fontsize{6pt}{6pt}\selectfont $\pm$ 0.4} &  16.5 {\fontsize{6pt}{6pt}\selectfont $\pm$ 0.4} &  81.2 {\fontsize{6pt}{6pt}\selectfont $\pm$ 0.2} &  57.9 {\fontsize{6pt}{6pt}\selectfont $\pm$ 0.6} & {84.3 {\fontsize{6pt}{6pt}\selectfont $\pm$ 1.8}}&  73.2 {\fontsize{6pt}{6pt}\selectfont $\pm$ 0.2} &  25.8 {\fontsize{6pt}{6pt}\selectfont $\pm$ 0.2} & 65.5\\
%AST-Fusion\#5\#12\cite{niizumi2022composing} &  \textcolor{black}{{94.2}} &  \textcolor{black}{85.5} &  \textcolor{black}{80.4} &  \textcolor{black}{24.9} &  \textcolor{black}{87.6} &  \textcolor{black}{60.7} &  \textcolor{black}{{82.9}} &  \textbf{77.6} &  \textcolor{black}{34.6} & \textcolor{black}{69.8} \\
HTS-AT\cite{Chen2022HTS-AT}$^{\dagger}$ & 31M & {95.7 {\fontsize{6pt}{6pt}\selectfont $\pm$ 0.7}}& 83.8 {\fontsize{6pt}{6pt}\selectfont $\pm$ 0.1} & 82.1 {\fontsize{6pt}{6pt}\selectfont $\pm$ 0.3} & 18.1 {\fontsize{6pt}{6pt}\selectfont $\pm$ 0.4} & 82.3 {\fontsize{6pt}{6pt}\selectfont $\pm$ 0.3} & 56.2 {\fontsize{6pt}{6pt}\selectfont $\pm$ 0.6} & 85.1 {\fontsize{6pt}{6pt}\selectfont $\pm$ 0.5} & 73.3 {\fontsize{6pt}{6pt}\selectfont $\pm$ 0.8} & 26.3 {\fontsize{6pt}{6pt}\selectfont $\pm$ 0.5} & 67.0\\
%ATST Base\cite{Li2022ATST} & 94.1 {\fontsize{6pt}{6pt}\selectfont $\pm$ 0.6} & 85.8 & 95.1 & 72.0 & 97.6 {\fontsize{6pt}{6pt}\selectfont $\pm$ 0.0} & 68.8 {\fontsize{6pt}{6pt}\selectfont $\pm$ 1.3} & 78.9 {\fontsize{6pt}{6pt}\selectfont $\pm$ 3.5} &\textbf{76.2}& 32.8 {\fontsize{6pt}{6pt}\selectfont $\pm$ 0.0} & 77.9\\
%BEATs$_\text{iter3}$\cite{chen2022beats} & 86.9 {\fontsize{6pt}{6pt}\selectfont $\pm$ 1.4} & 84.8 {\fontsize{6pt}{6pt}\selectfont $\pm$ 0.1} & 89.4 {\fontsize{6pt}{6pt}\selectfont $\pm$ 0.1} & 41.4 {\fontsize{6pt}{6pt}\selectfont $\pm$ 0.7} & 94.1 {\fontsize{6pt}{6pt}\selectfont $\pm$ 0.3} & 64.7 {\fontsize{6pt}{6pt}\selectfont $\pm$ 0.8} & 72.6 {\fontsize{6pt}{6pt}\selectfont $\pm$ 4.3} & 75.9 {\fontsize{6pt}{6pt}\selectfont $\pm$ 0.2} & 39.3 {\fontsize{6pt}{6pt}\selectfont $\pm$ 0.4} & 72.1\\
BEATs$_\text{iter3+}$\cite{chen2022beats}$^{\dagger}$ & 90M & 95.5 {\fontsize{6pt}{6pt}\selectfont $\pm$ 0.3} & 87.6 {\fontsize{6pt}{6pt}\selectfont $\pm$ 0.3} & 86.7 {\fontsize{6pt}{6pt}\selectfont $\pm$ 0.1} & 37.0 {\fontsize{6pt}{6pt}\selectfont $\pm$ 0.2} & 92.5 {\fontsize{6pt}{6pt}\selectfont $\pm$ 0.1} & 67.6 {\fontsize{6pt}{6pt}\selectfont $\pm$ 1.5} & 84.6 {\fontsize{6pt}{6pt}\selectfont $\pm$ 0.5} & 73.1 {\fontsize{6pt}{6pt}\selectfont $\pm$ 0.4} & 35.7 {\fontsize{6pt}{6pt}\selectfont $\pm$ 0.3} & 73.4\\
\midrule
%\multicolumn{10}{l}{\textit{(i) Baseline M2D}}  \\
%\addlinespace[-0.02cm] \hdashline \addlinespace[0.05cm]
%\multicolumn{10}{l}{\textit{(Ours)}}  \\
M2D-AS/0.7 & 86M & \textbf{96.1 {\fontsize{6pt}{6pt}\selectfont $\pm$ 0.6}}&\textbf{89.8 {\fontsize{6pt}{6pt}\selectfont $\pm$ 0.2}}& 95.7 {\fontsize{6pt}{6pt}\selectfont $\pm$ 0.0} & 68.3 {\fontsize{6pt}{6pt}\selectfont $\pm$ 0.4} &\textbf{98.4 {\fontsize{6pt}{6pt}\selectfont $\pm$ 0.0}}&\textbf{73.1 {\fontsize{6pt}{6pt}\selectfont $\pm$ 0.6}}&\textbf{86.9 {\fontsize{6pt}{6pt}\selectfont $\pm$ 0.6}}& 75.2 {\fontsize{6pt}{6pt}\selectfont $\pm$ 0.2} & 41.7 {\fontsize{6pt}{6pt}\selectfont $\pm$ 0.3} & \textbf{80.6}\\

\midrule
\multicolumn{10}{l}{\textit{(Reference: The M2D model pre-trained in Section \ref{sec:eval-general})}} \\
M2D/0.7 & 86M & 91.3 {\fontsize{6pt}{6pt}\selectfont $\pm$ 0.6} & 87.6 {\fontsize{6pt}{6pt}\selectfont $\pm$ 0.2} &\textbf{96.0 {\fontsize{6pt}{6pt}\selectfont $\pm$ 0.1}}&\textbf{73.4 {\fontsize{6pt}{6pt}\selectfont $\pm$ 0.2}}& 98.3 {\fontsize{6pt}{6pt}\selectfont $\pm$ 0.0} & 73.0 {\fontsize{6pt}{6pt}\selectfont $\pm$ 0.7} & 84.1 {\fontsize{6pt}{6pt}\selectfont $\pm$ 2.7} & 75.7 {\fontsize{6pt}{6pt}\selectfont $\pm$ 0.1} &\textbf{42.1 {\fontsize{6pt}{6pt}\selectfont $\pm$ 0.2}}& 80.2\\

\bottomrule
\addlinespace[0.05cm]
\multicolumn{10}{l}{$^{\dagger}$
The results were obtained in \cite{niizumi2023byol-a}, \cite{niizumi2022M2D}, and this study using publicly available pre-trained models.}\\
\end{tabular}
}
\vspace{-5pt}
\end{table*}

%%%%%%%%%%%%%%%%%%%%%%%%%%%%%%%%%%%%%%%%%%%%%%%%%%%%%%%%%%%%%%%%%%%%
% AudioSet
%%%%%%%%%%%%%%%%%%%%%%%%%%%%%%%%%%%%%%%%%%%%%%%%%%%%%%%%%%%%%%%%%%%%
\subsection{Evaluation: Specialized Representation for AudioSet}\label{sec:eval-audioset}
%本実験はM2D-XがAudioSetに特化した効果的な表現を学習できることを検証する。
In this experiment, we validated that M2D-X learns an effective representation specializing in AudioSet, conventionally trained in supervised learning using \textit{labels}, such as PANNs\cite{kong2020panns} and AST\cite{gong2021ast}.
These models demonstrated strong performance for AudioSet and similar tasks such as ESC-50 by learning training signals of multi-label classification with 527 classes.
%音声分野同様、これらのモデルはSOTA性能を激しく競っている。
As in the speech domain, these models intensely compete for SOTA AudioSet performance.

%\vspace{0.1cm}
%\noindent\textbf{Approach.}\hspace{0.2cm}
%我々はM2D-Xを使い、offline networkの追加タスクとしてラベルの教師あり学習を利用する。図3(b)のシナリオi)の構成を利用し、音響特徴量HHをaverage poolingにして線形層で変換したHHDを予測値logitとして得て、HHDとラベルのmulti-hot encodingであるYCHの間でBCEにより損失を計算する。
We use M2D-X and utilize supervised learning of labels as an additional task in the offline network. Using the configuration in scenario i) from Fig. \ref{fig:system-x}(b), we obtain the predicted logits $\hat{h'} \in R^{B\times 527}$ transformed by a linear layer with the temporal average pooling of audio features $\hat{h} \in R^{B\times N_T \times D}$ and calculate the binary cross-entropy (BCE) loss between the $\hat{h'}$ and the $\tilde{y}$, which is a multi-hot encoding of the label.
%実験方法はSection Xにほとんど準ずる一方、この実験ではAudioSetのラベルを事前学習に利用する。
The experimental details follow those in Section \ref{sec:eval-general}, while we used AudioSet labels for pre-training and used only a masking ratio of 0.7, suitable for environmental sound tasks.
We did not use the background noise in this experiment.
The M2D-X loss weights were set to $L_\text{off}=1.0$ and $L_\text{m2d}=1.0$. We specifically denoted M2D-X with these settings as \textit{M2D-AS}.

\begin{table}[tb!]
\vspace{-5pt}
\caption{Evaluation \ref{sec:eval-audioset} Specialized representation for AudioSet: Fine-tuning results with 95\% CI.}
\label{tab:exp-audioset:sota-ft}
\vspace{-5pt}
\centering
\resizebox{1.0\columnwidth}{!}{%
\begin{tabular}{llllll}
\toprule
  & AS2M & AS20K &     ESC-50 &  SPCV2 &       VC1\\
\vspace{-1pt} Model (/masking ratio), \# Params  & mAP & mAP &  acc(\%) &  acc(\%) &    acc(\%)\\
\midrule
\textcolor{black}{AST {\scriptsize (Single),-P/S}\cite{gong2021ast}}{\scriptsize , 86M}  & \textcolor{black}{45.9} & \textcolor{black}{34.7} & \textcolor{black}{95.6} & \textcolor{black}{98.11} & \textcolor{black}{-} \\
\textcolor{black}{HTS-AT\cite{Chen2022HTS-AT}}{\scriptsize , 31M} & \textcolor{black}{47.1} & \textcolor{black}{-} & \textcolor{black}{{97.0}} & \textcolor{black}{98.0} & \textcolor{black}{-}\\
%\textcolor{black}{EAT-S/M\cite{gazneli2022EAT}}  & \textcolor{black}{42.6} & \textcolor{black}{-} & \textcolor{black}{96.3} & \textcolor{black}{98.15} & \textcolor{black}{-} \\
%\textcolor{black}{PaSST\cite{Koutini2022passt}} & \textcolor{black}{47.1} & \textcolor{black}{-} & \textcolor{black}{96.8} & \textcolor{black}{-} & \textcolor{black}{-}\\
%ATST Base\cite{Li2022ATST} & - & {37.4} & - & 98.0 & 94.3 \\
%BEATs$_\text{iter3}$\cite{chen2022beats} & {48.0} & {38.3} & {95.6} & 98.3 & - \\
\textcolor{black}{BEATs$_\text{iter3+}$\cite{chen2022beats}}{\scriptsize , 90M} & \textcolor{black}{\textbf{48.6}} & \textcolor{black}{\textbf{41.8}} & \textcolor{black}{\textbf{98.1}} & \textcolor{black}{98.1} & \textcolor{black}{-} \\
\midrule
%\multicolumn{6}{l}{\textit{(Ours)}} \\
M2D-AS/0.7{\scriptsize , 86M} & {48.5 {\fontsize{6pt}{6pt}\selectfont $\pm$ 0.0}}&\textbf{41.8 {\fontsize{6pt}{6pt}\selectfont $\pm$ 0.1}}& {97.2 {\fontsize{6pt}{6pt}\selectfont $\pm$ 0.1}}& 98.4 {\fontsize{6pt}{6pt}\selectfont $\pm$ 0.1} & 95.6 {\fontsize{6pt}{6pt}\selectfont $\pm$ 0.2} \\

\midrule
\multicolumn{6}{l}{\textit{(Reference: The M2D model pre-trained in Section \ref{sec:eval-general})}} \\
M2D/0.7{\scriptsize , 86M} & 47.9 {\fontsize{6pt}{6pt}\selectfont $\pm$ 0.0} & 38.6 {\fontsize{6pt}{6pt}\selectfont $\pm$ 0.1} & 96.0 {\fontsize{6pt}{6pt}\selectfont $\pm$ 0.2} &\textbf{98.4 {\fontsize{6pt}{6pt}\selectfont $\pm$ 0.1}}&\textbf{96.3 {\fontsize{6pt}{6pt}\selectfont $\pm$ 0.2}}\\

\bottomrule\\
\end{tabular}
}
\vspace{-15pt}
\end{table}

\vspace{0.2cm}
\subsubsection{Comparison with SOTA models}\label{sec:exp-audioset:results}
%表5と6に示す結果は、AudioSet教師あり学習モデルAST、HTS-ATに加え、SSLに教師データも利用して我々に近い学習を行ったBEATs+を比較する。M2D-Xは目標タスクとするAudioSet(表6のAS2M, AS20K)と表5のESC-50において、M2Dの性能を向上させ、SOTAモデルの結果に比類する結果を示す。特に同じ評価条件でfrozenモデルを比較する線形評価で一番良い性能を示すことから、M2D-Xはtop-level性能を学習可能であることを示した。学習時間の観点で見ると、BEATsは複数繰り返しの学習を要し、事前学習に16GPUを利用するのに対して、M2D-Xは4 GPUによる3日間で学習可能である。この２つのモデルの性能は、AS2MではBEATs+の結果が上回り、AS20KではM2D-Xの結果が上回っており、同等の性能が得られたと考えられる。
The results shown in Tables \ref{tab:exp-audioset:sota-le} and \ref{tab:exp-audioset:sota-ft} compare ours with the AudioSet supervised learning models AST and HTS-AT, plus BEATs$_\text{iter3+}$, which used supervised data in SSL similar to ours.
M2D-AS improved M2D performance on the target task AudioSet (AS2M, AS20K in Table \ref{tab:exp-audioset:sota-ft}) and ESC-50 in Tables \ref{tab:exp-audioset:sota-le} and \ref{tab:exp-audioset:sota-ft} with results comparable to SOTA models, notably, a close result of AS2M/AS20K with BEATs$_\text{iter3+}$, suggesting the performance limit of a similar learning paradigm. The results demonstrate that the M2D-X framework is capable of learning specialized representations for AudioSet with top-level performance.

Unlike previous models, M2D-AS showed the best average result in Table \ref{tab:exp-audioset:sota-le}, demonstrating the general-purpose performance as in M2D.
The VC1 result exceptionally dropped by 5.1, but 68.3 \% is still a good performance.
Other models typically do not perform at the top level on non-target tasks, such as speech tasks, as shown in our AST/HTS-AT/BEATs$_\text{iter3+}$ results in Table \ref{tab:exp-audioset:sota-le} and our previous study \cite{niizumi2023byol-a}.
Our results indicate that combining supervised learning and M2D SSL can achieve specialized and general-purpose performance in a single representation.

In terms of training time, M2D-X can be pre-trained within 2.5 days on 4 A100 GPUs, while BEATs require multiple iterations and 16 GPUs for pre-training.
%The performance of the two models is considered comparable, with BEATs$_\text{iter3+}$ outperforming M2D-AS for AS2M and M2D-AS outperforming BEATs$_\text{iter3+}$ for AS20K.

%%%%%%%%%%%%%%%%%%%%%%%%%%%%%%%%%%%%%%%%%%%%%%%%%%%%%%%%%%%%%%%%%%%%
% Speech
%%%%%%%%%%%%%%%%%%%%%%%%%%%%%%%%%%%%%%%%%%%%%%%%%%%%%%%%%%%%%%%%%%%%
\subsection{Evaluation: Specialized Representation for Speech} \label{sec:eval-speech}
%本実験では提案法が競争の激しい分野で最先端性能の表現を学習できることを検証する。そのために、提案法を用いて音声表現を事前学習し、主流な評価ベンチマークSUPERBを用いて最先端の音声モデルとの比較を行う。
In this experiment, we validated that M2D and M2D-X learn effective representations with top-level performance in a highly competitive domain. We pre-trained our models to learn speech representations and compared them with SOTA speech models, expressly HuBERT~\cite{Hsu2021HuBERT} as a baseline, on the benchmark SUPERB\cite{yang2021superb}.

\vspace{0.1cm}
\noindent\textbf{Approach.}\hspace{0.2cm}
We first adapted M2D for speech, starting with ablation studies of M2D settings (training dataset, patch size, and input duration) to assess the best M2D performance in this domain.

%その後、ベースラインのHuBERTに近い性能向上のためのドメイン技術を組み込むために、M2DをM2D-Xに拡張した。HuBERT は、マスクされたタイムス テップのベクトル化された特徴を予測するように反復学習し、これらのターゲット特徴は、前の反復モデルからの特徴の k 平均クラスタリングによって作成されたコードブック内の少数のコード（500）である。従って、HuBERT がクラスタ化された特徴を出力すると仮定し、図 XX(b)からシナリオ ii)の構成のように HuBERT 出力特徴を抽出する。
We then extended M2D to M2D-X to incorporate knowledge learned from the domain techniques for obtaining performance closer to the baseline HuBERT.
HuBERT iteratively learns to predict vector-quantized features of masked timesteps, and these target features are a small number (500) of codes in a codebook created by k-means clustering of the features from the previous iteration model. Thus, we assume HuBERT to output clustered target features and distilled HuBERT output features as in the configuration in scenario~ii) from Fig. \ref{fig:system-x}(b).

%この実験ではデノイズタスクによる学習も同時に行うため、背景ノイズを利用する。従来手法WavLMはマスク予測に加えてデノイズタスクを学習することで性能を向上させており、このドメインテクニックを利用する。M2D-Xフレームワークにおいては背景ノイズを利用することで、M2Dにはノイジースピーチを、オフラインモデルにはクリーンスピーチを与えることで、デノイズタスクを構成する。
We also used background noise to jointly learn from a denoising task. The conventional method WavLM\cite{Chen2022WavLM} gains performance by training a denoising task in addition to masked prediction, and we utilized this domain technique.
The M2D-X framework provides noisy speech for M2D and clean speech for the offline model, composing a denoising task.

\vspace{0.1cm}
\noindent\textbf{Pre-training details.}\hspace{0.2cm}
%M2D-X のオフライン・ネットワークの蒸留構成では、線形レイヤーを使用して、音声特徴 HH を時間フレーム毎の特徴 HHD にマッピングする（NT は時間フレーム数、DH は HuBERT の特徴次元）。オフラインエンコーダはHuBERTであり、式(2)の損失計算、つまりl2正規化によるMSEを用いて、その出力特徴YHを抽出する。
In a distillation configuration of the offline network in M2D-X, we used a linear layer to map the audio feature $\hat{h} \in R^{B\times N_T \times D}$ to the feature $\hat{h'} \in R^{B\times N_T \times D_H}$, a per-time-frame feature, where $N_T$ is the number of time frames, and $D_H$ is the feature dimension of HuBERT.
We used the ninth transformer layer output of the second-iteration HuBERT Base model\footnote{\scriptsize\url{https://huggingface.co/facebook/hubert-base-ls960}} as an offline encoder, following WavLM\cite{Chen2022WavLM}, and distilled its output feature $\hat{y} \in R^{B\times N_T \times D_H}$ using the loss calculation of \eqref{eq:eq-byol-mse}, an MSE with $l_2$-normalization.

%我々は事前学習の設定を、EMA decayなどを含め基本的にセクションIII-Bと同じものを利用したが、masking ratioは0.6、入力音声長は4-6秒、パッチサイズを80X2として音声に適応した。M2D-Xでは、損失重みをLoff=0.5、Lm2d=1.0とし、オフラインエンコーダにはWavLMに倣いHuBERT Baseの第9層の表現を利用した。ターゲットデータセットには281,241サンプル(960h)のLibriSpeechを、背景ノイズデータセットにはAudioSetのbalancedとunbalanced train segmentsから2,005,132サンプル(5,569h)を利用した。このように音声に適応したM2D-Xを、この章では特にM2D-S (Speech)と呼ぶ。
We used the pre-training settings based on those in Section \ref{sec:eval-general} with a masking ratio of 0.6, number of epochs to 1000, warm-up epochs to 60, input speech length of 4, 5, and 6 s, and patch size of $80\times 2$ and $80\times 4$.
We used LibriSpeech\cite{Panayotov2015LibrispeechAA} with 281,241 samples (960 h) from all training splits (LS-960) for the pre-training dataset. For the background noise dataset, we used AudioSet\cite{gemmeke2017audioset} (AS) with 2,005,132 samples (5569 h).
We used M2D-X loss weights of $L_\text{off}=0.5$ and $L_\text{m2d}=1.0$. 
We denote M2D-X adapted to speech as \textit{M2D-S} (Speech).

\begin{table}[tb!]
%\vspace{-5pt}
\caption{Evaluation \ref{sec:eval-speech} Specialized Representation for Speech:\\
Dataset noise ratio ablations.\\
\footnotesize{(input duration $T=2.08\text{ s}$ and patch size $80\times 4$)}}
\vspace{-5pt}
\label{tab:results-data-mix}
\centering
\resizebox{\columnwidth}{!}{%
\begin{tabular}{llllll|ll}
\toprule
 &    PR &     KS &     IC &     SID &    ER & \textcolor{black}{ENV}  & \textcolor{black}{MUS} \\
\vspace{-1pt} Dataset noise ratio $\eta$       &  PER$\downarrow$ & Acc$\uparrow$& Acc$\uparrow$& Acc$\uparrow$& Acc$\uparrow$& \textcolor{black}{Acc$\uparrow$} & \textcolor{black}{Acc$\uparrow$} \\
\midrule
0.0 \textit{(LS-960 only)}  &\textbf{10.98}&  96.85 &\textbf{95.02}&  74.77 &\textbf{63.02}&\textcolor{black}{66.54}&\textcolor{black}{49.83}\\
0.1         &  11.97 &  97.06 &  94.38 &  77.89 &  61.44 &\textcolor{black}{72.77}&\textcolor{black}{51.68}\\
0.2         &  11.87 &  96.99 &  93.28 & \textbf{78.46} &  61.75 &\textcolor{black}{73.57}&\textcolor{black}{53.76}\\
0.3         &  12.19 &\textbf{97.23}&  94.46 &  78.25 &  61.55 &\textcolor{black}{74.67}&\textcolor{black}{54.44}\\
0.4         &  12.39 &  97.08 &  94.15 &  77.15 &  61.52 &\textcolor{black}{75.16}&\textcolor{black}{54.39}\\
0.5         &  12.53 &  96.82 &  92.14 &  76.58 &  61.07 &\textcolor{black}{76.38}&\textcolor{black}{54.73}\\
1.0 \textit{(AudioSet only)} &  27.04 &  95.60 &  82.78 &  68.54 &  60.85 &\textcolor{black}{\textbf{83.31}}&\textcolor{black}{\textbf{61.88}}\\
\midrule
HuBERT Base \cite{Hsu2021HuBERT}$^{\dagger}$ & 5.41 & 96.30 & 98.34 & 81.42 & 64.92 & \textcolor{EMgray}{\textit{62.76}} & \textcolor{EMgray}{\textit{46.26}} \\
\bottomrule
\multicolumn{8}{l}{$^{\dagger}$
ENV and MUS results were obtained using publicly available pre-trained models.}\\
\end{tabular}
}
\vspace{-5pt}
\end{table}

\begin{table}[tb!]
%\vspace{-5pt}
\caption{Evaluation \ref{sec:eval-speech} Specialized Representation for Speech:\\
Patch size ablations.\\
\footnotesize{(dataset noise ratio $\eta=0.2$ and input duration $T=2.08\text{ s}$)}}
\vspace{-5pt}
\label{tab:exp-speech:patch-size}
\centering
\resizebox{\columnwidth}{!}{%
\begin{tabular}{llllll|ll}
\toprule
 &    PR &     KS &     IC &     SID &    ER & \textcolor{EMgray}{ENV}  & \textcolor{EMgray}{MUS} \\
\vspace{-1pt} Patch size $\textit{Freq.}\times \textit{Time}$      &  PER$\downarrow$ & Acc$\uparrow$& Acc$\uparrow$& Acc$\uparrow$& Acc$\uparrow$& \textcolor{EMgray}{Acc$\uparrow$} & \textcolor{EMgray}{Acc$\uparrow$} \\
\midrule
$16\times16$ (M2D \cite{niizumi2022M2D}) &  77.92 &  96.36 &  83.23 &  79.65 &  58.88 &\textcolor{EMgray}{80.27}&\textcolor{EMgray}{\textbf{59.60}}\\
$40\times8$  &  29.28 &  96.89 &  89.67 &  77.51 &  59.39 &\textcolor{EMgray}{78.56}&\textcolor{EMgray}{56.62}\\
$40\times4$  &\textbf{11.49}&  96.82 &  90.93 &  81.64 &  60.30 &\textcolor{EMgray}{80.00}&\textcolor{EMgray}{57.48}\\
$40\times2$  &  15.14 &  96.79 &  82.73 &\textbf{85.44}&  61.49 &\textcolor{EMgray}{\textbf{80.98}}&\textcolor{EMgray}{58.67}\\
$80\times8$  &  30.28 &  96.30 &  90.51 &  75.66 &  58.97 &\textcolor{EMgray}{72.62}&\textcolor{EMgray}{54.28}\\
$80\times4$  &  11.87 &  96.99 &  93.28 &  78.46 &\textbf{61.75}&\textcolor{EMgray}{73.57}&\textcolor{EMgray}{53.76}\\
$80\times2$ ($=$ speech models) &  11.74 &\textbf{97.25}&\textbf{93.51}&  78.86 &  60.67 &\textcolor{EMgray}{74.98}&\textcolor{EMgray}{53.28}\\
\midrule
%\multicolumn{5}{l}{\textit{(Previous similar study using ViT-based \textit{Base} model)}} && \\
SSAST-Frame~\cite{gong2022ssast} & - & 96.7 & - & 80.8 & 60.5 & - & - \\
HuBERT Base \cite{Hsu2021HuBERT} & 5.41 & 96.30 & 98.34 & 81.42 & 64.92 & \textcolor{EMgray}{\textit{62.76}} & \textcolor{EMgray}{\textit{46.26}} \\
\bottomrule
\end{tabular}
}
%\vspace{-5pt}
\end{table}

\begin{table}[tb!]
%\vspace{-5pt}
\caption{Evaluation \ref{sec:eval-speech} Specialized Representation for Speech:\\
Input duration ablations.\\
\footnotesize{(dataset noise ratio $\eta=0.2$ and patch size $80\times 4$)}}
\vspace{-5pt}
\label{tab:exp-speech:input-dur}
\centering
\resizebox{\columnwidth}{!}{%
\begin{tabular}{llllll|ll}
\toprule
 &    PR &     KS &     IC &     SID &    ER & \textcolor{EMgray}{ENV}  & \textcolor{EMgray}{MUS} \\
\vspace{-1pt} Input duration $T$   &  PER$\downarrow$ & Acc$\uparrow$& Acc$\uparrow$& Acc$\uparrow$& Acc$\uparrow$& \textcolor{EMgray}{Acc$\uparrow$} & \textcolor{EMgray}{Acc$\uparrow$} \\
\midrule
2.08 s &  11.87 &  96.99 &  93.28 &  78.46 &  61.75 &\textcolor{EMgray}{73.57}&\textcolor{EMgray}{\textbf{53.76}}\\
3.04 s &   9.81 &  97.09 &  95.15 &  79.18 &  63.39 &\textcolor{EMgray}{74.36}&\textcolor{EMgray}{52.14}\\
4.00 s &   8.50 &\textbf{97.34}&  94.83 &\textbf{81.24}&  63.81 &\textcolor{EMgray}{74.11}&\textcolor{EMgray}{52.23}\\
5.12 s &   8.10 &  97.17 &  94.70 &  78.73 &\textbf{65.47}&\textcolor{EMgray}{\textbf{74.69}}&\textcolor{EMgray}{51.66}\\
6.08 s &\textbf{7.74}&  97.17 &\textbf{95.50}&  80.48 &  64.06 &\textcolor{EMgray}{72.42}&\textcolor{EMgray}{50.64}\\
\midrule
HuBERT Base \cite{Hsu2021HuBERT} & 5.41 & 96.30 & 98.34 & 81.42 & 64.92 & \textcolor{EMgray}{\textit{62.76}} & \textcolor{EMgray}{\textit{46.26}} \\
\bottomrule
\end{tabular}
}
%\vspace{-5pt}
\end{table}

\begin{table*}[tb!]
%\vspace{-10pt}
\caption{Evaluation \ref{sec:eval-speech} Specialized Representation for Speech: Comparison with SOTA speech models.}
%\vspace{-5pt}
\label{tab:exp-speech:results-sota}
\centering
\resizebox{0.8\textwidth}{!}{%
\begin{tabular}{lllllllll|ll}
\toprule
 & \# Params & Dataset &    PR &     KS &     IC &     SID &  ASV &    ER & \textcolor{EMgray}{ENV}  & \textcolor{EMgray}{MUS} \\
\vspace{-1pt} Model (/masking ratio) & Million   &  &  PER$\downarrow$ & Acc$\uparrow$& Acc$\uparrow$& Acc$\uparrow$& 
EER$\downarrow$ & Acc$\uparrow$& \textcolor{EMgray}{Acc$\uparrow$} & \textcolor{EMgray}{Acc$\uparrow$} \\
\midrule
\multicolumn{6}{l}{\textit{(SOTA models)}} &&&& \\
wav2vec2.0 Base \cite{baevski2020wav2vec2}$^{\dagger}$ & 95M & \footnotesize{LS-960} & 5.74 & 96.23 & 92.35 & 75.18 & 6.02 & 63.43 & \textcolor{EMgray}{\textit{37.66}} &  \textcolor{EMgray}{\textit{32.02}} \\
WavLM Base~\cite{Chen2022WavLM}$^{\dagger}$ & 95M & \scriptsize{LS-960+DNS} & \textbf{4.84} & 96.79 & \textbf{98.63} & \textbf{84.51} & \textbf{4.69} & 65.94 & \textcolor{EMgray}{\textit{54.45}} & \textcolor{EMgray}{\textit{40.98}} \\
%\multicolumn{6}{l}{\textit{(Baseline)}} &&& \\
HuBERT Base \cite{Hsu2021HuBERT}$^{\dagger}$ \textit{(Baseline)} & 95M & \footnotesize{LS-960} & 5.41 & 96.30 & 98.34 & 81.42 & 5.11 & 64.92 & \textcolor{EMgray}{\textit{62.76}} & \textcolor{EMgray}{\textit{46.26}} \\
\midrule
\multicolumn{6}{l}{\textit{(Ours: M2D-S with $\lambda_\text{off}=0.5$, $\lambda_\text{m2d}=1$, $\eta=0.2$, and patch size $80\times 2$)}} &&& \\
M2D-S/0.6 T=4.0s & 86M & \footnotesize{LS-960+AS} &   5.72 &  96.47 &  97.80 &  81.97 & 6.29 & \textbf{66.36} &\textcolor{EMgray}{53.22}&\textcolor{EMgray}{41.71}\\
M2D-S/0.6 T=5.12s  & 86M & \footnotesize{LS-960+AS} &   5.64 &  {96.87} &  97.65 &  80.69 & 7.07 &  65.35 &\textcolor{EMgray}{57.34}&\textcolor{EMgray}{43.23}\\
M2D-S/0.6 T=6.08s & 86M & \footnotesize{LS-960+AS} &   5.33 & {96.80} &  97.63 &  81.74 & 5.97 &   66.13 &\textcolor{EMgray}{54.77}&\textcolor{EMgray}{43.75}\\

\midrule

\multicolumn{6}{l}{\textit{(Reference: M2D with setting modifications only, $\eta=0.2$, and patch size $80\times 4$)}} &&& \\
M2D/0.6 T=4.00s & 86M & \footnotesize{LS-960+AS} &   8.50 &\textbf{97.34}&  94.83 & {81.24} & 7.13 &  63.81 &\textcolor{EMgray}{74.11}&\textcolor{EMgray}{52.23}\\
M2D/0.6 T=5.12s & 86M & \footnotesize{LS-960+AS} &   8.10 &  97.17 &  94.70 &  78.73 & 7.15 & {65.47}&\textcolor{EMgray}{{74.69}}&\textcolor{EMgray}{51.66}\\
M2D/0.6 T=6.08s & 86M & \footnotesize{LS-960+AS} & {7.74}&  97.17 & {95.50}&  80.48 & 6.84 &  64.06 &\textcolor{EMgray}{72.42}&\textcolor{EMgray}{50.64}\\

\multicolumn{6}{l}{\textit{(Reference: The M2D model pre-trained in Section \ref{sec:eval-general}) \scriptsize{$^\ast$ASV test fails with the patch size $16\times 16$.}}} &&& \\
M2D/0.6, T=6.08s, patch size $16\times 16$ & 86M & \footnotesize{AS} &  78.30 &  95.65 &  76.77 &  80.68 & N/A &  61.17 &  \textcolor{EMgray}{\textbf{88.63}}&\textcolor{EMgray}{\textbf{66.56}}\\

\multicolumn{6}{l}{\textit{(Reference: Previous similar study using ViT-based \textit{Base} model)}} &&& \\
SSAST-Frame~\cite{gong2022ssast} & 89M & \footnotesize{LS-960 $\cup$ AS}$^{\ddagger}$ & - & 96.7 & - & 80.8 & - & 60.5 & - & - \\
SSAST-Patch~\cite{gong2022ssast} & 89M & \footnotesize{LS-960 $\cup$ AS}$^{\ddagger}$ & - & 94.8 & - & 57.1 & - & 56.8 & - & - \\

\bottomrule
\addlinespace[0.1cm]
\multicolumn{9}{l}{$^{\dagger}$
ENV and MUS results were obtained using publicly available pre-trained models.}\\
\multicolumn{9}{l}{$^{\ddagger}$ The LS-960 and AS samples were used without mixing one as background noise into others, unlike M2D-S.}\\
\end{tabular}
}
\vspace{-5pt}
\end{table*}

\vspace{0.1cm}
\noindent\textbf{Evaluation details.}\hspace{0.2cm}
%我々は事前学習した全てのモデルをSUPERBと線形評価で評価した。SUPERBの評価ではモデルの重みを凍結し、全てのtransformer layerから得られる特徴量を重み付き和した特徴量が評価に利用された。
We evaluated models using SUPERB\cite{yang2021superb} for the speech task performance and conducted a linear evaluation for supplemental assessment of the non-speech performance. In the SUPERB evaluation, the model weights were frozen, and the weighted sum of the features from all transformer layers was used in the evaluation.
We standardized spectrograms with statistics of the dataset of each SUPERB task.
The evaluation tasks included phoneme recognition (PR), keyword spotting (KS),  intent classification (IC), speaker identification (SID), and emotion recognition (ER); we chose them to form a balanced subset from each speaker/content/semantics/paralinguistics subcategories listed in Table I from WavLM\cite{Chen2022WavLM}. In addition, we also tested models on automatic speaker verification (ASV) in Table \ref{tab:exp-speech:results-sota}. %when compared to multiple SOTA modals.
%我々はセクションIII-Vと同じ線形評価の設定を利用し、2つの環境音タスクの平均結果(ENV)と3つの音楽タスクの平均結果(MUS)を報告する。SUPERBとは異なり、凍結したモデルの最終層特徴のみを利用して評価している。
For the linear evaluation, we used the same setup as in Section \ref{sec:eval-general} and assessed the average accuracies of two environmental sound tasks (ENV) and three music tasks (MUS). Note that, unlike SUPERB, the linear evaluation used only the final layer features of the frozen model.

\vspace{0.2cm}
\subsubsection{Adapting M2D to Speech}\label{sec:exp-speech:m2d-adaptation}
We adapted the M2D settings for the speech tasks. In the following ablation studies, we sought the best parameter with the defaults: dataset noise ratio of 0.2, input duration of 2.08 s, and patch size of $80\times 4$.

\vspace{0.1cm}
\noindent\textbf{Pre-training dataset ablations.}\hspace{0.2cm}
%データセットが与える事前学習の性能への影響を、ターゲットドメインのデータセットLibriSpeechと一般的なオーディオデータセットAudioSetを用いてM2Dで評価する。データセット混合比率Aを変化させた。
We assessed the impact of datasets on pre-training by changing the dataset noise ratio $\eta$ to control the mixing of the target data LibriSpeech and the background noise AudioSet, and we set $\lambda_\text{off}=0$.
%AudioSetのみを使う場合、LibriSpeechサンプル数を1エポックとし、AudioSetからランダムにサンプリングした。
When using AudioSet only, we defined one epoch as the number of LibriSpeech samples, and we randomly sampled from AudioSet.

%表2の結果は、音声タスクはLibriSpeechのみを使ったとき一番良く、AudioSetの比率が大きくなるほど性能が悪化する傾向を示す。逆に、非音声(ENV, MUS)タスクはAudioSetのみを使ったときに一番性能が良く、LibriSpeechの比率が大きくなるほど性能が悪化する傾向を示す。例外的にSIDは比率A=0.2で音声に背景ノイズがあるときに、一番性能が良い結果を示した。しかしながら音声ドメインのデータセットを使うことが総じて性能に寄与すると考えられる。
The results in Table \ref{tab:results-data-mix} show that the speech tasks perform best when using only LibriSpeech, and performance deteriorates as the ratio of AudioSet increases. Conversely, the non-speech tasks (ENV, MUS) perform best when using only AudioSet, and performance deteriorates as the ratio of LibriSpeech increases.
Except for SID, which performed best at around $\eta=0.2$, using a speech dataset only performs better in speech tasks.

Note that many AudioSet samples contain speech. While the speech quality is typically bad compared to a speech corpus, it may provide a performance boost when used with LibriSpeech.

\vspace{0.1cm}
\noindent\textbf{Patch size ablations.}\hspace{0.2cm}
%汎用モデルと音声モデルの大きな違いである、入力スペクトログラムのパッチ分割が性能に与える影響を調べた。
We examined the impact on the performance of the patch splitting of the input spectrogram, which is a key difference between the general-purpose model and the speech model.
%表3はパッチサイズの結果は、speech modelと等価な設定(80x2)がSIDを例外として最も性能のバランスが良いことを示している。パッチサイズ80x2(Frequency vs. Time-step)は、F=80は周波数軸に分割せず、T=2は1フレームあたり20msを示す。周波数方向にパッチを二分割する40x4、40x2はそれぞれPR、SIDで最も良い結果を示したものの、その他のタスクでの結果が劣る。これらは、パッチを周波数軸で分割しないほうがバランスよく音声タスクに有効であることを示している。
The results in Table \ref{tab:exp-speech:patch-size} indicate that the setting equivalent to the speech model ($80\times2$) has the best performance balance except for SID; for patch size $80\times2$ (Frequency $\times$ Time-step), 80 indicates no splitting on the frequency axis, and 2 corresponds to 20 ms per frame.
Notably, this result aligns with SSAST \cite{gong2022ssast} using a similar patch size of $128 \times 2$.
While $40\times2$ performs the best on SID, it significantly degrades on PR and IC.
The results also show that the longer time steps (lower frame rates) degrade the performance, especially in PR, similar to the results in \cite{Meng2023Compressing}.
%The patch sizes of $40\times4$ and $40\times2$, which assign two patches along the frequency axis, showed the best results for PR and SID, but inferior results on some other tasks. Overall, these results indicate that no patch split along the frequency axis provides a balanced better performance on the speech tasks.

\vspace{0.1cm}
\noindent\textbf{Input duration ablations.}\hspace{0.2cm}
%表Xに示されるように、入力長を長くすることで、ベースラインHuBERTの性能にさらに近づけることができる。
Table \ref{tab:exp-speech:input-dur} shows that increasing the input duration brings the performance closer to the baseline HuBERT.
%これらの結果は、KS, SID, ERの性能はHuBERT同程度に到達した一方で、PR, ICの性能を近づけることが困難であることを示す。
These results show that while the performance on KS, SID, and ER reached the same level as HuBERT, it is difficult to approach that level of performance on PR and IC.

\vspace{0.2cm}
\subsubsection{Comparison with SOTA Models}\label{sec:exp-speech:m2d-s}
With the best parameters in the previous section, we extended M2D to M2D-S to assess further performance gain, especially on PR and IC, using the speech domain techniques.
M2D-S learns from a multi-task of M2D, distilling HuBERT features for learning clustered features and denoising as in WavLM. Table \ref{tab:exp-speech:results-sota} summarizes the results of M2D-S, M2D, and SOTA models.

The table shows M2D-S results comparable to HuBERT except for IC and ASV, with a performance gap of $-0.54$ to $-0.71$ on IC and $-0.86$ to $-1.96$ on ASV.
Compared to wav2vec2.0, M2D-S results are better except for ASV, for which they are comparable.
Compared to WavLM, M2D-S performed worse with gaps on PR of $-0.49$ to $-0.88$, IC of $-0.83$ to $-1.0$, SID of $-2.54$ to $-3.82$, and ASV of $-1.28$ to $-2.38$. While M2D-S learns from a denoising task as in WavLM, we did not observe a performance boost such as WavLM's SID with $84.51$, suggesting the M2D-S's denoising did not work as well as WavLM's.

Overall, M2D-S performed better than wav2vec2.0 and closer to HuBERT. However, it did not show robust results, especially on ASV, and performed worse than WavLM.
Although the M2D-S results were not comparable with all speech model results, the comparison among M2D-S, wav2vec2.0, and HuBERT indicated that M2D-S achieved learning a domain representation with top-level performance.

\begin{table}[tb!]
\vspace{-5pt}
\caption{Evaluation \ref{sec:eval-speech} Specialized Representation for Speech: Pre-training task ablations.\\
\footnotesize{(input duration $T=2.08\text{ s}$, and patch size $80\times 4$)}}
\vspace{-5pt}
\label{tab:exp-speech:tasks}
\centering
\resizebox{\columnwidth}{!}{%
\begin{tabular}{lllllllll|ll}
\toprule
&\multicolumn{2}{c}{Offline} & Online &    PR &     KS &     IC &     SID &    ER & \textcolor{EMgray}{ENV}  & \textcolor{EMgray}{MUS} \\
 \cmidrule(lr){2-3} \cmidrule(lr){4-4}
&\vspace{-1pt} denoise & distill & M2D &  PER$\downarrow$ & Acc$\uparrow$& Acc$\uparrow$& Acc$\uparrow$& Acc$\uparrow$& \textcolor{EMgray}{Acc$\uparrow$} & \textcolor{EMgray}{Acc$\uparrow$} \\
\midrule
(a)&&&\checkmark        &  11.87 &\textbf{96.99}&  93.28 &\textbf{78.46}&  61.75 &\textcolor{EMgray}{\textbf{73.57}}&\textcolor{EMgray}{\textbf{53.76}}\\
(b)&&\checkmark&           &   8.10 &  96.05 &  94.81 &  65.51 &  61.27 &\textcolor{EMgray}{56.96}&\textcolor{EMgray}{44.16}\\
(c)&\checkmark&\checkmark& &   7.80 &  94.66 &  93.38 &  70.03 &  61.58 &\textcolor{EMgray}{42.42}&\textcolor{EMgray}{35.36}\\
(d)&&\checkmark&\checkmark                      &   8.78 &  95.72 &  91.35 &  62.68 &  60.85 &\textcolor{EMgray}{57.15}&\textcolor{EMgray}{43.75}\\
(e)&\checkmark&\checkmark&\checkmark      &\textbf{7.02}&  95.48 &\textbf{97.10}&  78.12 &\textbf{64.09}&\textcolor{EMgray}{55.49}&\textcolor{EMgray}{41.51}\\
\bottomrule
\end{tabular}
}
\vspace{-10pt}
\end{table}

\vspace{0.1cm}
\noindent\textbf{Pre-training task ablations.}\hspace{0.2cm}
%音声拡張としてオフラインネットワークを追加して実現したM2D-Sのdenoising distillationタスクの効果を実験により検証する。表5に実験結果を示す。offline networkとM2Dのタスクは、それぞれ式eq:eq-m2ds-lossのWO、WMを0または1にすることによって切り替えた。
%We validate the effect of the M2D-S denoising distillation task, implemented by adding an offline network.
%我々はM2D-Sの各タスク、denoising distillation、M2Dの有効性を検証する。我々はdenoisingを行わない場合も検証する。
We further assessed the contribution of the tasks learned in M2D-S.
%Table \ref{tab:exp-speech:tasks} shows the experimental results.
We switched the tasks in the offline and M2D networks by setting $\lambda_\text{off}$ and $\lambda_\text{m2d}$ in \eqref{eq:eq-m2ds-loss} to 0 or 1.0, respectively, and switched the denoising task by setting $\eta$ to 0 (disabled) or 0.2 (enabled).
%さらに，すべてのタスクを使う場合の最適なLMoffの結果を最下段に示す。
%Additionally, we show in the bottom row the result of the best settings when using all tasks.

%結果は、(b)distillationが(a)M2DやM2Dにdistillationを加えた(d)の結果よりも良いことから、clustered featureを学習したモデルをdistillationすることが音声タスクに対して効果が高いことを示す。distillationだけの(b)に対してdenoisingを追加した(c)は性能を向上させ、denoisingタスクの効果を示した。M2D-Sの構成である(e)は、すべてのタスクを合同で行うことで単にdistillationを行う(b)やM2Dと組み合わせた(d)より性能がよく、denoisingをdistillationやM2Dと組み合わせることが非常に効果的であることが示される。
The results in Table \ref{tab:exp-speech:tasks} show that (b) distillation only is better than (a) M2D or (d) M2D plus distillation, indicating that the distillation of a speech model trained with clustered features is effective.
Compared to (b) with only distillation, (c) with additional denoising performs better except for KS and IC, indicating the effectiveness of the denoising task.
M2D-S configuration (e) shows that learning all tasks together effectively improves performance.

%Section Xの表XXではLibriSpeechだけをM2Dに入力するα=0の結果が一番性能が良いのに対して、表YYでは(e)のα=0.2の結果が一番良い性能を示している。ランダムノイズを付加した音声入力を直接学習するタスクからdenoisingタスクに切り替えることで、同じ入力データの持つ役割が変化することは興味深く示唆的である。
%Table \ref{tab:results-data-mix} in Section \ref{sec:exp-dataset} shows the best performance for $\eta=0$, where only LibriSpeech is input to M2D, while Table \ref{tab:exp-speech:tasks} shows the best performance for $\eta=0.2$ in (e).
%It is interesting and suggestive that switching from a task that learns directly from speech input with random background noise to a denoising task changes the role of the same input data.
%It is interesting and suggestive that switching tasks changes the role of the same input data.

%%%%%%%%%%%%%%%%%%%%%%%%%%%%%%%%%%%%%%%%%%%%%%%%%%%%%%%%%%%%%%%%%%%%%%%
%%%%%%%%%%%%%%%%%%%%%%%%%%%%%%  SECTION  %%%%%%%%%%%%%%%%%%%%%%%%%%%%%%
\subsection{Evaluation: Specialized Representation for Medical Application} \label{sec:eval-icbhi}
%本実験では、提案手法に少データタスクに特化した表現を学習させて性能を評価した。マスク予測による事前学習は通常大量のデータを必要とすることから、少データで事前学習することはチャレンジングである。そのために、医療分野から呼吸音分類アルゴリズム評価用データベース ICBHI2017(ICBHI=International Conference on Biomedical and　Health Informatics)を利用する。このデータセットは79人分539録音(3.33h)しかなく、事前学習が困難である。
This experiment simulated cases of private and propriety data, where the data size is typically limited and its distribution differs from a large pre-training dataset; we validated our models for learning a specialized representation with top-level performance in these cases.
We used the Respiratory Sound Database from ICBHI (International Conference on Biomedical and Health Informatics) 2017 Challenge (ICBHI2017)\cite{ICBHI2017}, a dataset from the medical field for evaluating respiratory sound classification algorithms, containing only 920 recordings (5.5 h) for 128 subjects. %, which makes pre-training very difficult.
Since pre-training with masked prediction usually requires a large dataset\cite{Xie2023OnDataScalingMasked,zhang2023mae-survey}, pre-training on a small amount of data is challenging.
%なお、事前実験でICBHI2017データセットだけを使った事前学習を試みたが、ランダム初期化済みのモデルよりも悪い性能となり、事前学習が成功しないことを確認した。%またそれらの結果は、Mohammand et al.[M]と比べて非常に悪い結果であった。
We tried to pre-train an M2D using only ICBHI2017 in a preliminary evaluation; however, as shown in Table \ref{tab:exp-icbhi:preliminary}(2), it underperformed the randomly initialized models, confirming that the pre-training had failed. %Also, these results were very poor compared to Moummad et al.

%事前学習の効果を比較するために、既存法Mohammand et al.[M]のコードベースを利用して、音の特徴量を抽出するエンコーダだけを入れ替える方法で評価する。このコードベースはPANNsの事前学習モデルを用いており比較に適している。Supervised Contrastive LearningとCross Entropy損失それぞれを使った解決があるが、比較を容易にするため後者のみを利用した。
To facilitate the assessment of the contribution of pre-training, we used a codebase from Moummad et al.\cite{moummad2022icbhi}, in which we only replaced the encoder that extracts the audio features. This codebase uses PANNs\cite{kong2020panns} pre-trained models as an encoder and is suitable for comparison.
Solutions of the codebase use Supervised Contrastive Learning (SCL)\cite{Khosla2020SCL} and learning based on Cross Entropy (CE) loss for the fine-tuning algorithm; we used only the CE loss for ease of comparison.

%少リソース環境での適用可能性をさらに広げるため、実験条件において計算やデータリソースも少リソースに制限した。事前学習やその後のfine-tuningはメモリ24GBのGPU 1枚で実行し、背景ノイズには多様な音を含みながらも入手が容易でデータサイズが取り扱いやすいFSD50Kを利用した。学習時間もそれぞれ一日以内で実行できることを目安とした。
We also limited the computational and data resources in the experiments to assess the applicability to a real-world scenario. We used a single GPU with 24 GB of memory (e.g., NVIDIA GeForce RTX 3090 Ti) for pre-training and subsequent fine-tuning and FSD50K\cite{fonseca2020fsd50k} for background noise, which is practical in terms of availability and size while providing a wide variety of sounds. We also set a goal of completing the pre-training and fine-tuning within a day for each.

\vspace{0.1cm}
\noindent\textbf{Approach.}\hspace{0.2cm}
%我々はfurther pre-training戦略を取り、AudioSet事前学習済みモデルを更に目標データセットICBHI2017でM2D-Xを使って事前学習した。このときM2D-XのOffline teacher modelにはAudioSet事前学習済みモデルを使う。背景ノイズを使うことによるデータ拡張とノイズ除去、更にAudioSet事前学習済み表現の蒸留を組み合わせることで、目標データ(ICBHI2017)の効果的な表現の学習を行うことができる。
We adopted a \textit{further pre-training}\cite{Sun2019FurPT} strategy, which pre-trains a pre-trained model once again, using M2D-X on the target dataset ICBHI2017.

In the further pre-training of M2D-X, we used the configuration in scenario iii) from Fig. \ref{fig:system-x}(b), employing the M2D model pre-trained on AudioSet as the offline teacher model and as the initial M2D online encoder weights.
The combination of the data augmentation effect and denoising task brought by the use of background noise promotes an effective representation, and distilling the features from the frozen AudioSet pre-trained model regularizes learned features from overfitting to the small target dataset ICBHI2017.

We also evaluated further pre-training of M2D by setting $L_\text{off}=0$ in M2D-X, in addition to M2D-X configuration, to assess the contribution of the regularization.

\begin{table}[tb!]
%\vspace{-5pt}
\caption{Evaluation \ref{sec:eval-icbhi} Specialized Representation for ICBHI2017: Priliminary evaluation results with 95\% CI.}
\vspace{-5pt}
\label{tab:exp-icbhi:preliminary}
\centering
\resizebox{\columnwidth}{!}{%
\begin{tabular}{lllll}
\toprule
Model (/masking ratio) & Sp & Se & Score \\
\midrule
\multicolumn{5}{l}{\textit{(1) Patch size/masking ratio ablations}} \\
16$\times$16/0.6 (T=6.08s) & 77.67 {\fontsize{6pt}{6pt}\selectfont $\pm$ 1.82} &  42.54 {\fontsize{6pt}{6pt}\selectfont $\pm$ 1.59} &       60.10 {\fontsize{6pt}{6pt}\selectfont $\pm$ 0.37} \\
16$\times$16/0.7 (T=6.08s) &  79.47 {\fontsize{6pt}{6pt}\selectfont $\pm$ 2.10} &  42.93 {\fontsize{6pt}{6pt}\selectfont $\pm$ 1.84} &       61.20 {\fontsize{6pt}{6pt}\selectfont $\pm$ 0.46} \\
%16$\times$8 (T=4.00s) & 80.41 {\fontsize{6pt}{6pt}\selectfont $\pm$ 1.57} &  40.54 {\fontsize{6pt}{6pt}\selectfont $\pm$ 2.02} &       60.47 {\fontsize{6pt}{6pt}\selectfont $\pm$ 0.56} \\
80$\times$4/0.7 (T=6.08s) &\textbf{79.52 {\fontsize{6pt}{6pt}\selectfont $\pm$ 1.44}}&  40.22 {\fontsize{6pt}{6pt}\selectfont $\pm$ 1.43} &       59.87 {\fontsize{6pt}{6pt}\selectfont $\pm$ 0.28} \\
16$\times$4/0.7 (T=2.00s) & 79.48 {\fontsize{6pt}{6pt}\selectfont $\pm$ 1.47} &\textbf{44.38 {\fontsize{6pt}{6pt}\selectfont $\pm$ 1.55}}&\textbf{61.93 {\fontsize{6pt}{6pt}\selectfont $\pm$ 0.25}}\\
%16$\times$2 (T=1.00s) &  74.76 {\fontsize{6pt}{6pt}\selectfont $\pm$ 1.79} &\textbf{44.58 {\fontsize{6pt}{6pt}\selectfont $\pm$ 1.78}}&       59.67 {\fontsize{6pt}{6pt}\selectfont $\pm$ 0.36} \\
\multicolumn{5}{l}{\textit{(2) Pre-training ablations}} \\
No pre-training (random initial weights) & 57.02 {\fontsize{6pt}{6pt}\selectfont $\pm$ 9.61} &  32.39 {\fontsize{6pt}{6pt}\selectfont $\pm$ 5.71} &       44.71 \\ % {\fontsize{6pt}{6pt}\selectfont $\pm$ 3.18} \\
M2D pre-training on ICBHI2017 only &  45.36 {\fontsize{6pt}{6pt}\selectfont $\pm$ 5.55} &  38.29 {\fontsize{6pt}{6pt}\selectfont $\pm$ 3.52} &       41.83 \\ %{\fontsize{6pt}{6pt}\selectfont $\pm$ 1.37} \\
\bottomrule\\
\end{tabular}
}
\vspace{-15pt}
\end{table}

\vspace{0.1cm}
\noindent\textbf{Further pre-training details.}\hspace{0.2cm}
%表Xが示す予備実験の結果から、パッチサイズ16x4、入力音長 2.0s、masking ratio 0.7のAudioSet事前学習済みM2Dモデルを利用した。このモデルをoffline teacherとして利用すると同時に、オンラインencoderの初期重みとして利用した。M2D-Xのloss weights Lm2d、Loffはどちらも1.0とした。
Based on the preliminary experiments shown in Table \ref{tab:exp-icbhi:preliminary}(1), we used the AudioSet pre-trained M2D model with a patch size of $16 \times 4$, input audio duration of 2.0 s, and masking ratio of 0.7. We used this model as an offline teacher and also as the initial online encoder weights.
We used the M2D-X offline network based on the distillation configuration in Section \ref{sec:eval-speech} and additionally reshaped the offline teacher output using \eqref{eq:msm-mae} to obtain the training signal per timestep $\tilde{y} \in R^{B \times N_T \times D}$.
We set the M2D-X loss weights $L_\text{m2d}$ and $L_\text{off}$ to $1.0$.
%学習プロセスでの取り扱いを容易にするため、データリストを繰り返し、端数を切り捨て総数10000サンプルとして取り扱い、これを1エポックとした。FSD50Kを背景ノイズとして利用し、noise_ratio 0.3とした。

While we used 539 ICBHI training samples, we virtually increased samples by replicating the data list and rounding fractions to 10,000 samples, treated as one epoch, to make the training process manageable. We used FSD50K as background noise and set the $\eta$ to $0.3$.
%24GB GPUで学習を行うために、バッチサイズ64、gradient accumulationをiteration 2 (実効バッチサイズ128)とした。learning rate と EMA decay rate は基本設定に準じた。我々はwarm upを24エポックとして600エポック学習させ、100エポックごとにチェックポイントを保存して評価した。
To train on a 24 GB GPU, we set the batch size to 64 and the gradient accumulation\cite{hermans2017accumulated} to two iterations (effective batch size of 128). We used the same learning rate and EMA decay as in Section \ref{sec:exp:basic}. We pre-trained 600 epochs with 24 warm-up epochs and saved and evaluated checkpoints every 100 epochs.

\vspace{0.1cm}
\noindent\textbf{Fine-tuning evaluation details.}\hspace{0.2cm}
%安定した実験結果を得るため、learning rateを0.00005、バッチサイズ64に固定して150エポック学習した。また小さなGPU上で実行可能にするため、gradientの蓄積を可能にするようコードベースを最小限変更し、推論時のミニバッチを4分割した。
For stable experimental results, we fixed the learning rate to 0.00005 and the batch size to 64 and trained 150 epochs. To accommodate execution on a small GPU, we minimally modified the codebase to allow for gradient accumulation, and the mini-batch was split into four during inference.
%ICBHI2017は、学習、テストセットそれぞれ539、XX録音から切り取った合計6,898サンプルのrespiratory cyclesで構成される。我々は学習にTT、validationにVV、テストにTTサンプルを利用する標準の4クラス分類タスクで評価した。
ICBHI2017 consists of 4142 training and 2756 test samples, for a total of 6898 sample respiratory cycles cropped from 539 training and 381 test recordings.
We evaluated it with a standard four-class classification task and 
%ICBHI2017のテスト評価指標であるSensitivity (Se), Specificity (Sp), これらの平均であるScoreを利用としてモデルを比較した。我々はこのScoreを主な比較指標として利用する。
compared models using the ICBHI2017 test metrics of Sensitivity (Se), Specificity (Sp), and Score, which is the average of Se and Sp. We use Score as the primary metric for comparison.
%結果の変動が大きいため、我々は4つの事前学習済みモデルをそれぞれ5回ファインチューニングして得た合計20回の結果の95%CI付き平均を結果とした。
Due to the high variability of the results, we averaged with 95\% CI the 20 results obtained by fine-tuning each of the four pre-trained models five times.

\begin{table}[tb!]
%\vspace{-5pt}
\caption{Evaluation \ref{sec:eval-icbhi} Specialized Representation for ICBHI2017: Comparison with SOTA models with 95\% CI.}
\vspace{-5pt}
\label{tab:exp-icbhi:sota}
\centering
\resizebox{\columnwidth}{!}{%
\begin{tabular}{lllll}
\toprule
Model (/masking ratio) & Backbone & Sp & Se & Score \\
\midrule
\multicolumn{5}{l}{\textit{(Baseline models with 4.3M params)}} \\
Moummad et al. (CE)\cite{moummad2022icbhi} & CNN6\cite{kong2020panns} & 70.09 & 40.39 & 55.24 \\
Moummad et al. (SCL)\cite{moummad2022icbhi} & CNN6\cite{kong2020panns} & 75.95  & 39.15 & 57.55 \\
\multicolumn{5}{l}{\textit{(SOTA models with 86M params)}} \\
Bae\&Kim et al. (CE)\cite{bae2023patchmix_icbhi} & AST\cite{gong2021ast} & 77.14 & 41.97 & 59.55 \\
Bae\&Kim et al. (MixCL)\cite{bae2023patchmix_icbhi} & AST\cite{gong2021ast} & \textbf{81.66} & 43.07 & 62.37 \\
\midrule
\multicolumn{5}{l}{\textit{(i) Ours: Original M2D (pre-trained on AudioSet only) with 86M params}} \\
M2D/0.7 (16$\times$16, T=6.08s) & M2D ViT &  79.47 {\fontsize{6pt}{6pt}\selectfont $\pm$ 2.10} &  42.93 {\fontsize{6pt}{6pt}\selectfont $\pm$ 1.84} &       61.20 {\fontsize{6pt}{6pt}\selectfont $\pm$ 0.46} \\
M2D/0.7  (16$\times$4, T=2.0s) & M2D ViT &{79.48 {\fontsize{6pt}{6pt}\selectfont $\pm$ 1.47}}&{44.38 {\fontsize{6pt}{6pt}\selectfont $\pm$ 1.55}}&{61.93 {\fontsize{6pt}{6pt}\selectfont $\pm$ 0.25}}\\
\multicolumn{5}{l}{\textit{(ii) Ours: further pre-trained on ICBHI2017 (16$\times$4, T=2.0s) with 86M params}} \\
M2D/0.7 ($\eta=0.3$)  & M2D ViT & 80.70 {\fontsize{6pt}{6pt}\selectfont $\pm$ 1.66} &  44.76 {\fontsize{6pt}{6pt}\selectfont $\pm$ 1.50} &       62.73 {\fontsize{6pt}{6pt}\selectfont $\pm$ 0.30} \\
M2D-X/0.7 ($\eta=0.3$) & M2D ViT &{81.51 {\fontsize{6pt}{6pt}\selectfont $\pm$ 1.03}}& {45.08 {\fontsize{6pt}{6pt}\selectfont $\pm$ 1.10}}&\textbf{63.29 {\fontsize{6pt}{6pt}\selectfont $\pm$ 0.22}}\\
\multicolumn{5}{l}{\textit{(iii) Ablations: further pre-trained without BG noise data (16$\times$4, T=2.0s) with 86M params}} \\
M2D/0.7 ($\eta=0.0$)  & M2D ViT & 78.66 {\fontsize{6pt}{6pt}\selectfont $\pm$ 1.60} &  43.67 {\fontsize{6pt}{6pt}\selectfont $\pm$ 1.49} &       61.17 {\fontsize{6pt}{6pt}\selectfont $\pm$ 0.35} \\
M2D-X/0.7 ($\eta=0.0$) & M2D ViT & 77.39 {\fontsize{6pt}{6pt}\selectfont $\pm$ 1.69} &\textbf{46.04 {\fontsize{6pt}{6pt}\selectfont $\pm$ 1.66}}&       61.71 {\fontsize{6pt}{6pt}\selectfont $\pm$ 0.34} \\
\bottomrule\\
\end{tabular}
}
\vspace{-15pt}
\end{table}

\vspace{0.2cm}
\subsubsection{Comparison with SOTA Models}\label{sec:exp-icbhi:sota}
%表Xに示す我々の結果とSOTAモデルの結果を比較する。(ii)のM2Dの結果はM2D-XのLoffを0にすることで得た。
Table \ref{tab:exp-icbhi:sota} compares the results.
%(i)に示すM2Dの標準的なパッチサイズ16x16のモデルは、SOTAであるBae\&Kim et al.のCE学習モデルの結果59.55を上回る。Bae\&Kim et al.はASTをencoderに利用し、Moummad et al.はPANNs CNN6を利用することから、これら教師あり学習により得られた表現と比較して、M2Dの自己教師あり学習表現が効果的であると考えられる。一方でfine-tuningアルゴリズムであるPatch-Mix Contrastive Learning (MixCL)を利用したBae\&Kim et al. (MixCL) のScoreは62.37で、M2Dを上回る。fine-tuningアルゴリズムが性能向上に貢献している。M2Dはパッチサイズを16x4に変更することで61.93に性能を向上させられるが、Bae\&Kim et al. (MixCL)に比べてまだ性能を下回る。
For the (i) original M2D results, the default patch size $16\times 16$ of $61.20$ outperforms the CE model of Bae\&Kim et al.\cite{bae2023patchmix_icbhi} of $59.55$ and Moummad et al.\cite{moummad2022icbhi} results.
Bae\&Kim et al. and Moummad et al. used AST\cite{gong2021ast} and PANNs CNN6\cite{kong2020panns} trained by supervised learning of AudioSet.
The results suggest that ICBHI2017 has a different data distribution from AudioSet and that the M2D representation learned by SSL is more applicable to the task than supervised learning representations.
%Since Bae\&Kim et al. use AST\cite{gong2021ast} as an encoder and Moummad et al.\cite{moummad2022icbhi} use PANNs CNN6\cite{kong2020panns}, the results indicate the effectiveness of the representation learned by M2D SSL over learned by supervised learning methods.
Meanwhile, Bae\&Kim et al. (MixCL) outperforms M2D with a SOTA score of $62.37$, which uses the fine-tuning algorithm Patch-Mix Contrastive Learning (MixCL), showing the effectiveness of the fine-tuning technique for the task. M2D patch size $16\times 4$ improves performance to $61.93$, but it still underperforms Bae\&Kim et al. (MixCL).

%(ii)に示すM2DとM2D-Xの結果は、背景ノイズデータを利用してfurther pre-trainingを行うことでSOTAを上回り、蒸留を利用することでさらに性能が向上することを示す。M2D (n=0.3) の結果は62.73でSOTAであるBae\&Kim et al. (MixCL)を0.36上回り、M2D-X (n=0.3) の結果は63.29となり、SOTAを0.92上回る。このように、M2D-Xを用いたfurther pre-trainingが、元のM2Dに対する性能向上に寄与している。
The (ii) results show that further pre-training improves the performance of our models to outperform SOTA, and M2D-X improves performance even more.
The M2D/0.7 ($\eta=0.3$) result of $62.73$ outperforms the SOTA by $0.36$, while the M2D-X/0.7 ($\eta=0.3$) result of $63.29$ outperforms SOTA by $0.92$.
The difference between M2D-X and M2D (M2D-X with $L_\text{off}=0$) indicates the regularization in the offline network contributes to gaining the performance from $62.73$ to $63.29$.

%総じてこれらの結果から、M2D-Xで導入した背景ノイズを利用したノイズ除去蒸留タスクを使うことで、小さいサイズの目標データセットでも効果的な表現が学習できることを示した。
Overall, these results show that M2D and M2D-X learn top-level representations even on a small target dataset under a further pre-training setting.

\vspace{0.2cm}
\subsubsection{Background Noise Ablations}\label{sec:exp-icbhi:abl}
%M2D-Xの背景ノイズデータの有無による性能への影響を評価する。(iii)のM2Dの結果もM2D-XのLoffを0にして得た。
We assessed the performance contribution of the use of background noise data.
%We also obtained the M2D results in (iii) by setting the $L_\text{off}$ of M2D-X to 0.

%背景ノイズデータを利用しない(iii)の結果は、M2D、M2D-Xどちらもfurther pre-trainingが失敗し、性能が悪化することを示す。特にM2D(n=0.0)の結果は61.17となり、further pre-training前の61.93からの悪化幅が大きい。一方でM2D-X(n=0.0)の結果は61.71であり、further pre-training前からの悪化幅は小さく、offline teacherの蒸留により表現が無益になることを防ぐ効果を確認した。
The (iii) ablations without background noise data show that both M2D and M2D-X further pre-training degrades, resulting in worse performance. In particular, the result for M2D/0.7 ($\eta=0.0$) is $61.17$, degraded by $0.76$ from $61.93$ before further pre-training. On the other hand, the result for M2D-X/0.7 ($\eta=0.0$) was $61.71$, showing a smaller decline by $0.22$ from $61.93$, confirming the effect of the regularization task in preventing the representations from overfitting.

%SSらは、NLPにおいてサイズの小さな目標データセットによる追加事前学習は害をなす一方、サイズが十分な目標と同じドメインのデータセットを使うことが有効であると報告している。M2D-Xでは、背景ノイズを利用することによるデータ拡張の効果が、サイズが大きな同じドメインのデータセットに近い働きをもたらしていると考えられる。
Sun et al.\cite{Sun2019FurPT} have reported that in the NLP domain, further pre-training on a small target dataset does harm, but that it is effective on a sufficiently large dataset of the same domain.
In M2D-X, we think that the effect of data augmentation through the use of background noise works similarly to the use of a dataset of the same domain with a larger size.

%総じて、M2D-Xで導入したoffline networkは小さなデータを事前学習する際に無益な表現になることを防ぐ効果をもたらし、背景ノイズの導入はデータ拡張効果により小さなデータの表現を学習可能にすることで、M2D-Xは少データシナリオでも効果的な表現を可能にすることを示す。
Overall, the offline network introduced in M2D-X indicates the effect of preventing overfitting, and the use of background noise suggests the effect of data augmentation, demonstrating that the M2D-X framework is effective for learning representations in the experimental scenarios.

\vspace{0.2cm}
\subsubsection{Performance Transition in Training Progress}\label{sec:exp-icbhi:perf-progress}
\begin{figure}[tp]
  \vspace{-5pt}
  \centering
  \includegraphics[width=1.0\columnwidth]{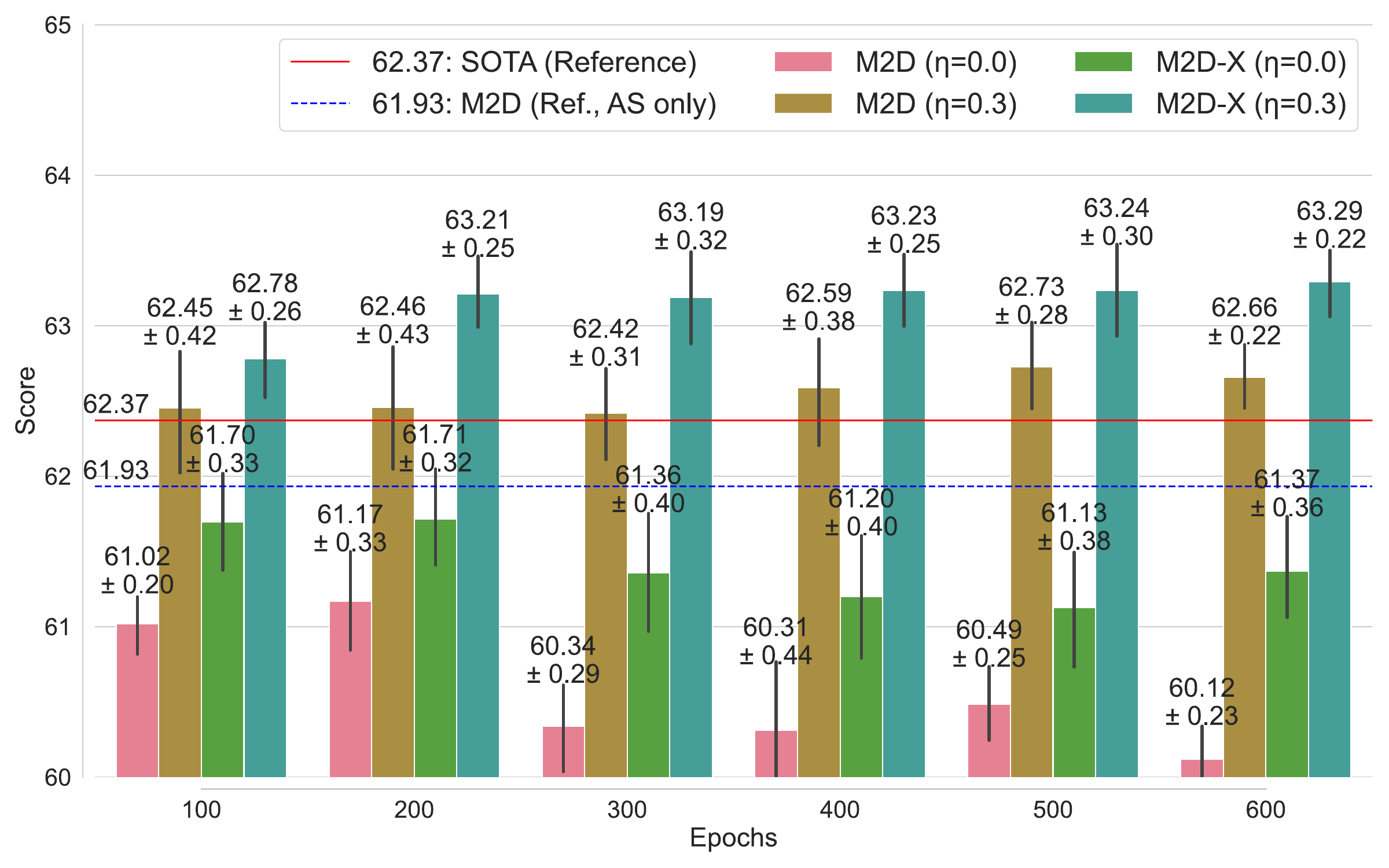} 
  \vspace{-20pt}
  \caption{Evaluation \ref{sec:eval-icbhi} Specialized Representation for ICBHI2017: Comparing the progress of Score (\%) among further pre-training settings.}
  \label{fig:exp-icbhi:progress}
  \vspace{-5pt}
\end{figure}

%図6は学習の途中で保存したそれぞれのcheckpointを評価した結果のグラフで、背景ノイズを使うことで(n=0.3)学習の進捗とともに性能が向上することを示す。一方で背景ノイズを使わない場合(n=0.0)、進捗とともに性能が悪化しており、学習した表現が役に立たなくなっていくことを表す。この結果は背景ノイズの利用によりfurther pre-trainingを可能にしていることを示すとともに、最適な学習エポック数の探索の必要性を小さくしていると考えられる。総じて、背景ノイズの利用は事前学習のロバスト性を高めていると考える。
We evaluated the performance of checkpoints saved during the training process as shown in Fig. \ref{fig:exp-icbhi:progress}. The graph shows that when background noise is used ($\eta=0.3$), the performance improves with the progress of the training process. On the other hand, when background noise is not used ($\eta=0.0$), performance deteriorates with progress, indicating that the learned representations become less useful as epoch progresses.

%これらの結果は、小さなデータだけを使うと進捗とともに学習される表現の性能が劣化していくことを、背景ノイズの導入により効果的な表現の学習に変えることができたことを示す。性能の変化に着目すると、M2Dでは性能の向上が400エポック以降で始まるのに対して、M2D-Xではより早く200エポックから始まっている。また、M2D-Xはより大きな性能の向上幅を示しており、背景ノイズや追加タスクの効果により小さなデータにおけるfurther pre-trainingを効果的に実現できることを示している。
These results show that introducing background noise changed the performance of learned representations from deteriorating with progress to improving for both M2D and M2D-X.
Focusing on the performance change, the improvement for M2D begins after 400 epochs, whereas for M2D-X, it begins earlier, at 200 epochs. In addition, M2D-X shows more significant performance gains, indicating that it can effectively achieve further pre-training in a small-data scenario with the help of the background noise and additional tasks.
%これらの結果は、背景ノイズを利用がM2D/M2D-Xどちらにもfurther pre-trainingの有効性を高め、学習エポックの進行とともに表現の有用性が高まることを示す。M2Dでは400エポック以降で向上するのに対して、M2D-Xでは200エポックで性能が顕著に向上させており、M2D-Xによる学習はより早く性能向上を実現している。 またM2D-Xはより大きな性能向上を果たしており、M2Dにノイズ除去蒸留タスクを追加することで、小さなデータからの表現学習が効果的に行われることを示す。 総じて、背景ノイズの利用により小さなサイズのデータでも効果的なfurther pre-trainingが可能になることを示すとともに、M2D-Xはその事前学習学習に背景ノイズを効果的に利用していることを示す。

\begin{figure}[tp]
  \vspace{-5pt}
  \centering
  \includegraphics[width=1.0\columnwidth]{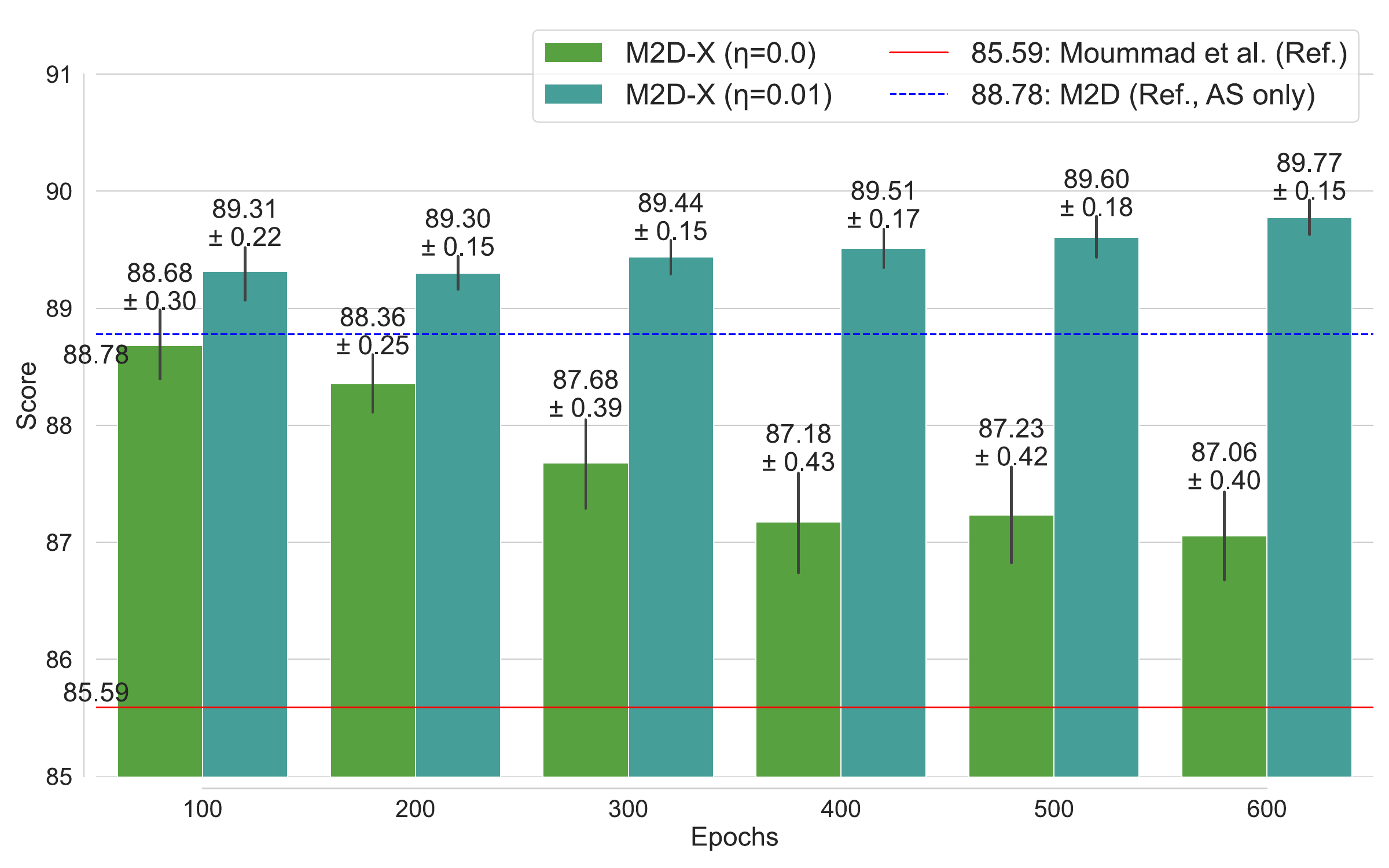} 
  \vspace{-20pt}
  \caption{Evaluation \ref{sec:eval-icbhi} Dataset Ablation using SPRSound: Comparing the progress of Score (\%).}
  \label{fig:exp-sprs:progress}
  \vspace{-10pt}
\end{figure}

\vspace{0.2cm}
\subsubsection{Dataset Ablation using SPRSound}\label{sec:exp-icbhi:sprsound}
%ICBHI2017と類似したデータセットSPRSoundを用いて異なるデータセットへのM2D-Xの適用性を検証した。SPRSoundは292被験者の2683録音(8.2 h)からなる呼吸音データセットであり、ICBHI2017(920録音/5.5 h/128被験者)と同様に小データである。[X]同様inter-patientテストスプリットを利用した。我々はpre-trainingのbackground noise ratioを0.01、fine-tuningの学習率を5x10-6、訓練エポック数を50としたほか、その他の実験詳細はICBHI2017を使用した実験同様である。
We further verified the applicability of M2D-X to another dataset using SPRSound\cite{SPRSound}, a dataset similar to ICBHI2017.
SPRSound is a respiratory sound dataset comprising 2683 recordings (8.2 h) from 292 subjects and is as small as ICBHI2017 (920 recordings/5.5 h/128 subjects). We used the inter-patient test split as in \cite{moummad2022icbhi}.
We set the background noise ratio to $0.01$ for pre-training, the learning rate to $5\times 10^{-6}$, and the training epochs to 50 for fine-tuning. Other experimental details are the same as in the experiment using ICBHI2017.

As shown in Fig. \ref{fig:exp-sprs:progress}, we observe behavior similar to that in the ICBHI2017 experiment.
When background noise is not used ($\eta=0$), the performance deteriorates as progress proceeds, but it improves when the noise is used ($\eta=0.01$).
Compared to the baseline, Moummad et al.\cite{moummad2022icbhi}, our results are higher than their result of $85.59$ with the highest result of 89.77.
In summary, we confirmed that our models also achieve top-level performance on SPRSound, and M2D-X benefits from the background noise for successful further pre-training on a small dataset.

\section{Conclusion}
%本論文はマスク予測の自己教師あり学習を改善する手法M2Dと、このM2Dを拡張して任意の音のタスクに事前学習方法を提供することを目指すM2D-Xを提案した。
This paper proposed M2D, a masked prediction-based SSL, and M2D-X, extending M2D to a universal audio pre-training framework.
%M2Dは従来手法と異なり、マスクパッチだけを利用して教師信号を得ることで性能を向上した。M2D-XはM2Dを拡張して追加学習タスクと背景ノイズデータを追加し、データが小さい場合でもアプリケーション専用表現の事前学習を可能にした。
M2D improved performance by using only masked patches to obtain training signals, unlike previous methods that use all patches.
M2D-X extends M2D by adding an extra learning task and background noise data, enabling pre-training of representations specialized for an application even when data size is small and the distribution is different from that public audio datasets, as typically found in private and proprietary cases.

Our experiments confirmed that M2D and M2D-X learn effective representations, including a general-purpose audio representation, specialized representations for the highly competitive AudioSet and speech domain, and a specialized representation for a medical application with small data. These representations all achieved top-level performance, demonstrating their potential for use as a universal framework to provide effective audio pre-training for diverse applications.

We found that encoding only the masked parts of the input to obtain a training signal improves the average performance of the masked prediction-based SSL, that a multi-task of supervised learning and M2D SSL enhances specialized performance on top of the general-purpose performance, and that M2D-X configured for regularization using background noise enables successful further pre-training on small data.

%我々は少データ環境の実験を一枚の24GB GPUだけで行う実用的な例として提示した。公開するコードとAudioSet事前学習済みモデルが、将来の様々な音のタスクで事前学習を可能にし、性能の向上に寄与することを望む。
As a practical example, we presented small-data medical application experiments using only a single GPU. We hope that the published code and AudioSet pre-trained models, as well as the example, will enable pre-training and improve performance in future audio studies\footnote{Codes, pre-trained weights, and examples are available at:\\ \url{https://github.com/nttcslab/m2d}}.
%将来の課題として、設定やデータの選択、学習後のモデルの評価に指針を与える研究が挙げられる。特に検証データを使った下流タスク性能推定は新しい指針を与える可能性がある。
Future work may include studies that guide the data selection, post-training model selection, and hyperparameter settings, such as downstream task performance estimation using validation loss\cite{SimShamCollapse,Xie2023OnDataScalingMasked}.

\appendices

%%%%%%%%%%%%%%%%%%%%%%%%%%%%%%%%%%%%%%%%%%%%%%%%%%%%
\section{Experiment with Images}
%%%%%%%%%%%%%%%%%%%%%%%%%%%%%%%%%%%%%%%%%%%%%%%%%%%%

%M2Dは2D構造データ一般を入力できるため、ImageNetを使った標準的な性能評価により、画像に対するM2Dの有効性を検証する。
%%ImageNet-1Kを使って事前学習を行い、その後fine-tuningを行った。我々はMAE同様に画像サイズ224x224のtop-1 accuracyを報告する。
Since M2D accepts general 2D structural data input, we validated the effectiveness of M2D for images through a standard performance evaluation using ImageNet\cite{ImageNet}.
We pre-trained on ImageNet-1K, followed by fine-tuning. The results showed top-1 validation accuracy for a single crop image of $224 \times 224$, the same as in the MAE\cite{he2022masked}.

%この実験においても4章同様にMAEのコードをベースとして最小限の変更にとどめることで、実験対象のみの差分の比較を可能にした。バックボーンにも4章同様ViT-Baseを利用した。
%事前学習のパラメータはエポック数300、バッチサイズ2048、とした他はマスク率0.75を含めMAEと同じ設定とした。ターゲットネットワーク更新のEMAパラメータTAUも音響信号での実験同様、学習開始時0.99995から終了時0.99999まで線形補間した。
%fine-tuningはMAEのソースコードをそのまま利用しパラメータも同一に設定した。
We used the MAE code as a base, as in Section \ref{sec:exp:basic}, and made minimal changes to it, allowing comparison of differences only in the experimental subjects of interest. We also used ViT-Base\cite{ViT} for the backbone.
For pre-training, we set the number of epochs as 300 and the batch size as 2048. All other settings were the same as in the MAE, including the masking ratio of 0.75. 
The EMA decay rate $\tau$ was the same as in Section \ref{sec:exp:basic}.
%The EMA decay rate $\tau$ for the target network update was the same as in Section \ref{sec:exp:basic}, linearly interpolated from 0.99995 at the start of training to 0.99999 at the end.
For fine-tuning, we used the same code and parameters as in the MAE.

%Table 6は、MAE、M2D、M2Dのターゲットエンコーダに全てのパッチを入力した場合、3モデルの結果を示す。従来法であるターゲットに全てのパッチを入力するM2D variantではMAE同等の性能であり、我々のM2Dはこれらの性能を上回ることを確認した。これらの結果は、提案手法が画像においても有効であることを検証するものである。
We compared three models, MAE, M2D, and an M2D variant, that feed all patches to the target encoder. Table \ref{tab:results-imagenet} shows the results. The results show that the performance of the M2D variant with all patches input to the target, which is conventional, is comparable to MAE, and that our M2D outperforms these methods, validating that our method is also effective on images in addition to audio.

\begin{table}[t!]
%\vspace{-5pt}
\caption{Fine-tuning results on ImageNet-1K.}
\label{tab:results-imagenet}
\vspace{-5pt}
\centering
\resizebox{\columnwidth}{!}{%
\begin{tabular}{llll}
\toprule
Model & Target & Target & Top-1 \\
\vspace{-1pt}      & encoder &  input   &    acc(\%) \\
\midrule
MAE &  & - & 83.22 {\fontsize{6pt}{6pt}\selectfont $\pm$ 0.024}\\
%\addlinespace[-0.02cm] \hdashline \addlinespace[0.05cm]
M2D variant (conventional) & \checkmark  & All patches & 83.22 {\fontsize{6pt}{6pt}\selectfont $\pm$ 0.085}\\
M2D (ours) & \checkmark & Masked patches only & \textbf{83.35} {\fontsize{6pt}{6pt}\selectfont $\pm$ 0.211}\\
\bottomrule\\
\end{tabular}
}
\vspace{-5pt}
\end{table}

%%%%%%%%%%%%%%%%%%%%%%%%%%%%%%%%%%%%%%%%%%%%%%%%%%%%
\section{Computational Resources for Pre-training}\label{appendix:resources}
%%%%%%%%%%%%%%%%%%%%%%%%%%%%%%%%%%%%%%%%%%%%%%%%%%%%
%本章では各実験の事前学習に要した要求計算リソースを示す。全ての事前学習の有効バッチサイズは2048、追加事前学習では128である。
We show the details of the required computational resources for the pre-training for each experiment in Table \ref{tab:appendix:resource}. The effective batch size for all pre-training is 2048, and it is 128 for additional pre-training.

\begin{table}[t!]
%\vspace{-5pt}
\caption{Required resources for pre-training or further pre-training}
\label{tab:appendix:resource}
%\vspace{-5pt}
\centering
\resizebox{1.0\columnwidth}{!}{%
\begin{tabular}{cccccccc}
\toprule
\multicolumn{6}{c}{\textit{Training type and settings}} & \multicolumn{2}{c}{\textit{Resources}}\\
 \cmidrule(lr){1-6} \cmidrule(lr){7-8}
PT$^\ast$ & Input & Patch & Epochs & Batch & Grad. & Training & GPU \\
/Fur-PT & dur. & size   &      & size & accum. & time (h) & usage\\
\midrule
\multicolumn{8}{l}{\textit{Section \ref{sec:eval-general} M2D for General-purpose}}\\
PT & 6 s & $16\times16$ & 300 & 512 & 1 & 55 & $4\times$A100 80GB \\
%\multicolumn{5}{l}{\textit{Sec. \ref{sec:eval-general} General-purpose audio representation: Linear evaluation}}\\
%\multicolumn{5}{l}{\textit{Sec. \ref{sec:eval-general} General-purpose audio representation: Fine-tuning}}\\
\addlinespace[0.05cm]
\multicolumn{8}{l}{\textit{Section \ref{sec:eval-audioset} M2D-X for AudioSet}}\\
PT & 6 s & $16\times16$ & 300 & 512 & 1 & 55 & $4\times$A100 80GB \\
\addlinespace[0.05cm]
\multicolumn{8}{l}{\textit{Section \ref{sec:eval-speech} M2D-X for Speech}}\\
PT & 4 s & $80\times2$ & 1000 & 512 & 1 & 56 & $4\times$A100 80GB \\
PT & 5 s & $80\times2$ & 1000 & 256 & 2 & 65 & $4\times$A100 80GB \\
PT & 6 s & $80\times2$ & 1000 & 256 & 2 & 77 & $4\times$A100 80GB \\
\addlinespace[0.05cm]
\multicolumn{8}{l}{\textit{Section \ref{sec:eval-icbhi} M2D-X for ICBHI2017}}\\
PT & 2 s & $16\times4$ & 300 & 512 & 1 & 109 & $4\times$A100 80GB \\
Fur-PT & 2 s & $16\times4$ & 600 & 64 & 2 & 7 & $1\times$RTX 3090 Ti \\
% Online encoder & xx\\
% (Online) Predictor & xx\\
% Target encoder & xx\\
% Total incl. pos. embs. and mask token & xx\\
% Training time w/ 4 A100 GPUs & 2.5 days\\
% \addlinespace[0.05cm]
% \multicolumn{2}{l}{\textit{M2D-X for distillation in Section \ref{sec:eval-speech}}}\\
% M2D & xx\\
% Online mapper & xx\\
% Offline encoder & xx\\
% Total incl. pos. embs. and mask token & xx\\
% Training time w/ 4 A100 GPUs & 4 days\\
% \addlinespace[0.05cm]
% \multicolumn{2}{l}{\textit{M2D-X for further pre-training in Section \ref{sec:eval-icbhi}}}\\
% M2D & xx\\
% Online mapper & xx\\
% Offline encoder (M2D) & xx\\
% Total incl. pos. embs. and mask token & xx\\
% Training time w/ 1 RTX 3090 Ti GPU & 7 hours\\
% \addlinespace[0.05cm]
\bottomrule
\addlinespace[0.05cm]
\multicolumn{8}{l}{$^{\ast}$ Pre-training (PT) or further pre-training (Fur-PT).}\\
\end{tabular}
}
\vspace{-5pt}
\end{table}

\bibliographystyle{IEEEtran}
\bibliography{refs}

\vspace{-20pt}
\begin{IEEEbiography}[{\includegraphics[bb=0 0 640 706,width=1in,height=1.25in,clip,keepaspectratio]{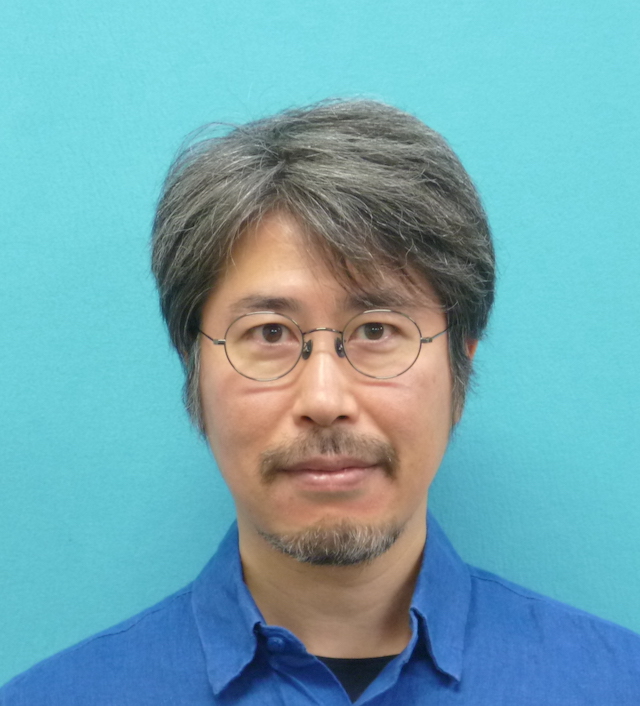}}]{Daisuke Niizumi}
Daisuke Niizumi received the B.S. and M.S. degrees from the Department of Computer Science and Systems Engineering of the Kyushu Institute of Technology in 1995 and 1997, respectively. From 1997 to 2020, he was a senior software and machine learning engineer/manager at several consumer electronics companies. He joined NTT Corporation in 2020. His research interests include representation learning, self-supervised learning, and multimodal deep learning.
\end{IEEEbiography}

\vspace{-20pt}
\begin{IEEEbiography}[{\includegraphics[bb=0 0 65 83,width=1in,height=1.25in,clip,keepaspectratio]{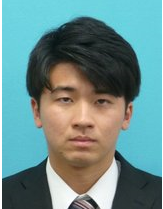}}]{Daiki Takeuchi}
Daiki Takeuchi received the B.S. and M.S. degrees from the Department of Intermedia Art and Science of Waseda University in 2018 and 2020, respectively.  He joined NTT Corporation in 2020. His research interests include signal processing and machine learning for multimodal information processing related to audio.
\end{IEEEbiography}

\vspace{-20pt}
\begin{IEEEbiography}[{\includegraphics[bb=0 0 602 741,width=1in,height=1.25in,clip,keepaspectratio]{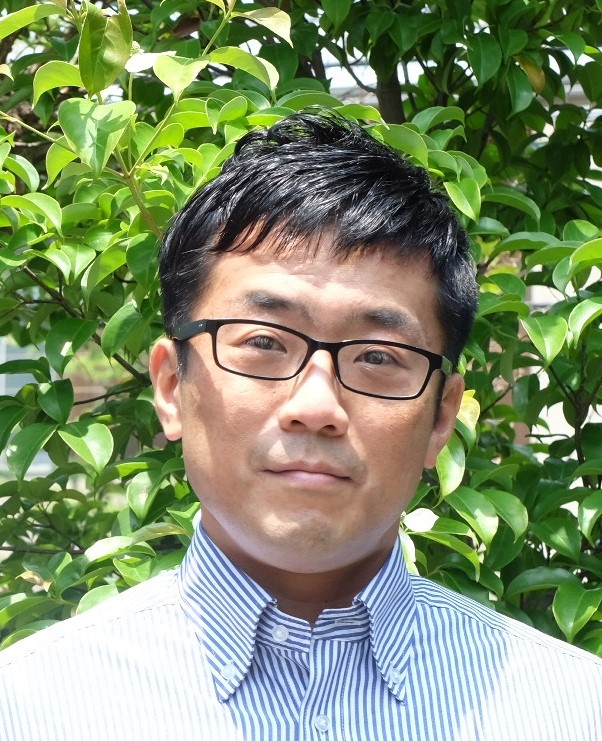}}]{Yasunori Ohishi}
Yasunori Ohishi (M '12) received the Ph.D. degree from Nagoya University in 2009. Since joining NTT in 2009, he has been researching speech and audio signal processing. His research interests generally concern audio event detection, music information retrieval, and crossmodal learning with audio applications. 
\end{IEEEbiography}

\vspace{-20pt}
\begin{IEEEbiography}[{\includegraphics[bb=0 0 224 289,width=1in,height=1.25in,clip,keepaspectratio]{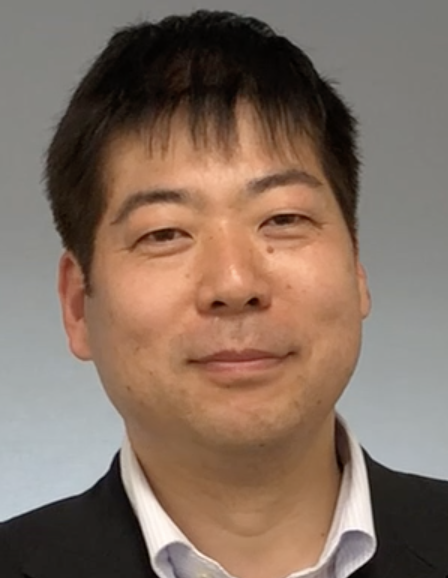}}]{Noboru Harada} Noboru Harada (M'99--SM'18) received the B.S. and M.S. degrees in computer science and systems engineering from the Kyushu Institute of Technology, Fukuoka, Japan, in 1995 and 1997, respectively, and the Ph.D. degree in computer science from the University of Tsukuba, Ibaraki, Japan, in 2017. He is currently a Senior Distinguished Researcher at Nippon Telegraph and Telephone Corporation and a visiting professor at Hokkaido University, Japan. Since joining NTT Corporation, Tokyo, Japan, in 1997, he has been involved with research on speech and audio signal processing, such as high-efficiency coding, lossless compression, and acoustic event detection, including anomaly sound detection.
\end{IEEEbiography}
% https://orcid.org/0000-0002-1759-4533

\vspace{-20pt}
\begin{IEEEbiography}[{\includegraphics[bb=0 0 1590 1909,width=1in,height=1.25in,clip,keepaspectratio]{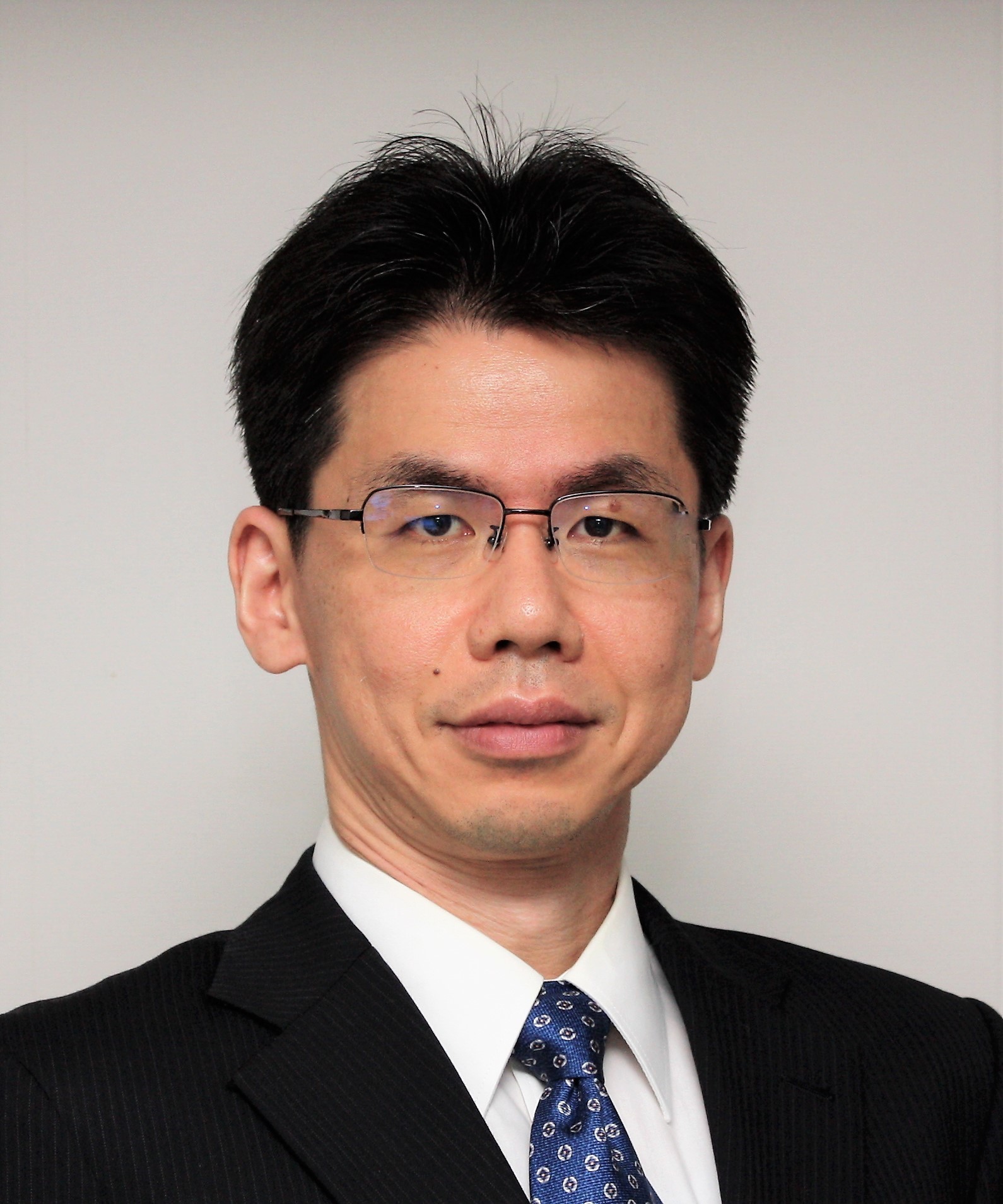}}]{Kunio Kashino}
Kunio Kashino (SM '05) received the Ph.D. degree from the University of Tokyo in 1995. He is currently a Fellow at Nippon Telegraph and Telephone Corporation and a visiting professor at Osaka University and the National Institute of Informatics, Japan. His research interests include audio and crossmodal information processing, media search, and biomedical informatics. He is a member of Association for Computing Machinery (ACM) and a Fellow of the Institute of Electronics, Information and Communication Engineers (IEICE). He received the IEEE Transactions on Multimedia Paper Award in 2004, the Maejima Award in 2010, the Kenjiro Takayanagi Achivement Award in 2016, the IEICE Achievement Award in 2002 and 2017, and the Commendation for Science and Technology by the Minister of Education, Culture, Sports, Science and Technology in 2019.
\end{IEEEbiography}

\vfill

\end{document}